\documentclass[%
 reprint,
 amsmath,amssymb,
 aps,
]{revtex4-2}

\usepackage{graphicx}
\usepackage{dcolumn}
\usepackage{bm}

\usepackage{fetamont}
\usepackage{hyperref}

\begin{document}

\preprint{}

\title{Single and merger soliton dynamics in scalar field dark matter \\with and without self-interactions}

\author{Matthias Stallovits$^{1}$}
\email{matthias.stallovits@gmail.com}
\author{Tanja Rindler-Daller$^{1,2,3}$}%
\email{tanja.rindler-daller@univie.ac.at}
\affiliation{%
	$^{1}$Institut f\"ur Astrophysik, Universit\"atssternwarte Wien,
	Fakult\"at f\"ur Geowissenschaften, Geographie und Astronomie,
	Universit\"at Wien, T\"urkenschanzstr.17, A-1180 Vienna, Austria\\
    $^{2}$Vienna International School of Earth and Space Sciences, Universit\"at Wien, Josef-Holaubek-Platz 2, A-1090 Vienna, Austria\\
    $^{3}$Wolfgang Pauli Institut, Oskar-Morgenstern-Platz 1, A-1090 Vienna, Austria
}%

\date{\today}

\begin{abstract}
Scalar field dark matter (SFDM) made of bosons has become a popular alternative to the cold dark matter (CDM) paradigm, especially for its potential to cure the so-called "small-scale problems" of CDM structure formation. Cosmological simulations have determined that SFDM halos exhibit a core-envelope structure, but they are computationally expensive. Halo cores have been found to be well approximated by "solitons", i.e. equilibrium structures of the underlying equations of motion. The study of single soliton and multiple soliton merger dynamics constitutes a more feasible approach to investigate in detail the genuine quantum dynamics of SFDM and its interplay with self-gravity for a multitude of free boson parameters. In this paper, we present dedicated simulations of single solitons and binary soliton mergers, for models without and with a 2-boson, repulsive, weak to intermediate self-interaction (SI), as well as multiple soliton mergers without SI. We adapt the open-source code {\ffmfamily PyUltraLight} to simulate solitons with SI and make our amended code available to the public. We derive numerical scaling relations between the central density and mass of solitons for several values of SI and find deviations from the monotonic relations known from fuzzy dark matter (no SI), or the strongly repulsive Thomas-Fermi regime. Solitons with SI exemplify larger cores and lower central densities, compared to solitons without SI. Using our simulations, we extract numerical density profiles for solitons and post-merger objects, and fit them to analytical functions used in previous literature. For the cases considered, we find a mild preference for Gaussian cores for objects with SI, while the envelopes of post-mergers can be fit to NFW profiles albeit with caution as we discuss. We find global, persistent oscillations for solitons as well as post-mergers, confirming that self-gravitating SFDM has very long relaxation times, although objects with repulsive SI exhibit oscillations of comparatively smaller amplitude. 

\end{abstract}

\maketitle


\section{\label{sec:sec1}Introduction}

The nature of dark matter (DM) remains one of the most pressing open problems in astronomy and physics.
Many particles beyond the established standard model of particle physics have been conceived as possible DM candidates. The current cold dark matter (CDM) paradigm invokes weakly-interacting massive particles (WIMPs), which are (much) heavier than the proton, or the very light quantum chromodynamics (QCD) axion with a mass of $m \sim 10^{-6}-10^{-5}$ eV/c$^2$; see e.g. \cite{Feng2010} for a review. In either case, the DM dynamics on galactic and cosmological scales has been numerically modelled by performing simulations of the N-body gravitational dynamics of minimum-size objects, which assemble structure from galaxies to the cosmic web hierarchically over cosmic time. While the minimum size of objects from Jeans instability within static or expanding backgrounds is determined by the concrete DM particle physics model, in practice, the minimum size used in simulations is dictated by limitations on resolution and computation times. As a result, that minimum size of objects in simulations is typically much larger than the physics-based Jeans scale (roughly one order of magnitude even smaller than the Jeans scale is the minimum size of objects in virial or hydrostatic equilibrium in a given DM model). This limitation is one reason why the interpretation of CDM simulations and their comparison with observations (before and after the addition of baryons) remains a challenge. Discrepancies have been reported between CDM simulations and observations of DM-dominated dwarf galaxies, which go under the header of "small-scale problems" for their appearance at scales of a few hundred parsecs to a few kpcs, see e.g. \cite{Bullock2017, Popolo_SmallScaleProblems2017} for more details. In particular, rotation curves of DM-dominated dwarf galaxies often hint towards (nearly-)constant central DM density "cores", rather than "cusps", while CDM universally predicts cuspy DM density centers, with the density going as $\sim r^{-1}$ for $r\to 0$, within virialized objects from dwarf galaxies to galaxy clusters. This issue has been called "cusp-core-problem" of CDM.
It remains unclear whether (the proper simulation and physical feasibility of) baryon feedback at all scales is able to "explain away" the empirically found discrepancy, within the CDM paradigm, but in light of several issues and discrepancies, along with the hitherto non-detections of above CDM particle candidates, the community has been considering alternative DM particles, altogether. In order to explain the small-scale problems purely from the DM side, a larger Jeans or characteristic scale, below which structure formation is suppressed, is required. One family of such models comprises scalar field dark matter ("SFDM"), which consists of (ultra-)light bosons in a Bose-Einstein-condensed many-particle state. This type of model has been considered since the $\sim 2000$s using many names: if the ultralight bosons have no self-interaction (SI), they have been known as fuzzy dark matter ("FDM") or (quantum-) wave DM, see \cite{Sin1994, HuFDM_2000, MatosSFDM_2000, Schive_Ursprung, Schive_CE, Hui2017, Lin_etal_ViralizedWaveDMHalos2017}, although the term "BECDM" has been also commonly used, see e.g.  \cite{Mocz_Solitonen2017, Mocz2020}. If an (attractive) axion-like SI is included with a much smaller boson mass than the QCD axion, the model is also called ultralight axion DM ("ULAs"), see e.g. \cite{MarshULDM_2010, HlozekMarsh2015}. If a strongly repulsive SI is considered, instead, the model has been also called "SFDM-TF" in \cite{Dawoodbhoy2021, Shapiro2022} or "SIBEC-DM" in \cite{Hartman_JCAP2022, Hartman2022}, to highlight the "Thomas-Fermi (TF) regime" or the BEC nature, respectively, although other terms for this model have been used by \cite{Peebles_2000}, \cite{GOODMAN2000}, \cite{BoehmerBEC_DM_2007}, \cite{Chavanis2011}, or \cite{TRDVortex2012}.  
 For models without SI, the characteristic scale is given by the de Broglie length of the bosons: \textit{if} a galactic minimum scale of $\sim 1$ kpc is desired, the bosons of SFDM have to have a tiny mass of $m \sim 10^{-23}-10^{-22}$ eV/c$^2$, i.e. SFDM is often associated with ultralight bosons. However, at the other extreme of the TF regime, the characteristic scale depends upon the parameter combination $g/m^2$, where $g$ denotes the SI coupling strength to be introduced below. In this regime, a minimum scale of order $1$ kpc is feasible for much higher boson masses $m\gg 10^{-22}$ eV/c$^2$. Overall, the potential of SFDM to make a cure to the CDM "small-scale problems", such as the "cusp-core problem", has been a motivation to consider the astrophysical implications of this family of models, using various theoretical and computational methods. For one thing, SFDM predicts cored DM halo centers, and the "solitons" of our focus in this paper, described shortly, serve as proxies for such SFDM halo cores.
 
 Generally, the equations of motion of SFDM are a coupled set of Klein-Gordon-Einstein equations which reduce in the Newtonian regime to a nonlinear Schrödinger equation (NLSE) or Gross-Pitaevskii (GP) equation, coupled to the Poisson (P) equation; we will refer to this set of differential equations as the GPP system. The Newtonian regime is appropriate for the study of single galactic SFDM halos (except in the vicinity of, say, central supermassive black holes), which are of interest to us for our study here.
 
Although the characteristic scales in SFDM can be much larger than in the CDM case, depending upon the SI regime and the choice of $m$, $g$ or their combination $g/m^2$, those scales are still small in the context of fully-fledged cosmological simulations. Therefore, issues related to resolution and computation time have not disappeared. In fact, in the TF regime of strongly repulsive SI, the SI scale given by $g/m^2$ is much larger than the de Broglie scale. For computational efficiency, any dynamics at that latter scale needs to be coarse-grained, typically resulting in fluid approximations of the fundamental equations of SFDM, or rather SFDM-TF; see \cite{Dawoodbhoy2021, Shapiro2022} for the theoretical and numerical foundational work, and follow-up studies in \cite{Hartman2022, Foidl2023}. 
While this approach in the TF regime is appropriate to study halo formation and evolution at the level of gross features, the addition of baryons represents a difficulty not yet pursued in the literature. On the other hand, in the FDM regime of large de Broglie lengths, more simulations have been performed to date, although cosmological simulations suffer the same issues of resolution and computational feasibility. The work by \cite{Mocz2020} is a rare example, where FDM structure formation simulations with baryons down to a redshift of $z \sim 5$ have been presented. The overwhelming number of studies have focused on FDM-only simulations, either in a cosmological setting, see e.g. \cite{Schive_Ursprung, Schive_CE, Mocz_Solitonen2017, NoriAndBaldi_AXGadget2018, SpringelMay2021, Schwabe2020} and e.g. the review by \cite{Zhang2019_CosmSims-Review}, or within static backgrounds by merging single so-called "solitons" to create larger halo-like structures, see e.g. \cite{Schwabe2016} and \cite{Hui_2021} and more references below. Solitons refer to minimum-energy objects in (near-)equilibrium, which occur generically in SFDM and which will be the focus of our work; for further studies see also e.g. \cite{ChavanisDelfini2011, Hui2017, Shukla2024, zhang2024}.
Given the fact that cosmological simulations of SFDM (with or without baryons) are extremely computationally expensive, the approach of studying gravitational dynamics at the level of single solitons remains an important means of investigation. Many details in the unfolding dynamics of gravitationally bound SFDM structures are not yet understood, and the few existing sophisticated simulations often lack the resolution to draw conclusions with respect to the full spatial range of those objects, let alone to probe the SFDM parameter space in a thorough way. Therefore, dedicated "small-scale" simulations are required to study various dynamical aspects in more detail, or probe a larger range of parameter space. 
Indeed, it is also important to understand the dynamics of individual solitons, given that simulations have found that they serve as approximate models for cores or central parts of evolving SFDM galactic halos. By the same token, mergers of solitons have been studied to investigate how structure formation at small scales may proceed. In general, merger simulations are computationally costly, and even in static backgrounds issues and open questions remain, as we will discuss. Thus, the work in this paper is motivated by all these considerations and shares its scope with similar previous investigations in the literature, where dedicated "small-scale" simulations are used to help shedding light on the dynamics of SFDM solitons. We will present our simulations of single soliton dynamics, as well as binary and multiple soliton merger dynamics in the forthcoming sections, and expand upon previous literature in several regards, notably the study of the density profiles of simulated solitons, and the inclusion of models with weakly repulsive SI.
Yet, this paper shall be considered explorative in nature: we are not concerned with model fitting to observational galactic data at this point. Instead, we hope to enhance our theoretical understanding of SFDM halos and halo cores to inform for the future more sophisticated procedures of fitting models to observational data. 

Finally, we think that our analysis here, as well as similar ones in the literature, albeit entirely theoretical in their approach, will also contribute towards a better understanding of \textit{self-gravitating quantum systems}, in general. Laboratory experiments are currently attempting to investigate the connection between quantum mechanics and gravity, but are limited to the environment of the Earth gravitational field (either co-moving on ground labs, or in free fall in space labs). On the other hand, theoretical investigations of the dynamics of self-gravitating solitons under various SI and gravity regimes enable to probe physical regimes, which are unattainable in the laboratory. The study of SFDM solitons has this further motivation, apart from the DM problem.        

This paper is organized as follows.
In Section \ref{sec:sec2}, we collect all the fundamental equations and the numerical routines, which are used to solve the equations of motion of soliton dynamics. It also includes a subsection on analytical density profiles for SFDM from the literature, which we will use to fit our density data. Another subsection is reserved for the code implementation. We use and modify the open-source code {\ffmfamily PyUltraLight} \footnote{The code is available at \url{https://github.com/auckland-cosmo/PyUltraLight}.} by \cite{PyUltraLight}, in order to include models with SI as well as to enable a fitting routine on the density data of our simulations. Section \ref{sec:sec3} presents our results of single-soliton dynamics, for models without and with SI, focusing on long-time behavior, density profiles and the relationship between central density and mass of solitons.
As of Section \ref{sec:sec4}, we move to merger simulations, where we first present binary mergers in that section, again, without and with SI included. We discuss various simulation settings and density profiles of post-merger objects. In Section \ref{sec:sec5}, we present results on multiple soliton mergers without SI, where we discuss long-time behavior, the impact of the initial configurations, and the issues that must be overcome, if SI is included. Finally, we summarize and discuss our results, and compare them with relevant previous literature in Section \ref{sec:sec6}. A discussion of further technical aspects, concerning our simulations and configurations, are deferred into three appendices.

\section{\label{sec:sec2}Fundamental equations and numerical routines}

\subsection{Equations of motion}

In the nonrelativistic regime, the dynamics of galactic halos or halo cores composed of a single species of ultralight bosonic SFDM is described by a single wave function, which obeys the coupled Schrödinger-Poisson (SP) equations, if the bosons of mass $m$ have no self-interaction (see e.g. the early account in \cite{WidrowSPS_1993} for a different motivation). The wave function is a complex variable $\psi(\Vec{x},t)$ of space and time, whose modulus squared describes the local probability (number) density of the SFDM particles; upon multiplication by $m$, we can interpret 
\begin{equation}\label{eq:rhoVONpsi}
    \rho(\Vec{x},t) = m |\psi(\Vec{x},t)|^2,
\end{equation}
as the corresponding local mass density of SFDM. If a particle self-interaction (SI) is included, the coupled system of differential equations has been also known as the Gross-Pitaevskii-Poisson (GPP) system. The Gross-Pitaevskii (GP) equation in (\ref{eq:Gross-Pitaevskii}) is essentially a "mean-field" equation for the field $\psi$ (see e.g. \cite{PitaevskiiGPPS_Book}), while the Poisson equation in (\ref{eq:Poisson}) has the usual form, determining the gravitational potential of the object self-consistently via its mass density as follows,
\begin{eqnarray}
    i \hbar \frac{\partial \psi(\Vec{x},t)}{\partial t} = - \frac{\hbar^2}{2m}\Delta \psi(\Vec{x},t) &&+ m \Phi(\Vec{x},t) \psi(\Vec{x},t) \nonumber \\ &&
    + g |\psi(\Vec{x},t)|^2 \psi(\Vec{x},t), \label{eq:Gross-Pitaevskii}
\end{eqnarray}
\begin{equation}\label{eq:Poisson}
    \Delta \Phi(\Vec{x},t) = 4 \pi G m |\psi(\Vec{x},t)|^2,
\end{equation}
where $\Phi$ is the gravitational potential of the object, and $G$ is the gravitational constant.
$\psi$ is supposed to represent the complex wave function of $N$ dark matter particles, having the same mass $m$ and being in the same state, which make up a gravitationally bound object of volume $V$, such that 
\begin{equation}\label{eq:norm}
    \int_V |\psi(\Vec{x},t)|^2 d^3\Vec{x} = N.
\end{equation}
  The GP equation in (\ref{eq:Gross-Pitaevskii}) has this form, upon using the standard first Born approximation to describe an essentially contact-like s-wave $2$-particle SI, whose strength is characterized by the constant self-coupling strength $g$. It is related to the s-wave scattering length $a_s$ via
\begin{equation}
    g = 4 \pi \hbar^2 \frac{a_s}{m}.
\end{equation}
The SI is of repulsive nature for $g > 0$, and attractive for $g<0$. Models without SI, for which $g=0$, have been the focus of the overwhelming part of the astrophysics literature on SFDM. Our paper is a rare exception, in that we will simulate the gravitational dynamics in the wave approach, i.e. by directly solving GPP (\ref{eq:Gross-Pitaevskii}-\ref{eq:Poisson}), for SFDM models with repulsive SI. We focus on zero and repulsive SI, because these models generically support gravitationally stable objects.

\subsection{Rescalings and code units}

Before the equations of motion are solved, it is recommended to write them in dimensionless units. We will use and modify for our purpose the open-source code {\ffmfamily PyUltraLight} by \cite{PyUltraLight}, hence we adopt its code units, as follows.  
To write the GPP system in dimensionless form, three scales for length (\ref{eq:lengthCU}), time (\ref{eq:timeCU}) and mass (\ref{eq:massCU}) are needed. These originate from \cite{Schive_CE} and are commonly used for SFDM (further examples are \cite{PAREDES2016, PyUltraLight, PySiUltraLight}). The code units are defined as
\begin{eqnarray}\label{eq:lengthCU}
    \mathcal{L} &&= \left(\frac{8 \pi \hbar^2}{3 m^2 H_0^2 \Omega_{m0}}\right)^{\frac{1}{4}} \approx 121 \cdot \left(\frac{10^{-23}\,\text{eV/c}^2}{m}\right)^{\frac{1}{2}}\,\text{kpc, } \\
    \label{eq:timeCU}
    \mathcal{T} &&= \left(\frac{8\pi}{3H_0^2 \Omega_{m0}}\right)^{\frac{1}{2}} \approx 75,5\,\text{Gyr and }
\end{eqnarray}
\begin{eqnarray}
    \mathcal{M} = \frac{1}{G}&&\left(\frac{8\pi}{3H_0^2 \Omega_{m0}}\right)^{-\frac{1}{4}} \left(\frac{\hbar}{m}\right)^{\frac{3}{2}} \nonumber \\
    \label{eq:massCU} &&\approx 7 \cdot 10^7 \cdot \left(\frac{10^{-23}\,\text{eV/c}^2}{m}\right)^{\frac{3}{2}}\,\text{M}_{\odot}.
\end{eqnarray}
Here, $m$ is again the boson mass, $G$ the gravitational constant, $H_0$ the present-day Hubble constant, $\hbar$ the reduced Planck's constant and $\Omega_{m0}$ the present-day matter density parameter. Constants and scaling parameters are shown in Table \ref{tab:Constants}. In fact, the global mass and length scales depend upon the particle mass, but since we will fix this mass to $m=10^{-22}$\,eV/c$^2$ in our simulations, these quantities are fixed, as well. We choose this particle mass in order to compare our results with previous ones in the literature. 
\begin{table}[b]
\caption{\label{tab:Constants} Constants and parameters. The global length and mass scale in code units (marked with a star) were calculated for a boson mass of $10^{-22}$\,eV/c$^2$.}    
\begin{ruledtabular}
    \begin{tabular}{ccc}
        Parameter & Value & Unit \\ \hline
        $\hbar$ & $1.0545718 \cdot 10^{-34}$ & Js \\
        $H_0$ & $67.7$ & $\frac{\text{km/s}}{\text{Mpc}}$ \\
        $\Omega_{m0}$ & $0.31$ & 1 \\
        $G$ & $6.67\cdot 10^{-11}$ & $\frac{\text{m}^3}{\text{kg s}^2}$ \\
        1\,pc & $3.0857\cdot 10^{16}$ & m \\
        $\text{M}_{\odot}$ & $1.989\cdot 10^{30}$ & kg \\
        $1$\,length-code unit$^\star$ & $\sim 38.26$ & kpc \\
        $1$\,mass-code unit$^\star$ & $\sim 2213594$ & $M_\odot$
    \end{tabular}
\end{ruledtabular}
\end{table}

Additionally to these re-scalings, we introduce a dimensionless self-coupling strength $\hat{\Lambda}$, which depends on the dimensional self-coupling strength $g$, according to
\begin{equation}
    \hat{\Lambda} = \frac{c^2 g}{4 \pi G \hbar^2}.
\end{equation}
Here, $c$ is the speed of light in vacuum. The derivation of this dimensionless self-coupling strength can be found e.g. in \cite{Padilla2021}, from which we also took the notation (except for the $4\pi$ difference in the Poisson equation). The dimensionless GPP equations read thus as
\begin{equation}\label{eq:DimlessGrossPitaevskii}
    i \frac{\hat{\partial} \hat{\psi}(\hat{\Vec{x}},\hat{t})}{\hat{\partial} \hat{t}} = - \frac{1}{2}\hat{\Delta} \hat\psi(\hat{\Vec{x}},\hat{t}) +  \hat{\Phi}(\hat{\Vec{x}},\hat{t}) \hat{\psi}(\hat{\Vec{x}},\hat{t}) + \hat{\Lambda} |\hat{\psi}(\hat{\Vec{x}},\hat{t})|^2 \hat{\psi}(\hat{\Vec{x}},\hat{t})
\end{equation}
\begin{equation}\label{eq:DimlessPoisson}
    \hat{\Delta} \hat{\Phi}(\hat{\Vec{x}},\hat{t}) = 4 \pi |\hat{\psi}(\hat{\Vec{x}},\hat{t})|^2.
\end{equation}
The hats refer to dimensionless quantities, but we will suppress them in the forthcoming for brevity, except for $\hat{\Lambda}$.

\subsection{Analytical density profiles for SFDM solitons and halos}

A main result in our paper will be the comparison of density profiles between our SFDM solitons from simulations and analytical models of the literature, which have been devised based upon various considerations. 
Importantly, it has been established from previous theoretical and simulation work that SFDM halos are characterized by a core-envelope structure, see e.g. \cite{GuzmanUrenaLopez2003, GU2004, TRDShapiro2014, Schive_CE, Mocz_Solitonen2017, Chavanis2019, Padilla2021, Dawoodbhoy2021, Shapiro2022, Hartman2022, Foidl2023}. As such, SFDM has been considered  to resolve several small-scale problems of CDM, notably the "cusp-core-problem" mentioned in the Introduction. It is the pressure contributions of SFDM, either in the form of the quantum pressure (for models without SI), or in the form of a $(n=1)$-polytropic pressure (for models with strongly repulsive SI, i.e. the TF regime) which avoid the central density cusps of CDM halos, and produce density cores, instead.

However, the outer halo parts ("envelope") of simulated SFDM halos have been found, on average, to be close to CDM halos, see e.g. \cite{Mocz2020, Dawoodbhoy2021, Shapiro2022, Foidl2023}. This similarity in the large-scale dynamics of SFDM and CDM halo envelopes was convincingly explained by \cite{Dawoodbhoy2021}. 

Now, the profiles found in the halo central parts differ, depending upon the SI regime of SFDM. As a result, different analytical profiles have been considered in the past.
In the regime of strongly repulsive SI, the TF regime of SFDM (i.e. SFDM-TF or SIBEC-DM), the central density profile is very close to a $(n=1)$-polytrope. Since we will not consider the TF regime in this paper, we refer the reader to the literature, e.g. \cite{GOODMAN2000, TRDVortex2012, Dawoodbhoy2021, Foidl2023}. 

Our focus in this paper are models without SI (such as FDM) and models with a weakly repulsive SI. In these cases, the de Broglie length is the characteristic scale which determines the size of galactic halo cores. In this regime, "solitons" are the minimum-energy solutions to the GPP equations, and they have been found to approximate the cores of fully evolved "core-envelope" halos (see cited literature above). These solitons or "solitonic cores" of SFDM halos can be modelled as coherent standing wave packets. Therefore, a Gaussian profile has been often adopted for simplicity, see e.g. \cite{AttrakSI_Chavanis2016, Schobesberger2021}. After all, the Gaussian profile has been used successfully for models with weak SI, especially as variational ansatz in energy calculations, in various quantum-mechanical systems. Assuming spherical symmetry, the radial density profile of a solitonic halo core as a Gaussian can thus be modelled as
 \begin{equation}\label{eq:Gauß}
    \rho(r) = a e^{-r^2/2\sigma^2},
\end{equation}
  with the central density $a = \rho(0)$ and the width of the Gaussian $\sigma$ (we use the notation "$a$" for the central density to avoid confusion with other models below).

On the other hand, the results of dedicated simulations of FDM (i.e. SFDM without SI) by \cite{Schive_Ursprung} prompted them to devise an "empirical" profile, as a fitting model and approximation to their numerical data of the central profiles of SFDM halos. For simplicity, we call it Schive profile; it is given by 
\begin{equation}\label{eq:Schive}
    \rho_S(r) = \frac{\delta_s}{\left(1+\left(\frac{r}{r_{sol}}\right)^2\right)^8},
\end{equation}
where $\delta_s$ denotes the central density, and $r_{sol}$ describes the width of the profile (the index $"s"$ stands for "Schive" to distinguish it from other models).
For a comparison between Schive and Gaussian profiles, we refer to \cite{MarshPop2015, Chavanis2019, Padilla2021}.
In this paper, we will use the Gaussian and Schive profiles as analytical fitting formulae for the density data of our simulated solitons at rest, as well as for the central parts of post-merger objects. For single solitons, we have found that the two core parameters are related roughly as 
\begin{equation}\label{eq:coreparams}
    r_{sol} \approx 4 \sigma.
\end{equation}
However, we stress that neither of these analytical "core profiles", Gaussian and Schive model, have been tailored to fit solitons in the first place. They should be considered as approximations to SFDM solitons and central halo cores, with all the limitations this statement entails. In fact, in this paper we show that deviations between numerical results and these fitting models exist, and these should be considered as a sign that the physics of solitons is more complicated than can be represented by these approximations, at least for the soliton mass regime considered in this paper.

Post-merger objects result upon colliding and merging of single solitons, and the final core of a post-merger object is expected to retain a solitonic nature, a finding of previous literature which we also confirm, using our simulations.
Our post-merger objects are found to develop a core-envelope structure, also in accordance with similar results in previous literature. In order to fit the envelope of these objects, we use a standard Navarro-Frenk-White (NFW) profile as analytical model. The NFW profile was first introduced in the context of fitting CDM halo density data from gravitational N-body simulations in \cite{NFW1997} and has the following form
\begin{equation}\label{eq:NFW}
    \rho_{NFW}(r) = \frac{\delta_{NFW}}{\frac{r}{r_s}\left(1+\frac{r}{r_s}\right)^2},
\end{equation}
 where $\delta_{NFW}$ is the characteristic density and $r_s$ is the scale radius. While the NFW profile is not adequate for SFDM halo central parts, for its cusp, it has been found to describe the outer parts of such halos in an averaged sense (see cited references above). Therefore, in our investigation, a combination of a core profile plus NFW will be used to fit the density data of our merger simulations. Again, we will see that NFW only serves as an approximation of SFDM envelopes, for the latter turn out to be very dynamical and non-relaxed, even after many Hubble times, for the soliton masses considered in this paper. This finding of long relaxation times is in accordance with previous literature and will be discussed in due course.

\subsection{Implementation}

To solve the GPP system of equations (\ref{eq:DimlessGrossPitaevskii}-\ref{eq:DimlessPoisson}), we use our modified version of the ultralight dark matter solver {\ffmfamily PyUltraLight}. This code uses a symmetric, pseudospectral Fourier-split-step-algorithm, where a discretization of the equations is done by using a simulation box with a $N \times N \times N$-grid. Here, $N$ is the number of points along an axis, which is determined by the resolution (not to be confused with the above number of particles also denoted $N$ - we will not need it in the forthcoming). The equations are solved with periodic boundary conditions, i.e. if matter would exit the box at one side it would reenter on the other side.
The original {\ffmfamily PyUltraLight} code was developed to solve the Schrödinger-Poisson system, i.e. ultralight SFDM without SI. In order to study models with (weak) SI, we modified the original code and created our own version, which we call {\ffmfamily PyUltraLight\_SI}. Indeed, while our work was in progress, the authors of \cite{PySiUltraLight} came up with their own modification of {\ffmfamily PyUltraLight} to include SI, but their code is not public. Therefore, we proceeded to create our own modification {\ffmfamily PyUltraLight\_SI}, which has also some additional features, notably the possibility to fit analytical functions to the numerical density data of the simulations. Our code is available at \footnote{\url{https://github.com/MStallovits/PyUltraLight_SI}}.  A new version of {\ffmfamily PyUltraLight} is also under development, see \footnote{See \url{https://github.com/Sifyrena/PyUL_NBody} and \cite{Wang2022}.}.

\subsubsection{Dynamical evolution}

The calculation of the dynamical evolution is described in \cite{PyUltraLight}, which is also the basis of the modification by \cite{PySiUltraLight}. In that paper, the SI-term is treated as being part of the potential term of the Schrödinger equation, "transforming" it basically into the GP equation. Since the SI can be viewed as another potential term, it can be treated in the pseudo-spectral method like the gravitational potential. Indeed, we came up independently with this very natural approach to include the SI. Therefore, the discretized  evolution can be schematically written as
\begin{eqnarray}
    \psi(\Vec{x},t+h) = &&\exp\left[-\frac{ih}{2}\Phi(\Vec{x},t+h)\right] \nonumber \\
    &&\cdot \exp\left[-\frac{ih \hat{\Lambda}}{2}|\psi(\Vec{x},t+h)|^2\right] \nonumber \\
    \cdot \mathcal{F}^{-1}&&\exp\left[-\frac{ih}{2}k^2\right] \mathcal{F} \exp\left[-\frac{ih}{2}\Phi(\Vec{x},t)\right] \nonumber \\
    &&\cdot \exp\left[-\frac{ih \hat{\Lambda}}{2}|\psi(\Vec{x},t)|^2\right]\psi(\Vec{x},t),
\end{eqnarray}
\begin{equation}
    \Phi(\Vec{x},t+h) = \mathcal{F}^{-1} \left(-\frac{1}{k^2}\right) \mathcal{F} 4 \pi |\psi(\Vec{x}, t_i)|^2.
\end{equation}
Here, $h$ is the time step, $\psi(\Vec{x}, t_i)$ the field configuration at half the full time step $t_i$, $k$ the wave number of the Fourier domain, $\mathcal{F}$ and $\mathcal{F}^{-1}$ denote the discrete Fourier transform and its inverse. The order of the operations goes from right to left. Thus, we start in position space, where a half step with the potential operator is taken. Then, we change to Fourier space, where a full step is taken. Afterwards, the inverse Fourier transform is used, to change back to position space, where the second half step of the potential operator is taken, completing a full iteration. However, our code uses an additional simplification, presented initially in \cite{PyUltraLight}, where the consecutive half steps of two iterations are combined to a single full step. Therefore, only the very first and the very last operation need an actual half step. Schematically this looks like
\begin{eqnarray}
    \psi(t+nh) = &&\exp\left[+\frac{ih}{2}\Phi\right] \exp\left[+\frac{i h}{2}\hat{\Lambda}|\psi(t)|^2\right] \nonumber \\
    &&\cdot \left(\prod^n \exp[-ih\Phi] \exp[-ih\hat{\Lambda}|\psi(t)|^2] \exp\left[\frac{ih}{2}\nabla^2\right]\right) \nonumber \\
    &&\cdot \exp\left[-\frac{ih}{2}\Phi\right] \exp\left[-\frac{i h}{2}\hat{\Lambda}|\psi(t)|^2\right] \psi(t).
\end{eqnarray}
We stress that this is only a schematic representation (there is no Fourier transformation in the equation above). Additionally, we want to point to the different signs in the beginning and the end, which are only needed for this representation. In the code itself, we need to add (not subtract like shown above) an additional half step, due to the way the execution of the algorithm is performed.

\subsubsection{Initial conditions}

Similar to the original {\ffmfamily PyUltraLight} code, simulation runs in {\ffmfamily PyUltraLight\_SI} require an initial soliton profile as an input. 
In {\ffmfamily PyUltraLight\_SI}, an initial soliton profile is loaded, which is stored in an external file.  
Together with the initial conditions for the soliton - including its mass in code units (CU) - specified in the associated Jupyter Notebook, the initial field configuration is created which is then passed on to the solution algorithm. This initial file is produced with an additional script called \textit{soliton\_solution.py}. This script was part of the original {\ffmfamily PyUltraLight} code (see \cite{PyUltraLight} for details), and we adapted it for our needs as follows. 
The initial soliton profile is calculated by imposing spherical symmetry onto the system of equations, and assuming that the gravitational potential and density are time-independent, i.e. by adopting a stationarity ansatz. The resulting ordinary differential equations can be solved with a 4th-order Runge-Kutta solver and the shooting method. We adopt the same ansatz, i.e. we assume
\begin{equation}\label{eq:InitialTrafo}
    \psi(\Vec{x},t) \rightarrow e^{i \beta t} f(r), ~~~ \Phi(\Vec{x},t) \rightarrow \varphi(r),
\end{equation}
where $r=|\Vec{x}|$, $\beta$ is a constant and $f(r)$ is the spatial part of the wave function. Using the Laplace operator in spherical symmetry and denoting the derivative with respect to $r$ with a prime, the dimensionless GP equation of (\ref{eq:DimlessGrossPitaevskii})
transforms to
\begin{equation}
     - \beta f(r) = -\frac{1}{2} f''(r) - \frac{1}{r}f'(r) + f(r) \varphi(r) + \hat{\Lambda}f^3(r).
\end{equation}
Using (\ref{eq:InitialTrafo}) and defining $\Tilde{\varphi}(r):= \varphi(r)+\beta$, the GP equation can be simplified to
\begin{equation}
    f''(r)  = - \frac{2}{r}f'(r) + 2 f(r) \Tilde{\varphi}(r) + 2\hat{\Lambda}f^3(r).
\end{equation}
With respect to the Poisson equation, we have 
\begin{equation}
    \Tilde{\varphi}''(r) = 4 \pi f^2(r) - \frac{2}{r} \Tilde{\varphi}'(r).
\end{equation}
The boundary conditions are as follows. In order to guarantee that the radial wave function is smooth, $f'(r)$ as well as $\Tilde \varphi'$ have to vanish at the origin, while the wave function and the gravitational potential go to $0$ for $r\rightarrow \infty$. In practice, the gravitational potential is varied at the origin until a solution for $f(r)$ is found, that approaches zero at some maximum radius.

This stationary GPP system of ordinary differential equations is solved
by \textit{soliton\_solution.py} \footnote{We note that in the code \textit{soliton\_solution.py} the variables have different names than here in the paper.}. After execution, the initial soliton profile is saved as numpy-file, which is loaded at the start of a simulation run. For details and how the user-specified initial conditions are incorporated, the reader is referred to Section 3.2 of \cite{PyUltraLight}.

Additionally, a text file is produced, which stores information about the calculation. While these are mostly incidental information, the information about \textit{Lambda\_hat} ($\hat{\Lambda}$) and \textit{beta} ($\beta$) are necessary for the main code {\ffmfamily PyUltraLight\_SI}, since they are needed for the evolution of the initial profile. This process of loading the parameters is done automatically, such that the user only has to specify $\hat{\Lambda}$ once. This is an improvement, since $\beta$ depends on $\hat{\Lambda}$ and before our adjustment, these parameters would have to be specified in the main part of the code
\footnote{The dependence of $\beta$ on $\hat{\Lambda}$ is indirect: inclusion of SI changes the overall potential term and hence affects $\beta$, see equation 3.12 in \cite{PyUltraLight}. In practice, the effect on $\beta$ is small: we checked to see that it changes little between $\hat{\Lambda} = 0$ and $\hat{\Lambda} \not= 0$, for the choice of $\hat{\Lambda}$ in this paper.}.

We also note that "high" positive values of $\hat{\Lambda}$ demand a smaller maximum radius for the calculation of the initial soliton profile. The reason is that the wave function profile is shallower at the maximum radius, due to the boundary conditions, and this effect is additionally enhanced, the higher $\hat{\Lambda}$. Without a proper adjustment of this maximum radius, the solver can grind and fail to converge to a solution (see also section \ref{sec:3B_SimSI}).
Finally, we point out that, for all simulations discussed in this paper, we choose periodic boundary conditions and we disregard phase shifts in the wave functions of binary or multiple solitons.

\subsubsection{Extraction and fitting of density profiles}

In order to analyze the density profiles of the simulated single solitons, as well as multiple soliton post-merger objects, a new tool was created in the code, called 
\textit{density\_profile}. This new additional feature of our {\ffmfamily PyUltraLight\_SI} package uses the 3D simulation data to create radially averaged density profiles. It is incorporated into the Jupyter notebook and can be controlled there \footnote{This is different to the boson mass, which has to be specified in the script \textit{density\_profile}.}. It is able to differentiate between single and multiple soliton simulations.

For the specified state of the simulation, the 3D density data from the main simulation is used to create the profile. In order to do so, the code determines the center of the soliton, or of the post-merger object, respectively, by searching for the grid point with the highest density. From this point, the minimum distance to the edge of the box is determined. This is necessary, since the center of a soliton or post-merger could be at any grid point, in principle, which would cause problems in the averaging, when more data is available in a certain direction, if the center of the object were misplaced from the center of the box. By truncating the density data at the maximum distance that is possible in all directions, it can be assured, that there will be the same amount of data in all directions. This maximum radial distance is then binned, depending on the resolution that was used. This creates spherical shells with different number of grid points. For each radial bin, the densities for all possible grid points in such a shell are averaged. As a result, a discrete radially-averaged density profile is created.

Our new package \textit{density\_profile} also offers the possibility of calculating the (cumulative) mass of the soliton or post-merger object from this profile, using two different routines: the trapezoid rule, or the integration with spherical shells. In most cases, the trapezoid rule proved to be better, which is why we recommend to use it for the mass determination. So, the routine can be used to calculate that radius, which contains a certain percent of the interior mass. This percentage can be specified by the user in the notebook. Especially the radius containing $99~\%$ of the soliton mass, denoted $R_{99}$, is a useful indicator for the success of a simulation, because it can provide clues as to whether the box size has been chosen too big or too small, for the problem at hand.

However, the main point of \textit{density\_profile} is the fitting of the discrete density profile \footnote{The discrete density profile is in the end only the raw density data for the fit, which is why will use both terms.} with analytical density models. We use the Python tool \textit{curve\_fit} from \textit{scipy.optimize}, which uses a non-linear least-squares method. As mentioned above, the radial density profile may consist of two different sections: The core and the outer envelope. For single solitons and for the core part of post-merger objects, we allow both the Gaussian model (\ref{eq:Gauß}) and the Schive model (\ref{eq:Schive}). Apart from being convenient analytical models, the Gaussian profile is motivated as a "fiducial" wave packet solution to the underlying differential equations, while the Schive model was found by fitting density profiles to halos from cosmological FDM simulations (see Section II C). Now, \textit{density\_profile} uses both to fit the numerical density data by using a score system to decide which core model fits best. This score system rates the profiles according to the following three criteria:
\begin{itemize}
    \item smallest residual
    \item standard deviation of the parameter for the central density
    \item standard deviation of the parameter which defines the width of the model
\end{itemize}
The fits are compared based on these criteria. That profile which "wins" in two of these three criteria is then used and displayed in our plots. These criteria have been chosen, in order to provide good fits of the central density and the width of the profile, which are important characteristics of solitons.

For a single soliton, it is appropriate to fit its density data using one of the above core profiles. However, for mergers an outer envelope is expected, as a result of acquiring a new equilibrium upon mass infall. The resulting density data require more than a core profile. For lack of a theoretically motivated global analytical model, we use a core and an envelope profile and fit them together; for the envelope we use the NFW model (\ref{eq:NFW}). Again, we stress that our choice of fitting models is based upon their use in previous literature, and in lack of theoretically better motivated profiles. In fact, as we will see, our results suggest that neither of the models used is perfect in fitting the data. 

For now, we have to answer an even more basic question, namely to determine where the core of our post-merger objects ends and where their envelope begins. We tested different approaches, but in the end only one turned out to be useful. To find the connection point, we use the standard deviation $\sigma$ from the fit to the Gaussian (even if the Gaussian is not the "best fit", according to our score system). In fact, we found that the numerical differences between the Schive and Gauss profile within the core region are actually small. In many cases, we found \textit{a posteriori} the relation (\ref{eq:coreparams}) but in the fitting we use as a "boundary" the model in \cite{Schobesberger2021}, where the spherical FDM halo core radius is defined by including 99 \% of the mass of the Gaussian (called the "Gaussian sphere"). Hence, we use the relation
\begin{equation}
    \text{size of soliton core} \approx 2.575829~\sigma,
\end{equation}
and pick that radial point that comes closest to this value as the connection point between core and outer envelope.
To fit the data beyond the connection point, the NFW profile is used.

Our code part \textit{density\_profile} discriminates automatically between simulations which involve one or more solitons. For a single soliton, the code will only use the core models to fit the radial density profile. For post-merger objects, it will determine the connection point and calculate the core as well as the NFW fit.

\section{\label{sec:sec3}Single-Soliton Dynamics}

One aspect of using {\ffmfamily PyUltraLight} is the study of the gravitational dynamics of SFDM under various scenarios, given the flexibility of the code environment. A general theme of interest concerns the long-time behavior of single solitons at rest with zero velocity. These solitons at rest serve as a test bed to study the fundamental nature and stability of ultralight FDM or SFDM objects under self-gravity, without the complications due to gravitational disturbances from other sources. In particular, the impact of different SI regimes (none, weak, strong; attractive vs. repulsive) can be studied in detail. Therefore, one pillar of our paper is the dedicated study of certain aspects of single-soliton stability and dynamics, and the comparison of our results with previous literature. Of course, another reason is the easy environment provided by single solitons, in order to test our modified version {\ffmfamily PyUltraLight\_SI}, before using it for more sophisticated simulations. In order to present our results in a coherent manner, we separate our discussion of simulations of models without and with SI.

\subsection{Simulations without self-interaction}

These models have $\hat{\Lambda}=0$, i.e. cases which traditionally belong to the FDM category. We place solitons at rest in the center of the box and study their evolution for different soliton masses (while the boson mass stays fixed). This way, we are able to investigate how low- and high-mass solitons behave, and how simulation parameters like resolution affect the results.
During our simulations, we encountered various problems which are related to issues of numerical resolution and box size, because these impact the overall spatial resolution of the simulations. We will mention some in passing, but defer details into Appendices \ref{appendix:Problems} and \ref{appendix:BC}.

\subsubsection{General setup}\label{sec:3A1}

During the first test runs with {\ffmfamily PyUltraLight}, we encountered already a problem which is well known in the community, let us call it the "high-mass problem". In FDM, the radius of (approximate) equilibrium objects, such as solitons, goes inversely proportional to their mass \footnote{Such inverse mass-size relationships are common in compact stellar objects as the end stages of stellar evolution.}, i.e. solitons get smaller with higher mass, leading to resolution problems for the central (= smallest) regions of solitons. Effectively, the central density regions of solitons suffer a cutoff, resulting in lower values than expected. The only way to avoid this problem seems to require a suite of dedicated simulations to find the "optimum" parameters, in order to observe the competing requirements between resolution and computation time, which change with changing soliton mass.  Generally, we initialized a soliton with a given mass in the center of the box with zero initial velocities. These solitons at rest were then simulated for $0.013$\,CU ($\sim 1$\,Gyr).
  
The spatial resolution of a simulation is based upon the numerical resolution and the box size. As a compromise, for a soliton of given mass, we increased the spatial resolution by increasing the numerical resolution and decreasing the box size, until the relative change in the central density was around $1$\,\%. Since this procedure would be too much effort for every single case, we performed simulations for representative cases, which are shown in Table \ref{tab:IdealResolBox}. For other masses, we used the "next best" parameters.

\begin{table}[b]
\caption{\label{tab:IdealResolBox} This table shows the lowest numerical resolution (second column) and the biggest possible box length (third column) for the investigated soliton masses, to ensure that we do not run into the "high-mass problem". We note that these parameters do not reproduce exactly the soliton masses as calculated from their density profile (see also main text). For soliton masses in between the listed ones, we always used the next best setup, e.g. for $30$\,CU we used the parameters for $40$\,CU. }
\begin{ruledtabular}
    \begin{tabular}{ccc}
        soliton mass [CU] & lowest resolution & biggest box size [CU] \\ \hline
        $20$ & $256$ & $8$ \\
        $40$ & $320$ & $8$ \\
        $60$ & $384$ & $7$ \\
        $80$ & $384$ & $5$ \\
        $100$ & $384$ & $4$ \\
        $200$ & $384$ & $2$ \\
        $300$ & $320$ & $1$ \\
    \end{tabular}
\end{ruledtabular}    
\end{table}

While this way, the central densities are correctly represented, the soliton masses, as calculated from their density profiles, deviate from the initial input values in the code. The culprit is the box size which needs to be large enough, given the periodic boundary conditions. As a result, the calculated soliton masses tend to be higher than the input numbers. However, the central density as a key quantity of interest to us is not affected by this offset in input vs. calculated soliton mass. 
In Appendix \ref{appendix:Problems}, we discuss a generic procedure, which can be followed to get a good approximation for the central density as well as for the soliton mass, given the competing constraints from resolution and computation time.

\subsubsection{Oscillations at rest}

\begin{figure*}
    \centering
    \includegraphics[scale=0.215]{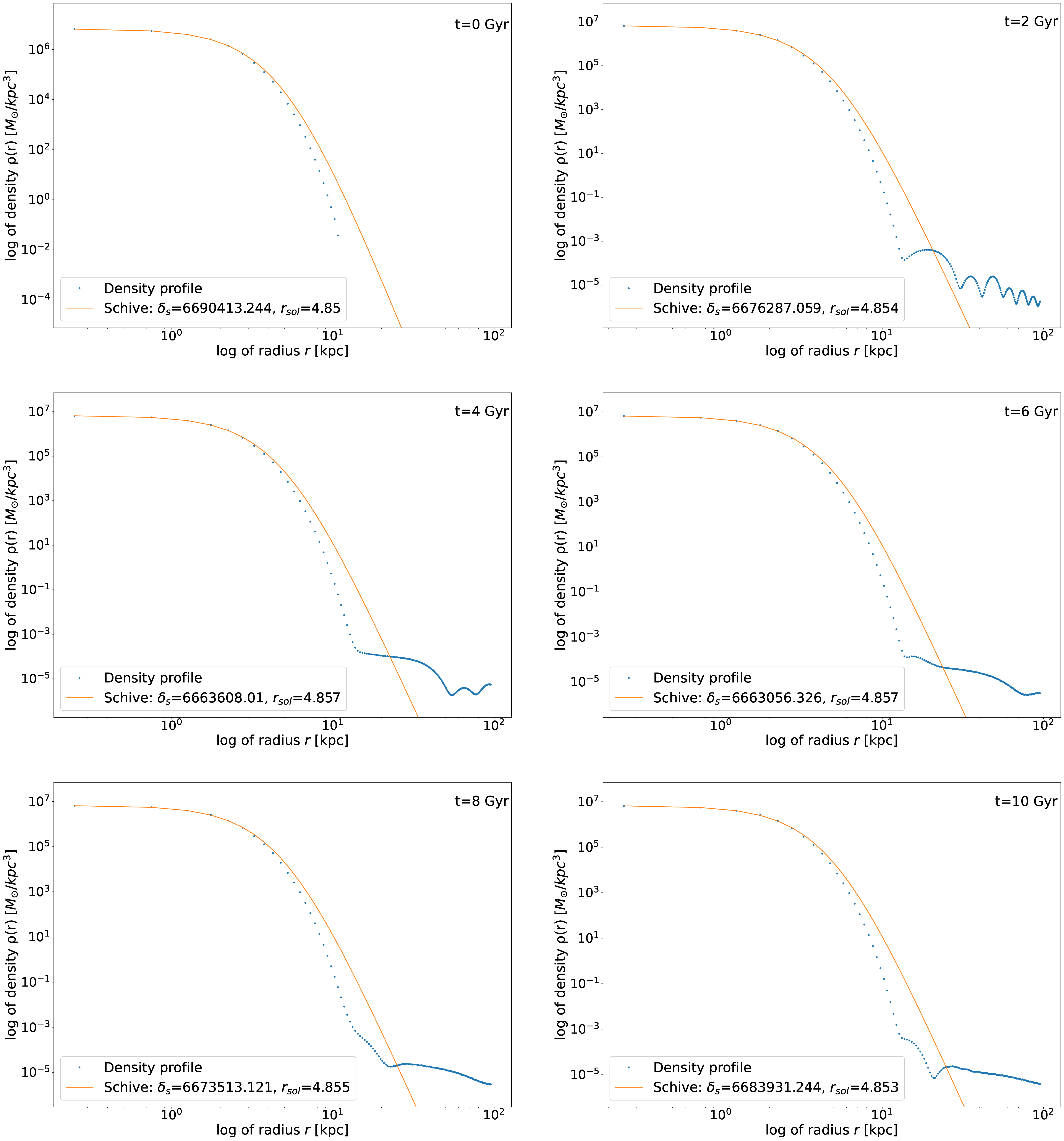}
    \caption{These plots show the density profiles for six selected snapshots (see legend) of a simulation of a $80$\,CU soliton ($\hat \Lambda = 0$), which was initialized at rest. The box size was $5$\,CU and the resolution $384$. The orange, solid curves are fits to the numerical data, according to equation (\ref{eq:Schive}), with units for $\delta_s$ in $M_\odot/\text{kpc}^3$ and $r_{sol}$ in kpc.}
    \label{fig:RhoProfiles_Oscillations}
\end{figure*}

We confirm previous results in the literature, concerning the oscillatory nature of solitons at rest (see e.g. \cite{Veltmaat2018, Hui_2021}), though we recognize it preferentially for higher soliton masses. We observe that solitons start oscillating, expanding and contracting over time. To demonstrate this phenomenon, we present a simulation of a soliton with $80$\,CU ($\sim 2\cdot 10^8\,M_\odot$), which ran for $0.13$\,CU ($\sim 10$\,Gyr) in a $5$\,CU box with a resolution of $384$. We show a selection of density profiles at different snapshots in figure \ref{fig:RhoProfiles_Oscillations}.
We can see how the numerical density profile changes over time, especially in the outer regions. The expansion and contraction can be also noticed based on the change of the fitting parameters of the Schive model, though the changes are small. An initial expansion, which causes the central density to decrease and the width parameter of the Schive model $r_{sol}$ to increase, is followed by a stagnation phase during $\sim 4-6$ Gyr. From $8$ Gyr onward, we see a contraction, causing an increase in the central density and a decreasing $r_{sol}$. At the end of the simulation, the soliton has become slightly broader, but the central density is lower in comparison to the initial snapshot.
The oscillations seem to be caused by the interplay between quantum pressure and self-gravity (first and second term on the right-hand side of (\ref{eq:DimlessGrossPitaevskii}), respectively), and the effect may be just more visible for higher-mass solitons than for lower-mass solitons, as follows. Solitons with higher mass have a higher density. As a result, the quantum pressure is higher as well (it can be shown that the equation of state for soliton models without SI is well approximated by a ($n=2$)-polytrope (see \cite{Chavanis_Polytrope2019, Schobesberger2021}), thus the pressure increases with increasing density). While the quantum pressure leads to expansion, self-gravity counteracts efficiently, the higher the mass, causing a contraction. 
Hence, we can see how the self-gravity of the soliton avoids its "dispersion", which would otherwise happen like for any other unconfined quantum-mechanical wave packet. This phenomenon of "gravitationally induced inhibition of dispersion" has been also predicted for laboratory quantum systems, see \cite{Giulini_Grav_ind_inh_SN_2011}. 

Let us also note that these oscillations due to competing dynamical terms as seen in figure \ref{fig:RhoProfiles_Oscillations} are different in nature from the oscillations which have been reported elsewhere, especially with regard to real scalar fields, or real SFDM such as axion-like DM, see e.g. \cite{CyncynatesTiron2021_Oscillation-Structure}. In these models, solitons are often called "oscillatons" or "oscillons" for this reason. Those oscillations refer to the inherent field oscillations about a mean, whose amplitude is higher, the lower the particle mass, and they should not be confused with the dynamical oscillations. Let us also note that in the presence of SI or for complex fields, the inherent field oscillations are much smaller in amplitude, compared to the real-field case for the same particle mass.   

Finally, we wish to comment on the density fits of the numerical profiles in figure \ref{fig:RhoProfiles_Oscillations}. In this particular case, the Schive model has turned out to have the higher score, according to our fit criteria of the previous section. The fit parameters may not change much between different snapshots in time, but this statement is case-specific. Generally, the fits to the central density are very good, which is important in order to draw conclusions with respect to the central density and its relationship to other soliton parameters, such as mass; see below. However, we recognize deviations between the analytical fit and the numerical data in the fall-off region of the profile. A larger box size leads to a more extended "tail" of the single soliton, than a smaller box. Nevertheless, we might conclude that the fit is not perfect in the fall-off region. Our numerical results thus suggest that the analytical models are subject to scrutiny, at least in the regime of soliton masses which we consider in this paper; an issue on which we will comment more in the next sections.

\subsubsection{Central-density-mass relation}

An important key quantity with respect to the nature of DM is the central density of self-gravitating DM structures. The run of the central density, as well as the presence or absence of constant central density cores, affect the dynamics of baryonic tracers, and so the DM density is subject to indirect observational constraints, e.g. from galactic velocity data. Since solitons have been recognized as approximate models for the core regions of simulated galactic FDM/SFDM halos, we are interested in their central density as a function of model parameters. 

\begin{figure}
    \centering
    \includegraphics[scale=0.55]{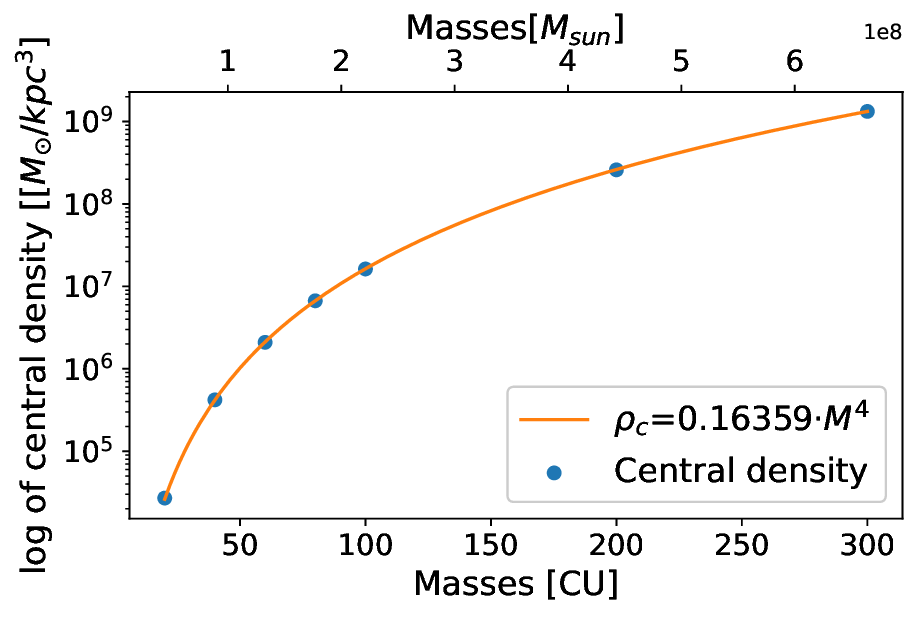}
    \caption{The central-density-mass relation without SI ($\hat \Lambda = 0$) from the simulation data (blue dots). The simulation parameters can be found in Table \ref{tab:IdealResolBox}. The central densities are extracted from the density profiles at the respective final snapshots of the evolution at $0.013$\,CU ($\sim 1$\,Gyr). The solid, orange curve is the fit, according to equation (\ref{eq:RhoMRel}).}
    \label{fig:RhoMRelWithoutSI}
\end{figure}

We use the density data we acquired during the analysis of the high-mass problem and extracted the central densities associated with the respective soliton mass, in order to quantify their scaling relation. This is useful, not only to compare our results with the literature, but also to predict, which central densities and thus which spatial resolution we need for certain soliton masses. It has been known from analytical considerations, confirmed later by simulations, that the central density of solitons for $\hat{\Lambda} = 0$ should scale with $M^4$ (see e.g. \cite{Dawoodbhoy2021}, \cite{Schobesberger2021} and references therein). In figure \ref{fig:RhoMRelWithoutSI}, we plot the central densities over the soliton masses (as blue dots) for the parameters from Table \ref{tab:IdealResolBox}.

We fit these data points with the theoretical expectation,
\begin{equation}\label{eq:RhoMRel}
    \rho_c = b M^4,
\end{equation}
with the central density $\rho_c$, the soliton mass $M$ and our fitting parameter $b$. The central density is in physical units, while the soliton mass is in mass code units, thus the unit of $b$ is $M_\odot/(\text{kpc}^3\text{CU}^4)$. In units of astronomical interest to us, our fit yields
\begin{equation} \label{anum}
    b \approx 0.16359 \frac{M_\odot}{\text{kpc}^3\text{CU}^4} \approx 6.81341 \cdot 10^{-27} \frac{1}{M_\odot^3\text{kpc}^3}.
\end{equation}


To compare this result with the literature, we need the exact definition of $b$, or rather the whole relation, which is quoted in \cite{Schobesberger2021} (see also \cite{Membrado1989}),
\begin{equation}
    \rho_c = \rho_j\left(\frac{G m^2}{\hbar^2}\right)^3 M^4,
\end{equation}
where $\rho_j$ is a dimensionless constant depending on the eigenstate and calculated numerically. Using the value for the ground state from the numerical calculation of \cite{Hui2017} which is $0.0044$ (see also \cite{ChavanisDelfini2011}), we get 
\begin{equation}
    b \approx 0.0044 \left(\frac{G m^2}{\hbar^2}\right)^3 \approx 7.05094 \cdot 10^{-27} \frac{1}{M_\odot^3 \text{kpc}^3},
\end{equation}
i.e. very close to (\ref{anum}).
We can see from figure \ref{fig:RhoMRelWithoutSI}, and the calculation of the fit parameter $b$, that our simulations with {\ffmfamily PyUltraLight\_SI} yield the expected scaling relation. Moreover, this result confirms that our modified code reproduces the soliton behavior for $\hat{\Lambda} = 0$.

\subsection{Simulations with self-interaction}\label{sec:3B_SimSI}

In this subsection, we consider single-soliton simulations for $\hat{\Lambda}\neq 0$. Because the work by \cite{PySiUltraLight} came out while ours was under progress, we took the opportunity to compare simulations performed with our modified code, with those of \cite{PySiUltraLight}. We reproduced their binary mergers, using the same simulation and SFDM parameters, for $\hat{\Lambda} = 0$, $\hat{\Lambda} = -0.02$ and $\hat{\Lambda} = 0.02$; see their figures 6, 7 and 8, respectively, in \cite{PySiUltraLight}. The results were in very good mutual agreement. We will present the results of our binary merger simulations in section \ref{sec:sec4} and multiple soliton mergers in section \ref{sec:sec5}, respectively.

\subsubsection{Range for $\hat{\Lambda}$}

As a matter of fact, previous literature which solved the GPP equations using the wave approach (e.g. by using pseudo-spectral methods), including \cite{PyUltraLight} and \cite{PySiUltraLight}, have mainly focused on models without SI, or on models with weak SI. Motivated by the QCD or string axions, attractive SI has been generally considered more frequently than repulsive SI. However, for strongly repulsive SI, there is another well-known regime, the so-called Thomas-Fermi (TF) regime, which has been explored as well, see Section \ref{sec:sec1} for references.
As described there, in that case, the stability of single "TF solitons" against gravitational collapse is being attained solely by the SI pressure, while quantum pressure is very much subdominant and can be neglected. However, quantum pressure does play a role in the envelopes of post-merger objects, such as fully-fledged galactic halos, also in the TF regime; these models being often called SFDM-TF or SIBEC-DM in the literature. Since models with $\hat{\Lambda} > 0$ offer another pressure contribution to oppose gravity (in contrast to attractive SI), we focus in this paper on simulations with positive, repulsive SI. Furthermore, our initial goal has been to carry out simulations in the TF regime, in order to compare the outcome with previous literature, which usually adopts a fluid approach to study SFDM-TF halo formation and evolution, as e.g. in \cite{Dawoodbhoy2021, Shapiro2022, Hartman2022, Foidl2023}.  

Now, in order to be in the TF regime, we require $\hat{\Lambda} \gg 17$; see \cite{Padilla2021}. However, we encountered serious issues with {\ffmfamily PyUltraLight\_SI} for sufficiently high $\hat{\Lambda}$, which prevented us from studying the TF regime. Given our quest of performing small, but dedicated simulations, we instead report in this paper our results of models with "intermediate" range in repulsive SI, having $\hat{\Lambda} \leq 10$. For one thing, this intermediate regime has not been explored in the previous literature. Furthermore,  we know from \cite{Dawoodbhoy2021} that the severe resolution limitations in SFDM-TF, where the de Broglie length is much smaller than the characteristic TF radius or length scale, lend themselves to the adoption of a fluid approach, rather than the wave equation approach which is used in {\ffmfamily PyUltraLight} and {\ffmfamily PyUltraLight\_SI}. It remains to be seen whether the numerical algorithms, or the specifics of the codes may be extended such that the TF regime can be covered, but this task turned out to be well beyond the scope of this paper.

\begin{figure}
    \centering
    \includegraphics[scale=0.25]{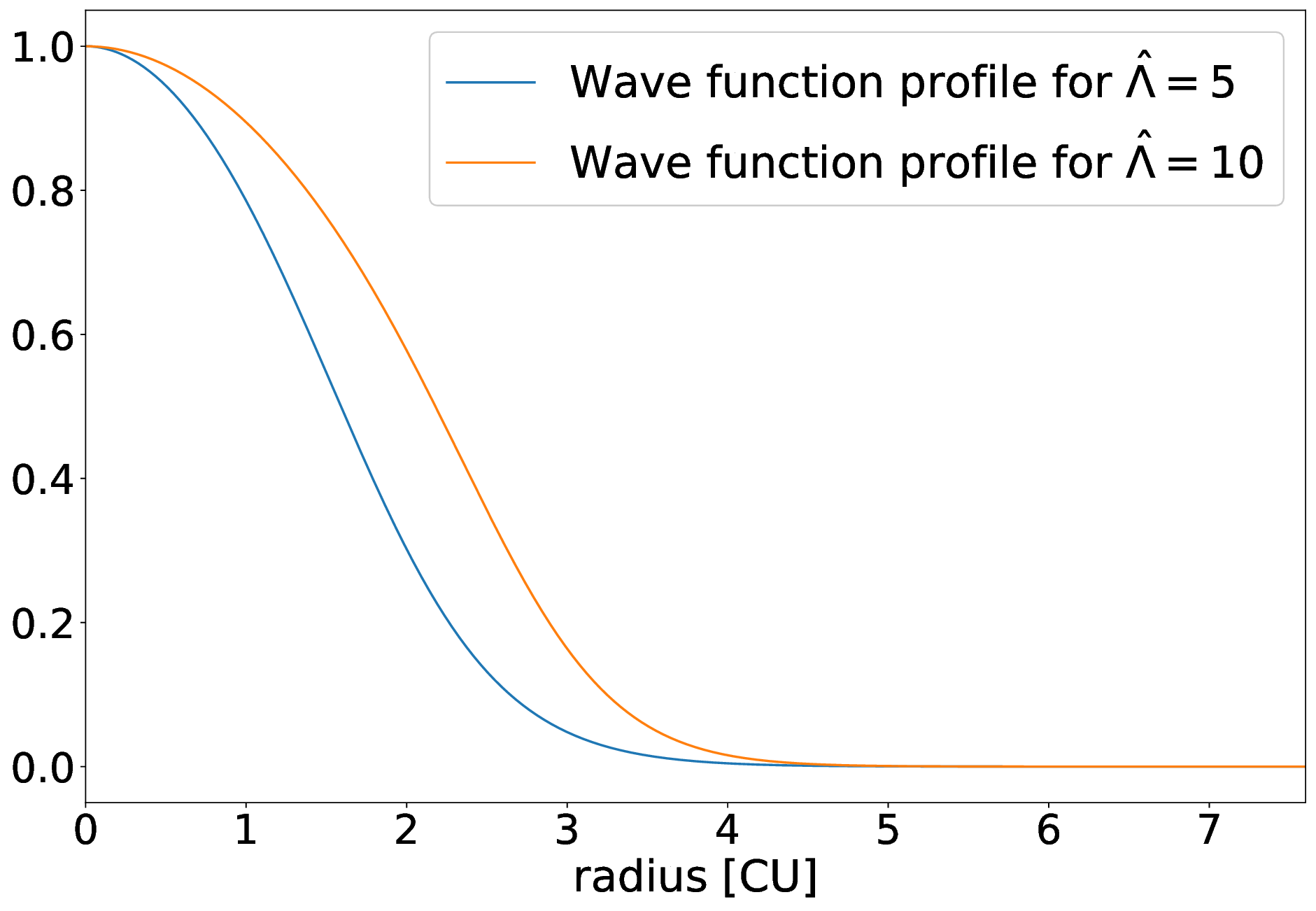}
    \caption{A comparison between two wave function profiles with $\hat{\Lambda}=5$ (blue curve), calculated with a maximum radius of $9$\,CU, whereas $\hat{\Lambda}=10$ (orange curve), calculated with a maximum radius of $7.6$\,CU. For comparison's sake, the blue curve is plotted to the maximum radius of the orange curve (notice the linear axis scales).}
    \label{fig:Wavefunction_Comparison}
\end{figure}

The mentioned issues come at play already at the level of the initial soliton profiles, as follows. We have described above how the initial profile is determined, via its boundary conditions and numerical method: in order to guarantee that the radial wave function is smooth, $f'(r)$ as well as $\Tilde \varphi'$ have to vanish at the origin, while the wave function and the gravitational potential go to $0$ for $r\rightarrow \infty$. In practice, the gravitational potential is varied at the origin until a solution for $f(r)$ is found, that approaches zero at some maximum radius. For small $\hat{\Lambda}$, no issue is encountered. However, for sufficiently high $\hat{\Lambda}$'s, and even for the intermediate cases considered here, the wave function becomes significantly broader (i.e. the associated density profile flatter), because the repulsive SI becomes stronger. Examples are presented in figure \ref{fig:Wavefunction_Comparison}, where two wave functions for different $\hat{\Lambda}$ are shown. The wave function for $\hat{\Lambda} = 10$ has a larger width than that with $\hat{\Lambda} = 5$. The solver requires to find solutions whose profiles go to zero at the maximum radius, but this process becomes increasingly difficult, the broader the profiles get. As a result, the would-be solution oscillates at a certain radius, ultimately causing the solver to grind.
To overcome this problem, we decrease the maximum radius for higher $\hat{\Lambda}$'s ($\hat{\Lambda} \geq 6$), such that the code never reaches that radius, where the solution starts oscillating. We deem this approach legitimate, because in the outer regions the profile goes quickly towards zero, and hence the introduced numerical error is small.

Because of this and other issues to be mentioned below, we discuss in the forthcoming two subsets of range in $\hat{\Lambda}$, and call them
\begin{itemize}
    \item low $\hat{\Lambda}$ in the range $0\leq \hat{\Lambda} \leq 1$
    \item intermediate $\hat{\Lambda}$ in the range $1< \hat{\Lambda} \leq 10$.
\end{itemize}

\subsubsection{Low $\hat{\Lambda}$}

As mentioned already, a repulsive SI leads to solitons with larger width and we expect them to spread more readily (or "disperse more readily") over time, given the repulsive SI pressure. This should lead, in turn, to decreased central densities, compared to solitons without SI. Since higher $\hat{\Lambda}$ leads to lower central density, the demands on the spatial resolution are less severe, and the requirements to avoid the "high-mass problem" are not so strict. On the other hand, more readily "dispersing" solitons take up more space over time within the simulation box. The latter is thus required to be chosen big enough to prevent the loss of scalar field mass beyond the box boundaries. At the same time, the choice of the box size must ensure that the calculated soliton mass is close to its input initial value (for resolution limits mentioned earlier).

\begin{figure}
    \centering
    \includegraphics[scale=0.25]{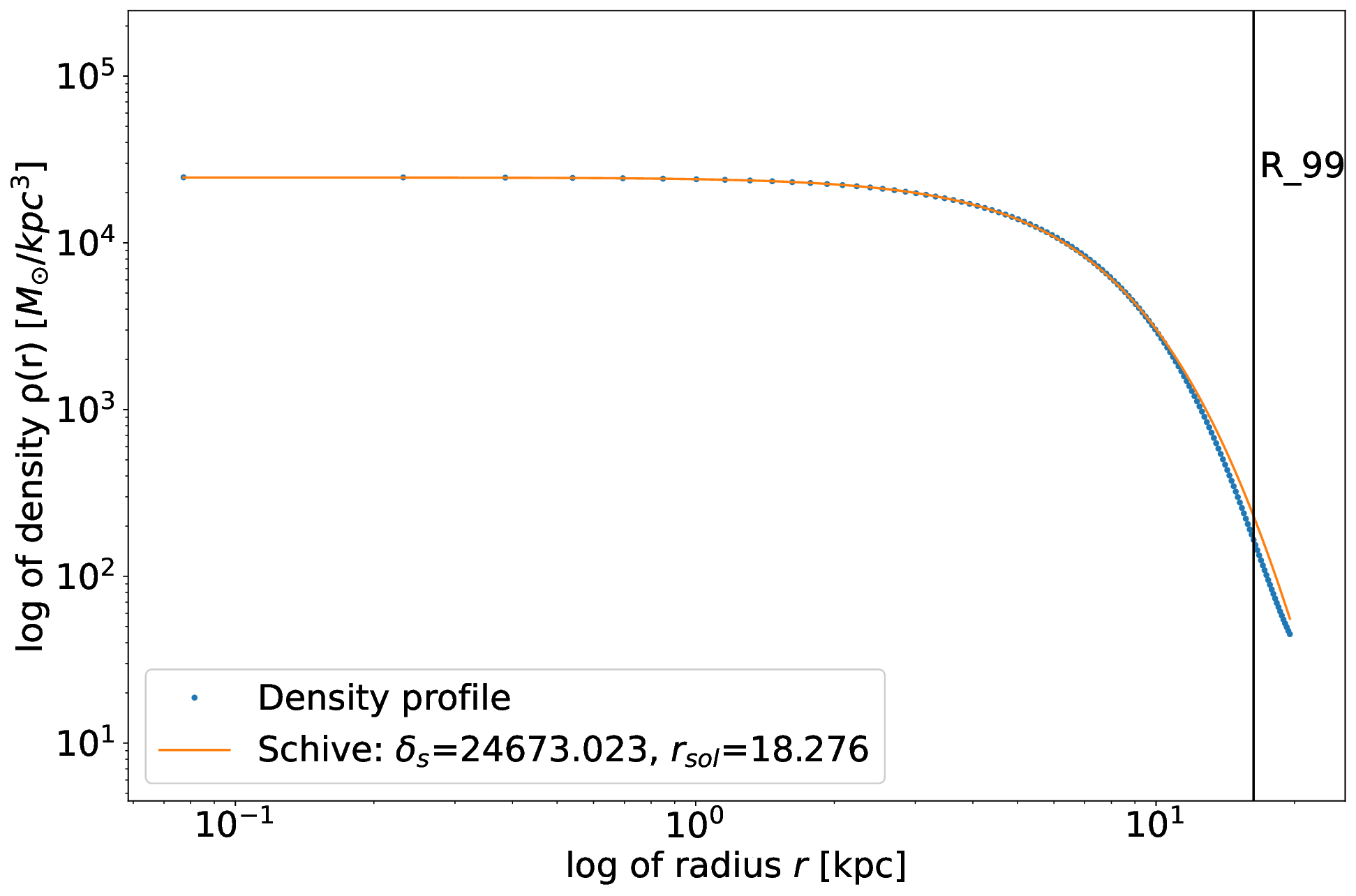}
    \caption{The density profile of the final state ($\sim1$\,Gyr) for a $20$\,CU soliton which was initialized in a $1.03$\,CU box with a resolution of $256$ and $\hat{\Lambda}=0.02$. The black vertical represents that radius, which contains $99$\,\% of the mass of the soliton, while the orange, solid curve is the fit according to equation (\ref{eq:Schive}), with units for $\delta_s$ in $M_\odot/\text{kpc}^3$ and $r_{sol}$ in kpc.}
    \label{fig:Profile_Single_002}
\end{figure}

To illustrate how the repulsive SI influences the density profiles, we set up a soliton at rest with a mass of $20$\,CU for $\hat{\Lambda}=0.02$. This value for the SI is inspired by the tests for the binary merger simulations made by \cite{PySiUltraLight}, which we reproduced. Furthermore, for this weak SI, we expected the outcome to be comparable to the case with $\hat{\Lambda}=0$ (see figure \ref{fig:RhoProfiles_Oscillations}, but note the difference in soliton mass). For the simulation parameters, we choose $0.013$\,CU ($\sim1$\,Gyr) and a resolution of $256$. For the box size, we first chose $2$\,CU to ensure that the soliton stays inside the box within the evolution time. Then, we reduced the box size, such that the calculated soliton mass resulted in $20$\,CU as it should be. This was accomplished, once $1.03$\,CU was chosen for the box size \footnote{Actually, a more detailed trial-and-error process enables an even more accurate mass adjustment, but since the change in the density profile would be minuscule, we did not pursue such fine-tuning}. The final density profile can be seen in figure \ref{fig:Profile_Single_002}. As expected, the density profile looks similar to the $\hat{\Lambda}=0$ case, except for the lower central density and greater width, exemplified by the larger core parameter of the associated Schive profile, and the fact that the soliton mass is different.

\begin{figure}
    \centering
    \includegraphics[scale=0.25]{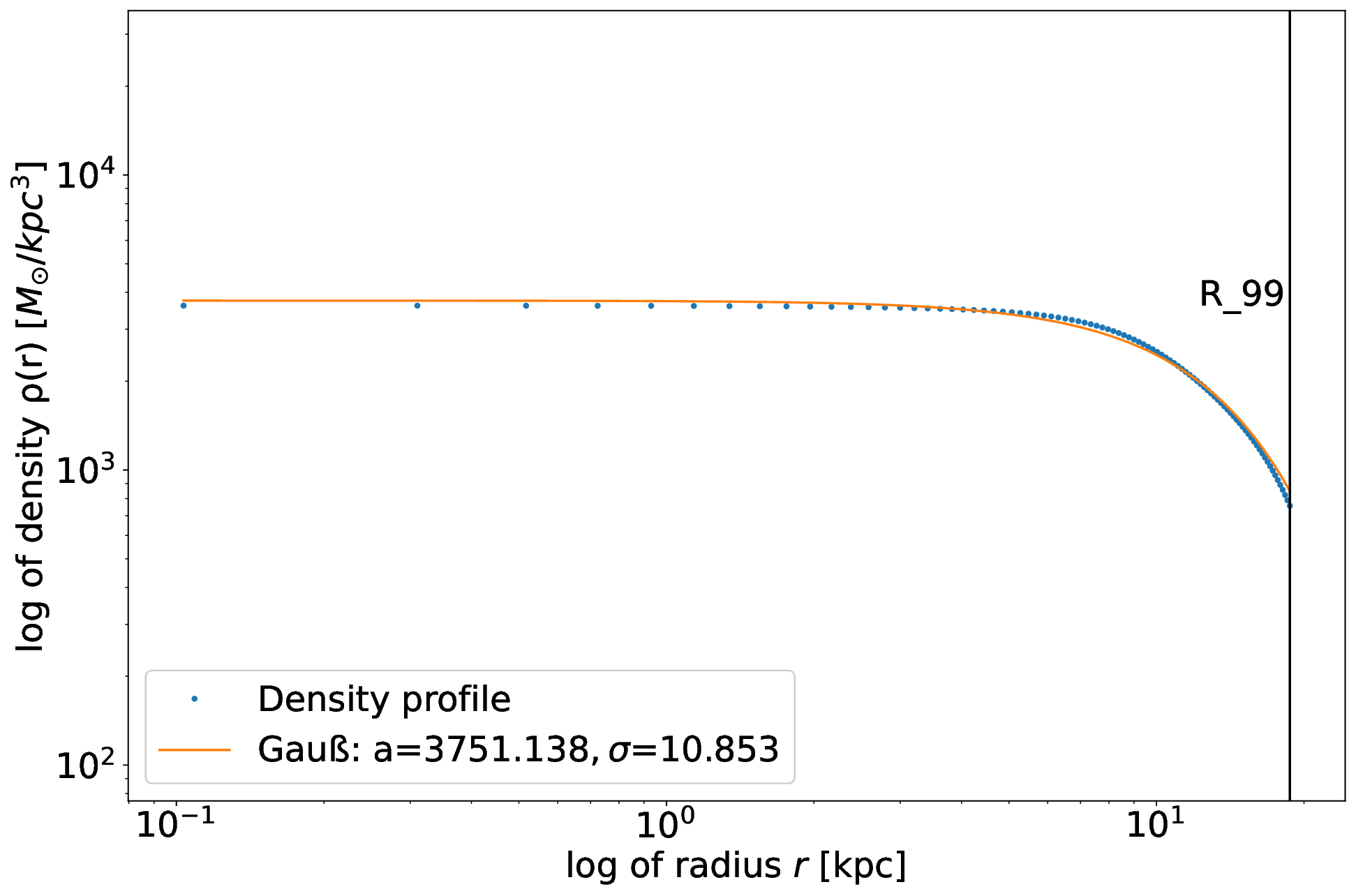}
    \caption{The density profile of the final state ($\sim1$\,Gyr) for a $20$\,CU soliton which was initialized in a $1.38$\,CU box with a resolution of $256$ and $\hat{\Lambda}=1$. The black vertical represents that radius, which contains $99$\,\% of the mass of the soliton, while the orange, solid curve is the fit according to equation (\ref{eq:Gauß}), with units for $a$ in $M_\odot/\text{kpc}^3$ and $\sigma$ in kpc. In comparison to figure \ref{fig:Profile_Single_002}, the core is much more extended, leading to a lower central density and a larger $R_{99}$.}
    \label{fig:Profile_Single_1}
\end{figure}

Now, we run the same soliton simulation, but we set $\hat{\Lambda}=1$. While the duration and the resolution stayed the same, the box size had to be increased to $1.38$\,CU due to the factor $50$ stronger SI. The density profile can be seen in figure \ref{fig:Profile_Single_1}. As expected from the trend seen in the wave functions, and the previous simulation, the profile has become even more uniform with a lower central density. Also, the radius $R_{99}$, which includes $99$\,\% of the mass of the soliton, is larger. This trend will play a role for the intermediate $\hat{\Lambda}$-case.

We conclude that a higher $\hat{\Lambda}$ yields a more uniform density distribution of matter within the soliton, compared to $\hat{\Lambda}=0$, as a higher repulsive SI pressure pushes against the central pileup of matter due to gravity, resulting also in a lower central density for the same soliton mass. For the $20$\,CU solitons plotted in figures \ref{fig:Profile_Single_002} and \ref{fig:Profile_Single_1}, we recognize a core parameter which is $\approx 2.4$ times larger in the case with $\hat{\Lambda}=1$, compared to the case with $\hat{\Lambda}=0.02$ (remember $r_{sol} \approx 4 \sigma$). However, we stress that the size of the fitted core parameter depends not only on $\hat{\Lambda}$, but also on the soliton mass.

\subsubsection{Intermediate $\hat{\Lambda}$}

We have seen above that the profiles become more uniform and favor larger core parameters for the fitting functions, as $\hat{\Lambda}$ increases. Also, $R_{99}$ gets larger.
This trend exacerbates with higher $\hat{\Lambda}$.
Thus, when we want to determine an accurate mass from the profile, in order to calculate $R_{99}$, we would need to reduce the box size and "cut into" the core-like, extended density profile, which affects the mass determination, in turn. In effect, too high or too low a calculated soliton mass results. 

This issue is one reason why we decided to split the range for $\hat{\Lambda}$, because for the simulated models with $\hat{\Lambda} > 1$ we have no accurate mass determination. Yet, we proceed with the evolution of the solitons and their density profiles over time with various $\hat{\Lambda}$'s. We simulated a $20$\,CU soliton at rest for $0.013$\,CU ($\sim1$\,Gyr), a box size of $8$\,CU and a resolution of $256$. We increased the time resolution by increasing the save number to $10$, such that we get $10$ snapshots with a resolution of $0.1$\,Gyr; we show $3$ snapshots of each simulation: the initial state at $t=0$\,Gyr, the middle state at $t=0.5$\,Gyr and the final state at $t=1$\,Gyr. For each snapshot, we look at the radial density profile as well as the 2D density profiles in the $x-y$-plane.
We ran simulations for $\hat{\Lambda} = 5$ and $\hat{\Lambda} = 10$, respectively.
 
\begin{figure}
    \centering
    \includegraphics[scale=0.24]{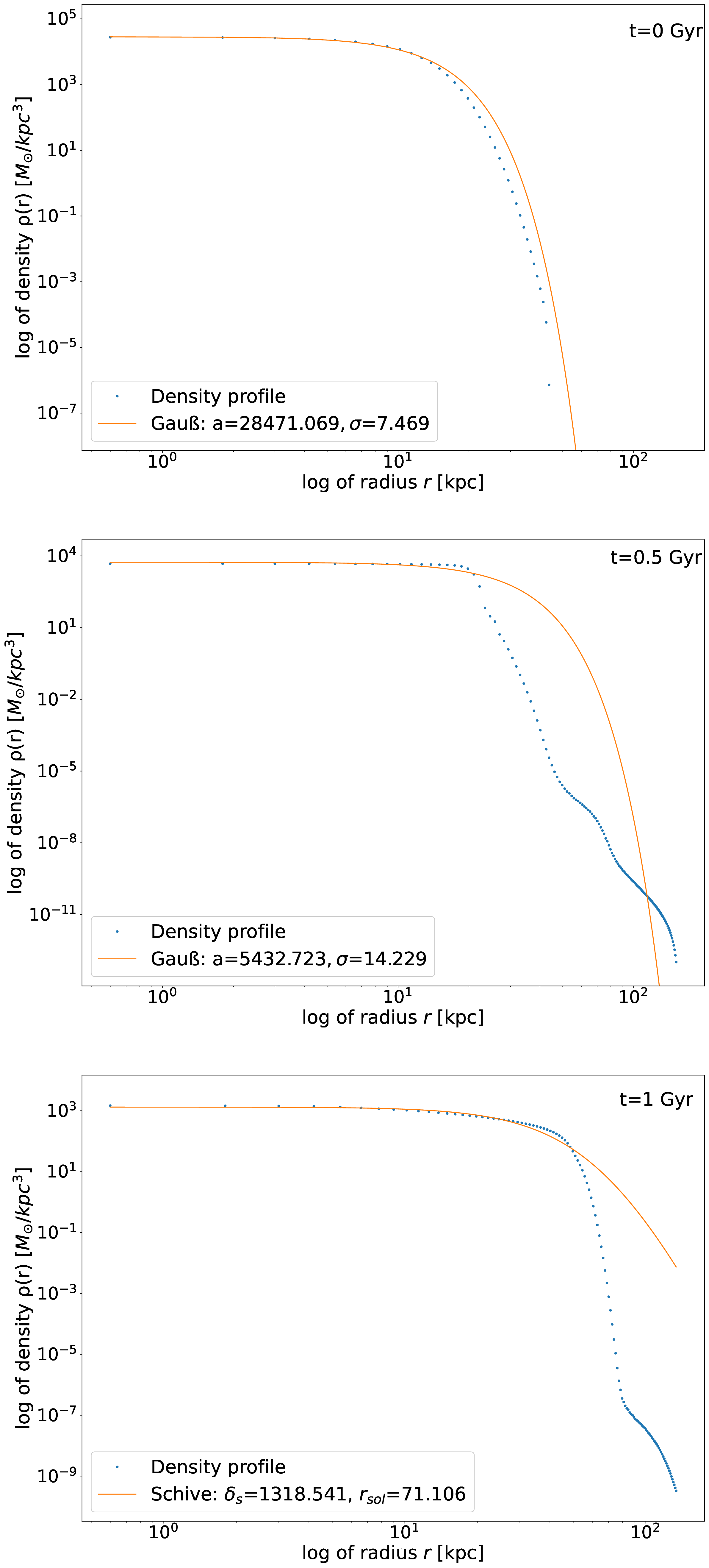}
    \caption{The evolution of the radial density profile of a $20$\,CU-soliton, which was initialized in a $8$\,CU box with a resolution of $256$ and $\hat{\Lambda}=5$. The orange, solid curves are fits according to equations (\ref{eq:Gauß}) and (\ref{eq:Schive}), with units for $a$ and $\delta_s$ in $M_\odot/\text{kpc}^3$ and $\sigma$ and $r_{sol}$ in kpc, respectively.}
    \label{fig:Profiles_SI5}
\end{figure}

\begin{figure}
    \centering
    \includegraphics[scale=0.38]{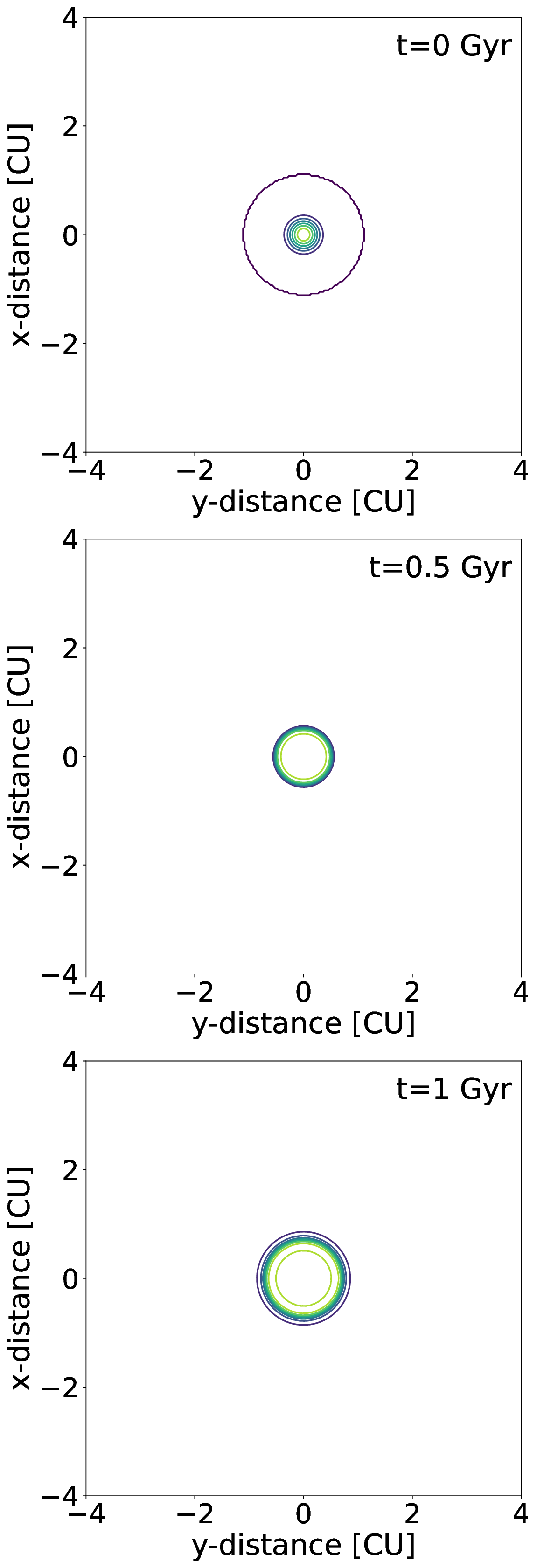}
    \caption{The corresponding 2D density profiles to figure \ref{fig:Profiles_SI5}. Shown are contour lines of constant density.}
    \label{fig:Density2D_SI5}
\end{figure}

In figure \ref{fig:Profiles_SI5}, we show the radial soliton density profiles for $\hat{\Lambda}=5$, while the associated 2D density profiles in the $x-y$ plane can be seen in figure \ref{fig:Density2D_SI5}. We recognize the same trend as seen in the previous runs with lower $\hat{\Lambda}$: wider, more uniform profiles with lower central density result. The middle snapshot seems to indicate the formation of a "ring", or rather a spherical "shell", which merely exemplifies the redistribution of matter in the soliton. 
Upon looking at the SI term in the GP equation (\ref{eq:DimlessGrossPitaevskii}), we see that it depends not only on the SI strength (given by $\hat{\Lambda}$), but also on the density (given by $|\hat{\psi}(\vec x, t)|^2$). Thus, for a given $\hat{\Lambda}$, the density affects the ensuing dynamics. Since the SI pressure is more prevalent in denser regions, it pushes matter outwards and this "expansion" is faster in the dense inner regions, compared to the outer regions. Eventually, the inner "shells" catch up with the outer layers, forming the "ring-like" structure in the 2D plot.  


\begin{figure}
    \centering
    \includegraphics[scale=0.24]{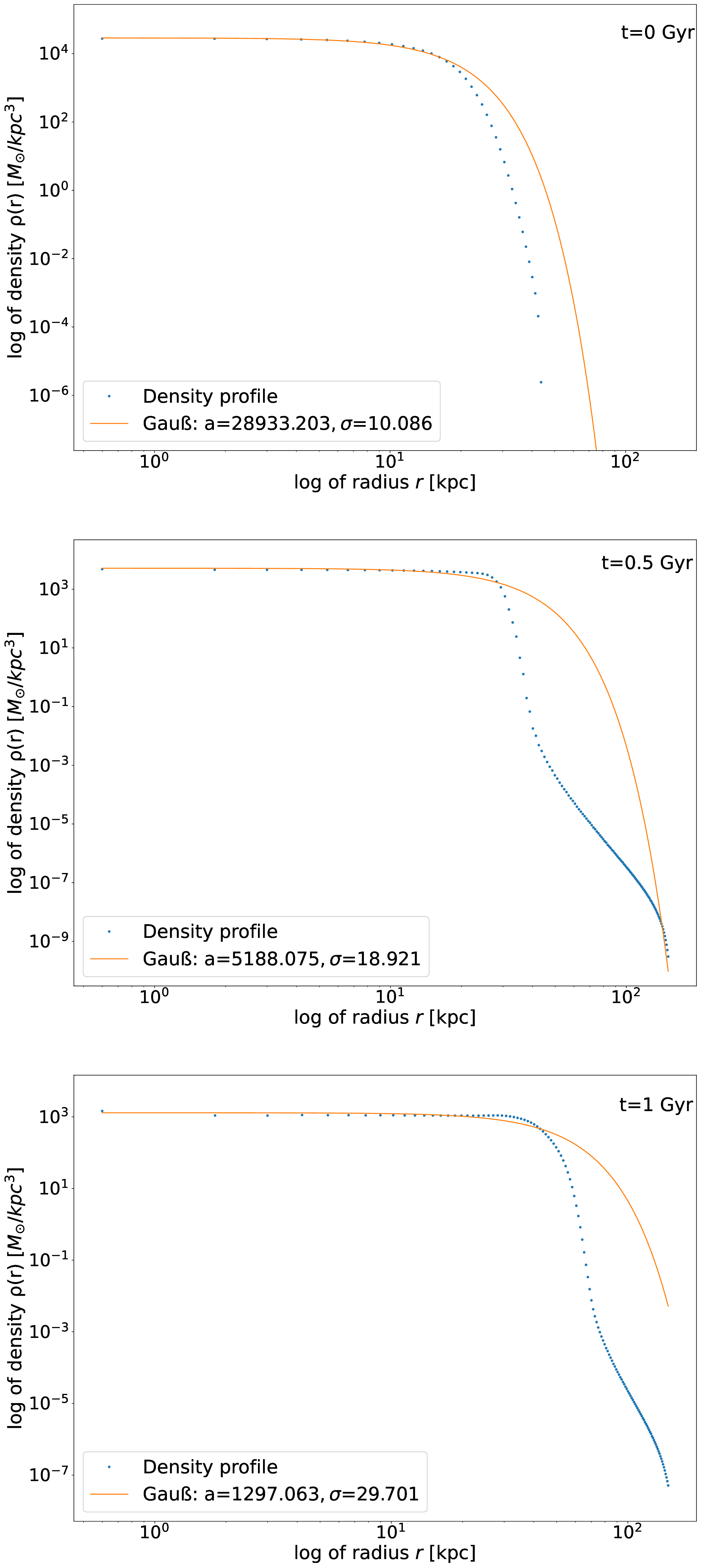}
    \caption{The evolution of the radial density profile of a $20$\,CU-soliton, which was initialized in a $8$\,CU box with a resolution of $256$ and $\hat{\Lambda}=10$. The orange, solid curves are fits according to equation (\ref{eq:Gauß}), with units for $a$ in $M_\odot/\text{kpc}^3$ and $\sigma$ in kpc.}
    \label{fig:Profiles_SI10}
\end{figure}

\begin{figure}
    \centering
    \includegraphics[scale=0.38]{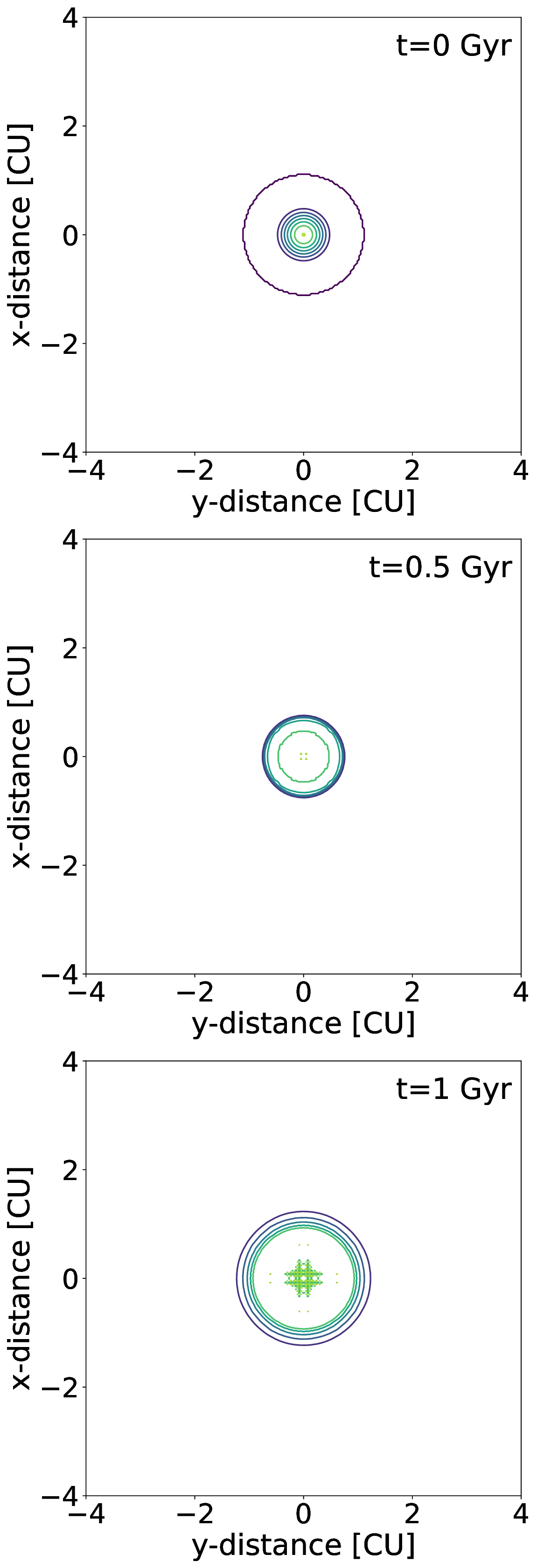}
    \caption{The corresponding 2D density profiles to figure \ref{fig:Profiles_SI10}. Shown are contour lines of constant density. The symmetric structure in the center of the bottom panel is a numerical artifact.}
    \label{fig:Density2D_SI10}
\end{figure}

Now, in figure \ref{fig:Profiles_SI10}, we show the radial soliton density profiles for $\hat{\Lambda}=10$, while the associated 2D density profiles in the $x-y$ plane can be seen in figure \ref{fig:Density2D_SI10}. We stress that the maximum radius for the initial soliton profile was reduced from $9$\,CU (the initial maximum radius for $\hat{\Lambda}<6$) to $7.6$\,CU (see the explanation to Fig.\ref{fig:Wavefunction_Comparison}). The overall evolution of the profiles is very similar to the ones for $\hat{\Lambda}=5$. Here, we encounter some numerical artifacts in the 2D profiles which, however, have no consequence for the radial profiles. Their origin remains unclear, although they arise generically, see e.g. figure 8 in \cite{Hui_2021}, or figure 10 in \cite{Navarrete_SpatialSolitons_2017}.  


We conclude that solitons exhibit a very uniform density profile for increasing values of $\hat{\Lambda}$, which we can understand as the interplay of repulsive SI pressure with self-gravity. Also, it appears that the outer parts of solitons experience fewer oscillations, the higher $\hat{\Lambda}$, maybe as a result of being "washed out" due to the matter redistribution. Yet, this statement also depends upon the soliton mass. In general, oscillations in the tails seem to be more prominent for higher soliton mass, independent of the given value of $\hat{\Lambda}$.  

Finally, we add that the same cautionary remark applies for models with $\hat{\Lambda} > 0$, with respect to the analytical density profiles and their "goodness of fit" to the numerical data, as already elaborated above for SFDM without SI. While the general trend of more uniform profiles is well represented by larger values of the fitted core parameters, the oscillatory behavior can change the fitting parameters in case-specific ways. Viewed in this light, our finding of a mild preference of Gaussian over Schive profiles in the cases with $\hat{\Lambda} > 0$ displayed in our paper might be more of a coincidence, rather than a real trend. To acquire more certainty, further simulations will be required in the future.

\subsubsection{Central-density-mass relation for $\hat{\Lambda}>0$}

In the previous subsection, we have seen that the central density gets lower with increasing $\hat{\Lambda}$, for fixed value of the soliton mass.
Now, we want to study how the central density might change with varying soliton mass at fixed $\hat{\Lambda}$, in order to produce a similar plot as in figure \ref{fig:RhoMRelWithoutSI}. However, the issues described above prevented us from studying this central-density-mass relation for intermediate values of $\hat{\Lambda}$. Instead, we focus on two cases, $\hat{\Lambda}=0.02$ and $\hat{\Lambda}=1$. 
Given the lack of literature to guide us, our expectations were as follows. For very small $\hat{\Lambda}$, we would expect no big qualitative change, compared to $\hat{\Lambda}=0$. On the other hand, since our range of studied $\hat{\Lambda}$'s is well below the TF regime of strongly repulsive SI, we would not expect to find a scaling similar to SFDM-TF, either. In fact, in the latter case, analytical considerations predict that the central density is proportional to the soliton mass, $\rho_c \propto M$, see e.g. \cite{Dawoodbhoy2021} and references therein, i.e. a much milder growth with mass, compared to equation (\ref{eq:RhoMRel})\footnote{In fact, this milder growth of the central density with soliton (resp. \textit{halo core}) mass may help SFDM-TF to better fit with galactic velocity data, than the FDM regime with $\hat{\Lambda}=0$, see \cite{Dawoodbhoy2021} for a comparison.}.

In our simulation runs, we set up solitons at rest with different masses and use the simulation parameters according to Table \ref{tab:IdealResolBox}, just as for the $\hat{\Lambda}=0$ case. Yet, our runs do not cover the same range of soliton masses. The difficulties with respect to the broadening ("dispersion") of solitons with $\hat{\Lambda} > 0$ described above become exacerbated for higher-mass solitons. In many cases, solitons became unstable, prohibiting the extraction of meaningful radially averaged density profiles from the simulations.
As a result, we present our findings for a different range in soliton mass. 

\begin{figure}
    \centering
    \includegraphics[scale=0.57]{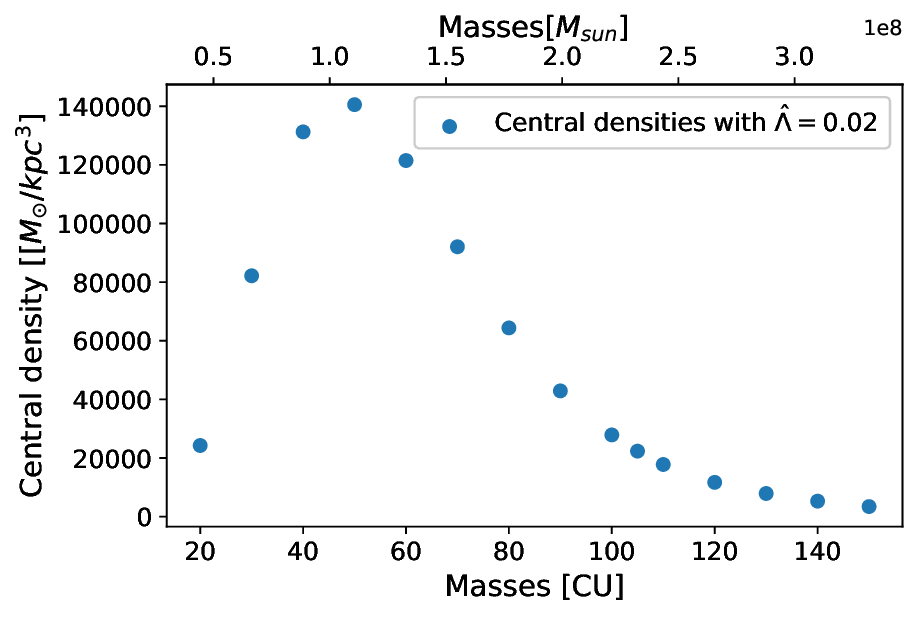}
    \caption{The central-density-mass relation for $\hat{\Lambda}=0.02$. The central densities are extracted from the density profiles at the respective final snapshots of the evolution (at $\sim 1$\,Gyr).}
    \label{fig:RhoMRelation_002}
\end{figure}

Figure \ref{fig:RhoMRelation_002} shows the central-density-mass relation at fixed $\hat{\Lambda}=0.02$ for solitons with a mass of up to $150$\,CU. The central densities were extracted from the density profiles at the final snapshot ($\sim 1$\,Gyr). We can recognize that $\rho_c$ increases monotonically with soliton mass with a slope that is shallower than in the case without SI.
However, at around $50$\,CU the central density peaks, beyond which $\rho_c$ decreases with mass. In light of the results of the previous subsection, we think we can understand this behavior as follows.
The broadening of the solitons is caused by the SI pressure, or SI term in the GP equation (\ref{eq:DimlessGrossPitaevskii}), which depends on $\hat{\Lambda}$ and on the density (wave function squared).
This term opposes the self-gravity in the equation, which attempts to contract the soliton. The gravitational potential $\Phi(r)$ also depends on the density via the Poisson equation (\ref{eq:DimlessPoisson}). Thus, for fixed $\hat{\Lambda} > 0$, the dynamical importance of each term (self-gravity vs SI repulsion) is affected by the density of the soliton in characteristic ways \footnote{In the TF regime, closed-form expressions for $\rho(r)$ and $\Phi(r)$, and their limits for $r \to 0$, can be derived which make the dependence more explicit.}. Our simulations suggest that the initially monotonic increase of $\rho_c$ with soliton mass can be attributed to the subdominance of the SI term for the chosen time frame \footnote{With that we mean, that the density is too low for the SI term to be able to counter the increase in density for higher masses in the given time span of the simulation. If we were to choose a later time to extract the central density, the turnover point would happen at even lower soliton masses.}. Once the soliton mass is chosen big enough, however, the SI term "takes over" or rather is "fast enough", such that we notice the broadening and uniformation of the soliton profile within the given simulation time in form of a decrease in the central density. This statement applies for the static and dynamic case: an initialized soliton with high mass will already have a lower central density, than a corresponding low-mass soliton, at given $\hat{\Lambda}$. During the course of its "evolution" -i.e. sitting in the box at rest-, the tendency of soliton broadening will make its impact onto the final snapshot of the simulation from which the central density is extracted. As a result, beyond the turnover to the prevalence of the SI term, we get lower central densities for bigger soliton masses.
 
\begin{figure}
    \centering
    \includegraphics[scale=0.57]{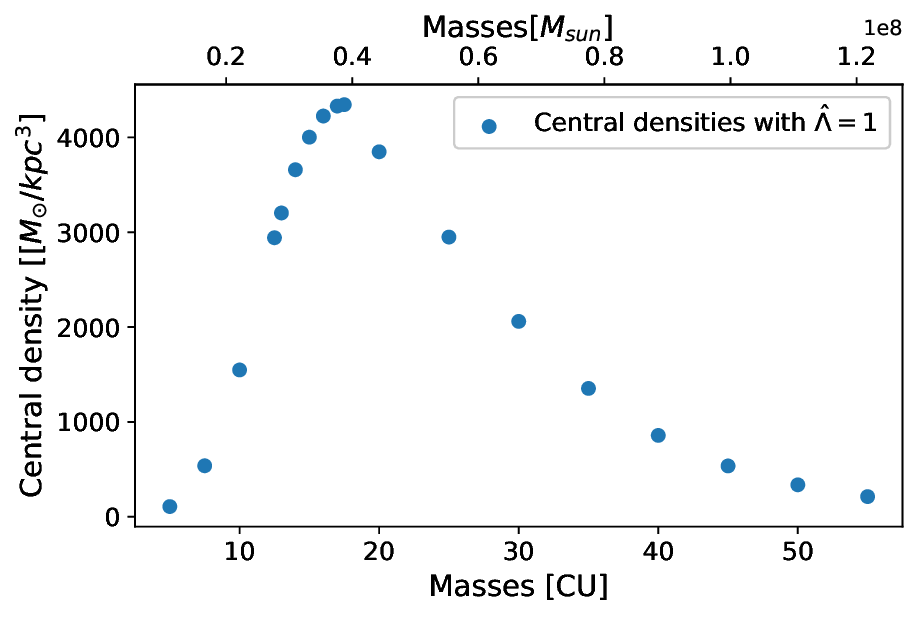}
    \caption{The central-density-mass relation for $\hat{\Lambda}=1$. The central densities are extracted from the density profiles at the respective final snapshots of the evolution (at $\sim 1$\,Gyr). More data points (i.e. simulation runs) at lower soliton mass are included, compared to figure \ref{fig:RhoMRelation_002}. }
    \label{fig:RhoMRelation_1}
\end{figure}

In order to investigate this further, we also studied the central-density-mass relation for a stronger SI with $\hat{\Lambda}=1$. In this case, the maximum soliton mass with useful density profiles was around $55$\,CU. Because of this limitation, we ran more simulations for solitons at lower masses, $<20$\,CU, using a higher resolution. The other simulation parameters were the same as for the previous runs of the central-density-mass relation with $\hat{\Lambda}=0.02$. In figure \ref{fig:RhoMRelation_1}, we plot the obtained central densities of the final states (again at $\sim 1$\,Gyr) versus soliton mass. As in the previous $\hat{\Lambda}=0.02$ case, we get a similar behavior: for sufficiently low soliton mass, the central densities grow with mass, peak at a maximum and then decrease with increasing soliton mass. We added more simulations, i.e. data points, at the lower-mass side of the prospective maximum, in order to improve the visibility of the trend that we see.

To conclude, we find a very similar run in the central-density-mass relation of solitons at low, repulsive SI: the central density grows with mass, reaches a maximum and then declines with increasing soliton mass. This behavior is very different from the monotonic relationships known in the FDM regime (no SI) and the TF regime (highly repulsive SI).
In lack of a firm analytical prediction, we refrained from presenting a fit \footnote{It appears that a Gaussian multiplied by a low-order polynomial might fit both numerical data sets, but more analytical investigations and simulations would be required to investigate this further. Surely, an analytical relationship would also help to predict the soliton mass at which the central density becomes a maximum, and its dependence on $\hat{\Lambda}$.} to our simulation data. Yet, we think we can understand our findings as a result of the dynamical competition between self-gravity and SI pressure, which depends on the chosen values for $\hat{\Lambda}$, soliton mass and simulation time. The importance of the SI term is further illustrated by the fact that the central densities of solitons of a given mass decrease markedly, for increasing $\hat{\Lambda}$. For example, a soliton of $50$\,CU $\sim 10^8\,M_\odot$ has a central density of $\rho_c \sim 5\times 10^5\,M_\odot \,\text{kpc}^{-3}$ if $\hat{\Lambda} = 0$ (figure \ref{fig:RhoMRelWithoutSI}), while $\rho_c \sim 10^5\,M_\odot \,\text{kpc}^{-3}$ for $\hat{\Lambda} = 0.02$ (figure \ref{fig:RhoMRelation_002}) and $\rho_c \sim 500\,M_\odot \,\text{kpc}^{-3}$ for $\hat{\Lambda} = 1$ (figure \ref{fig:RhoMRelation_1}), respectively. Of course, details depend on the fitting profiles which have been used to extract the values for $\rho_c$. By comparing figure \ref{fig:RhoMRelation_002} and figure \ref{fig:RhoMRelation_1}, we also recognize that the soliton mass, at which $\rho_c$ peaks, is shifted toward lower values if $\hat{\Lambda}$ is increased, in accordance with an ever more important SI term that counteracts self-gravity already at lower soliton masses.

\section{\label{sec:sec4}Binary-Soliton Mergers}

Apart from the single-soliton dynamics studied in the previous section, one further focus of this paper is the merger dynamics of solitons. In this section, we discuss our binary merger results, while the next section will present mergers consisting of more than two solitons.

The most basic interaction between two solitons is a direct collision. This kind of scenario is a main testbed for analysing the behavior of SFDM, see e.g. \cite{GonzalesGuzman2011_Collisions_Interference}. Collisions have been also performed using {\ffmfamily PyUltraLight} in the original work by \cite{PyUltraLight}, as well as in \cite{PySiUltraLight}. In \cite{PySiUltraLight}, such collisions were simulated, for models with and without SI. However, their choice of parameters did not lead to successful mergers. In contrast, we are particularly interested in creating mergers, which meant to run many simulations in order to test setup and model parameters. We investigated the unfolding merging process, and studied the density profiles of the post-merger objects at the final snapshots of our simulations.

For the mergers we consider two basic scenarios: i) The solitons are supposed to collide head-on along two different lines, one is the x-axis and the other along the main diagonal line in the x-y-plane. These scenarios are basically the same, for we could show, that the results of mergers is the same when the whole initial configuration is rotated about the z-axis. Therefore, it would be enough to only look at the x-axis-case. However, we decided to use the diagonal variant for a different reason. Using the longer diagonal of our square box is useful to evade the ejection of SFDM mass from the main point of interaction (i.e. to prevent exiting and re-entering of mass in the box), 
given the requirements of box size on an accurate mass determination.   
  
The simulation parameters were set to avoid running into the "high-mass problem". The simulation time was set to $0.18$\,CU ($\sim 13.59$\,Gyr), i.e. roughly one Hubble time. All solitons were placed $0.4$\,CU from the center of the box in the x-y-plane. The velocities were varied until a merger happens and a final end product was visible within the simulation time. Since we are working within the x-y-plane, the z-component of all quantities is zero. 

The relative velocity between solitons is a key parameter, which determines whether a merger is successful, or not. As in previous literature, our choice made in this paper may appear somewhat artificial, in that we make sure that mergers happen. The chosen values also reflect computational demands, in order to guarantee mergers within a Hubble time, given our simulation setup. Cosmological simulations of FDM, using the wave approach, as well as using fluid variants for SFDM-TF, have established that mergers do happen and that structure formation above the characteristic Jeans scales proceeds in similar fashions than in CDM, see e.g. \cite{Mocz_Solitonen2017, Hartman2022, Foidl2023}.
Nevertheless, the many details of the merger dynamics in different SI regimes are far from being fully understood and are subject of ongoing research. Importantly, the low and intermediate SI regime has not been simulated previously. Therefore, we investigate mergers in the framework of dedicated simulations here to probe fundamental characteristics, deferring more realistic scenarios to future work.
In the following sections, we will discuss the different configurations and our results.

\subsection{Mergers without self-interaction}

First, we consider mergers without self-interaction ($\hat{\Lambda}=0$). We ran a lot of test simulations to determine those relative velocities which work with our requirements. Anticipating our later results, we note that higher velocities are possible, compared to cases with repulsive SI, without running the risk of not accomplishing a merger within a Hubble time.
For the soliton masses considered, we found that a relative velocity of $5$\,km/s works well, which are $5$\,CU for the absolute value for the initial velocity for each soliton. We discuss two kinds of simulations: mergers with equal mass ratio and mergers with a mass ratio of $2:1$.

\subsubsection{Mergers with equal mass ratio}

\begin{figure}
    \centering
    \includegraphics[scale=0.25]{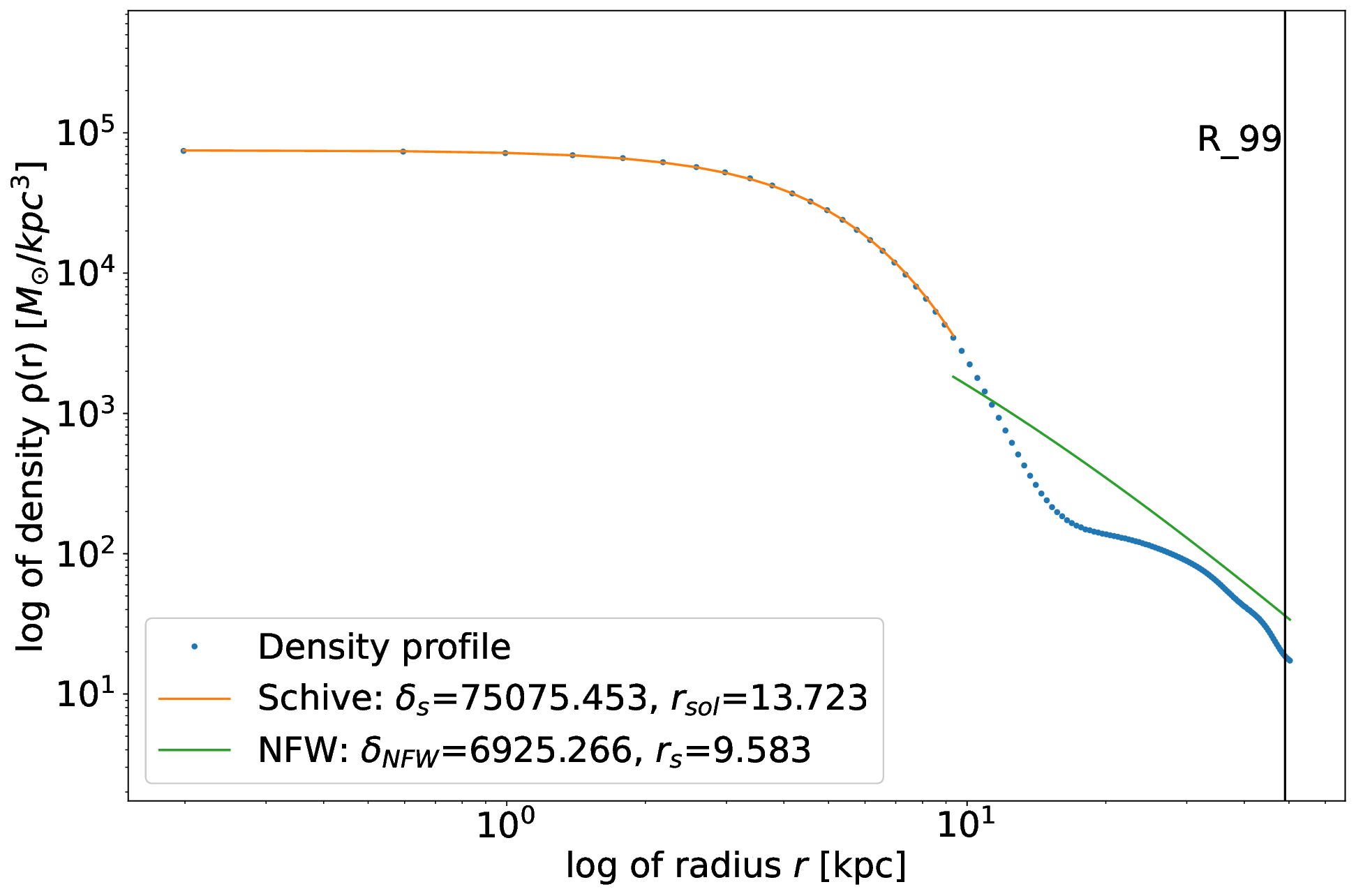}
    \caption{The density profile of the final snapshot ($\sim 13.59$\,Gyr) of a merger of two $20$\,CU-solitons which collide head-on along the x-axis: $\hat{\Lambda} = 0$ and the remaining parameters can be found in tables \ref{tab:SimPar_Binary_Kx_Equal} and \ref{tab:SolPar_Binary_Kx_Equal}. The black vertical marks that radius, which contains $99$\,\% of the mass of the post-merger object. The orange, solid curve is the fit according to equ.(\ref{eq:Schive}), with units for $\delta_s$ in $M_\odot/\text{kpc}^3$ and $r_{sol}$ in kpc, while the green, solid curve is the fit according to equ.(\ref{eq:NFW}), with units for $\delta_{NFW}$ in $M_\odot/\text{kpc}^3$ and $r_{s}$ in kpc.}
    \label{fig:Profile_Binary_Kx_Equal}
\end{figure}

Here the solitons have the same mass of $20$\,CU. The simulation parameters and the soliton parameters for the x-axis variant can be found in tables \ref{tab:SimPar_Binary_Kx_Equal} and \ref{tab:SolPar_Binary_Kx_Equal}. The resulting density profile of the post-merger object is shown in figure \ref{fig:Profile_Binary_Kx_Equal}. We can see that the core is well described with the Schive model, while the outer parts are not well described with a NFW profile. This comes with no surprise, because post-merger objects made of a small number of solitons typically end up with density profiles which look very similar to single solitons, namely core-dominated. The NFW profile can fit fully-developed post-merger envelopes, which arise for multiple ($N > 2$) soliton mergers, as we will see in the next section.

\begin{table}
\caption{\label{tab:SimPar_Binary_Kx_Equal} The simulation parameters for the equal-mass merger without self-interaction, which was initialized, such that the two solitons collide head-on along the x-axis. These simulation parameters are also used for the case, where the two solitons collide along the main diagonal of the x-y-plane. The box size is $\sim 100$\,kpc on a side, the simulation time is $\sim 13.59$\,Gyr, which results -together with the saving number that gives the number of snapshots- in a temporal resolution of $1.359$\,Gyr.}
\begin{ruledtabular}
\begin{tabular}{ccc}
Parameter & Value & Unit \\ \hline
box size & $2.65$ & CU \\
resolution & $256$ & / \\
simulation time & $0.18$ & CU \\
saving number & $10$ & / \\ \hline
\end{tabular}
\end{ruledtabular}
\end{table}

\begin{table}
\caption{\label{tab:SolPar_Binary_Kx_Equal} The soliton parameters for the equal-mass merger without self-interaction, which was initialized such that the two solitons collide head-on along the x-axis, each with mass of $20$\,CU $\sim 44.5\cdot10^6\,M_\odot$. The initial separation of the solitons is $\sim 30.7$\,kpc and the relative velocity is $\sim 5$\,km/s.}
\begin{ruledtabular}
\begin{tabular}{cccccc}
soliton nr. & mass & x & y & v$_{\text{x}}$ & v$_{\text{y}}$ \\ \hline
$1$ & $20$ & $0.4$ & $0$ & $-5$ & $0$ \\
$2$ & $20$ & $-0.4$ & $0$ & $5$ & $0$
\end{tabular}
\end{ruledtabular}
\end{table}

\begin{figure}
    \centering
    \includegraphics[scale=0.25]{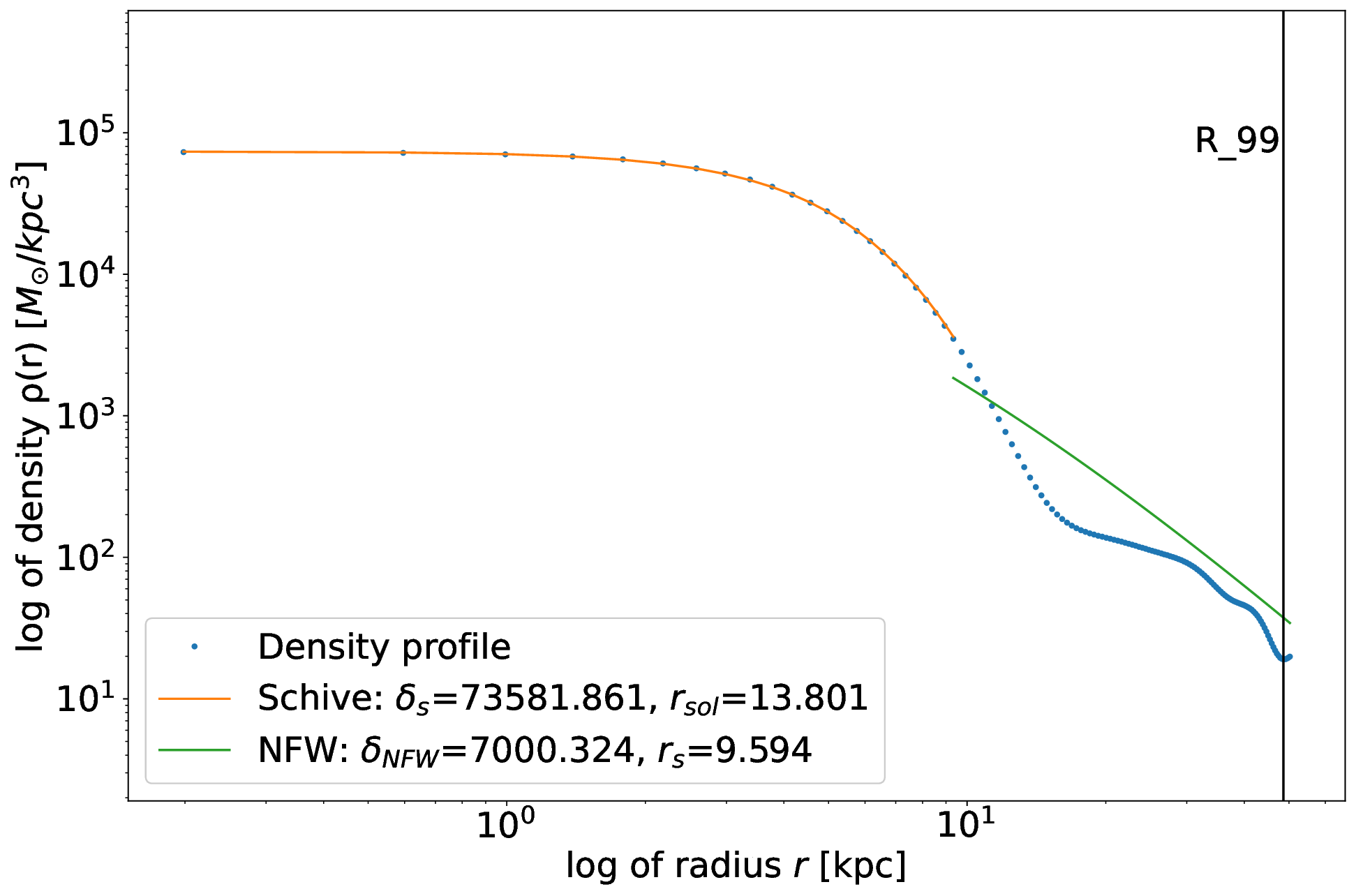}
    \caption{The density profile of the final snapshot ($\sim 13.59$\,Gyr) of a merger of two $20$\,CU-solitons which collide head-on along the main diagonal of the x-y-plane. $\hat{\Lambda} = 0$ and the remaining parameters can be found in tables 
    \ref{tab:SimPar_Binary_Kx_Equal} and \ref{tab:SolPar_Binary_Diagonal_Equal}. The black vertical marks that radius, which contains $99$\,\% of the mass of the post-merger object. The orange, solid curve is the fit according to equ.(\ref{eq:Schive}), with units for $\delta_s$ in $M_\odot/\text{kpc}^3$ and $r_{sol}$ in kpc, while the green, solid curve is the fit according to equ.(\ref{eq:NFW}), with units for $\delta_{NFW}$ in $M_\odot/\text{kpc}^3$ and $r_{s}$ in kpc.}
    \label{fig:Profile_Binary_Diagonal_Equal}
\end{figure}

Now, we look at the diagonal variant of the merger, with the simulation parameters in table \ref{tab:SimPar_Binary_Kx_Equal} and the soliton parameters in table \ref{tab:SolPar_Binary_Diagonal_Equal}. The density profile of the post-merger object is shown in figure \ref{fig:Profile_Binary_Diagonal_Equal}.  We mentioned already that the density profile should in principle look the same as in the x-axis head-on collision, due to rotational symmetry. However, the fit parameters are different, and also the density tail of the data look a little different. The reason are the rounding errors which are introduced when the irrational initial parameters for position and velocity are truncated. Nevertheless, the density profiles look very much the same, with a dominant Schive-fit core, surrounded by a poorly developed envelope.

\begin{table}
\caption{\label{tab:SolPar_Binary_Diagonal_Equal} The soliton parameters for the equal-mass merger without self-interaction, which was initialized such that the two solitons collide head-on along the main diagonal of the x-y-plane, each with mass of $20$\,CU $\sim 44.5\cdot10^6\,M_\odot$.  
The initial separation of the solitons is $\sim 30.5$\,kpc and the relative velocity is $\sim 5$\,km/s. Note that due to the separation in the x- and y-component the values look different to the x-axis head-on collision, but are the same if we look at the absolute value (and ignore rounding errors).}
\begin{ruledtabular}
\begin{tabular}{cccccc}
soliton nr. & mass & x & y & v$_{\text{x}}$ & v$_{\text{y}}$ \\ \hline
$1$ & $20$ & $0.28284$ & $0.28284$ & $-3.53553$ & $-3.53553$ \\
$2$ & $20$ & $-0.28284$ & $-0.28284$ & $3.53553$ & $3.53553$
\end{tabular}
\end{ruledtabular}
\end{table}

\subsubsection{Mergers with $2:1$-mass ratio}\label{sec:UnequalMerger}

\begin{figure}
    \centering
    \includegraphics[scale=0.25]{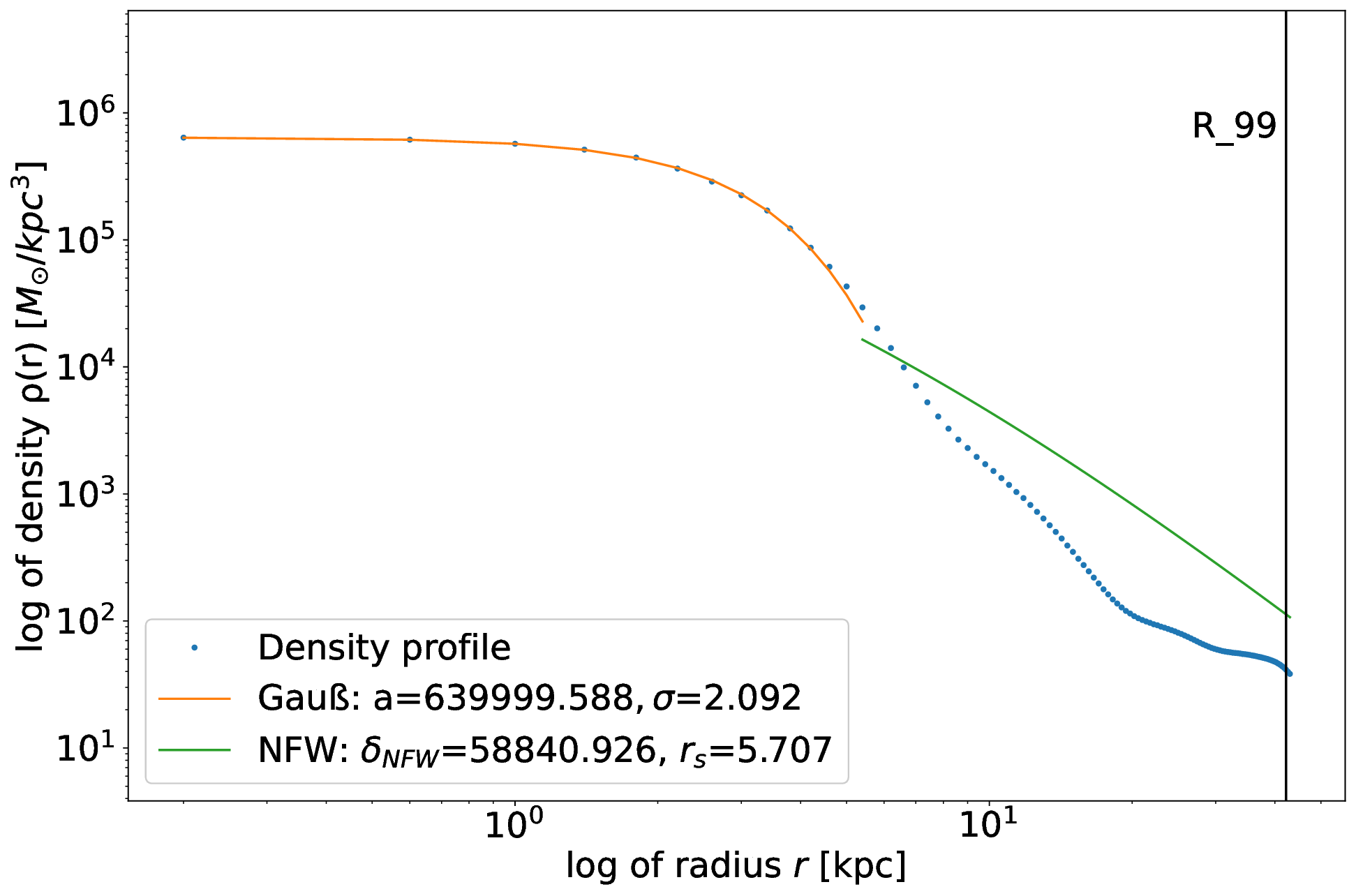}
    \caption{The density profile of the final snapshot ($\sim 13.59$\,Gyr) of a merger with mass ratio of $2$:$1$, where the solitons collided head-on along the x-axis. $\hat{\Lambda} = 0$ and the remaining parameters can be found in tables 
     \ref{tab:SimPar_Binary_Kx_Unequal} and \ref{tab:SolPar_Binary_Kx_Unequal}. The black vertical marks that radius, which contains $99$\,\% of the mass of the post-merger object. The orange, solid curve is the fit according to equ.(\ref{eq:Gauß}), with units for $a$ in $M_\odot/\text{kpc}^3$ and $\sigma$ in kpc, while the green, solid curve is the fit according to equ.(\ref{eq:NFW}), with units for $\delta_{NFW}$ in $M_\odot/\text{kpc}^3$ and $r_{s}$ in kpc.}
    \label{fig:Profile_Binary_Kx_Unequal}
\end{figure}

Here, one soliton has twice the mass of the other soliton. Since the total mass is now bigger, this would enable us to use higher velocities. But to compare the different merger simulations with each other, we decided to leave the relative velocity at $5$\,km/s. We start again with the x-axis head-on collision. The simulation and soliton parameters for this case can be found in tables \ref{tab:SimPar_Binary_Kx_Unequal} and \ref{tab:SolPar_Binary_Kx_Unequal}, respectively. We note that we used a bigger box, in order not to miss any SFDM mass that might otherwise leave and reenter the box. The density profile of the post-merger object for this merger variant is shown in figure \ref{fig:Profile_Binary_Kx_Unequal}.
At first glance, the profile looks very similar to equal-mass mergers, yet there are some notable differences. First, the connection point between core and envelope, that was determined by the fitting routine, lies closer toward the origin. This makes sense, since a soliton with a higher mass is smaller in size (for $\hat{\Lambda}=0$). As a result, the core, which still dominates the profile, is also smaller. Furthermore, the central density is now much higher, since more mass is able to accumulate in the center. While the smaller core "leaves more room" for a comparatively larger envelope, the latter is still not developed well enough to make the NFW model a good fit to the data. However, also the core is not fit as well as in the equal-mass merger case. One explanation could be again the "high-mass problem", i.e. the spatial resolution for this merger may not be good enough to resolve the central regions of the final object. We were prevented from increasing the spatial resolution, since the box size was set such that the calculated mass agreed with the initial mass. The highest resolution we tested was $512$, but much higher resolution would be required to resolve the central densities. Given the purpose for which {\ffmfamily PyUltraLight} - and our code modification of it-, was designed in the first place (to investigate SFDM dynamics on laptop workstations), we did not pursue further HPC resources for this project. For the dedicated simulations here, we can say that the limited spatial resolution in the unequal-mass mergers (at least with the masses we used) will tend to produce smaller values for the central densities than expected. Still, the analysis of the density profiles is not much affected by this limitation. Finally, we recognize that the envelope does not show as many global oscillations, as in the equal-mass merger cases. We think that the higher total mass helps the merger to get more rapidly to a (near-)equilibrium state.

\begin{table}
\caption{\label{tab:SimPar_Binary_Kx_Unequal} The simulation parameters for the merger without self-interaction, where the solitons had a mass ratio of $2$:$1$. They were initialized such that the two solitons collide head-on along the x-axis. The box size is $\sim 101.5$\,kpc on a side, the simulation time is $\sim 13.59$\,Gyr, which results together with the saving number (gives the number of snapshots) in a temporal resolution of $1.359$\,Gyr.}
\begin{ruledtabular}
\begin{tabular}{ccc}
Parameter & Value & Unit \\ \hline
box size & $2.67$ & CU \\
resolution & $256$ & / \\
simulation time & $0.18$ & CU \\
saving number & $10$ & / \\
\end{tabular}
\end{ruledtabular}
\end{table}

\begin{table}
\caption{\label{tab:SolPar_Binary_Kx_Unequal} The soliton parameters for the merger without self-interaction, where the solitons had a mass ratio of $2$:$1$. They were initialized such that the two solitons collide head-on along the x-axis, with mass of $20$\,CU $\sim 44.5\cdot10^6\,M_\odot$ and $40$\,CU $\sim 89\cdot10^6\,M_\odot$, respectively, an initial separation of $\sim 30.7$\,kpc and a relative velocity of $\sim 5$\,km/s.}
\begin{ruledtabular}
\begin{tabular}{cccccc}
soliton nr. & mass & x & y & v$_{\text{x}}$ & v$_{\text{y}}$ \\ \hline
$1$ & $40$ & $0.4$ & $0$ & $-5$ & $0$ \\
$2$ & $20$ & $-0.4$ & $0$ & $5$ & $0$
\end{tabular}
\end{ruledtabular}
\end{table}

\begin{figure}
    \centering
    \includegraphics[scale=0.25]{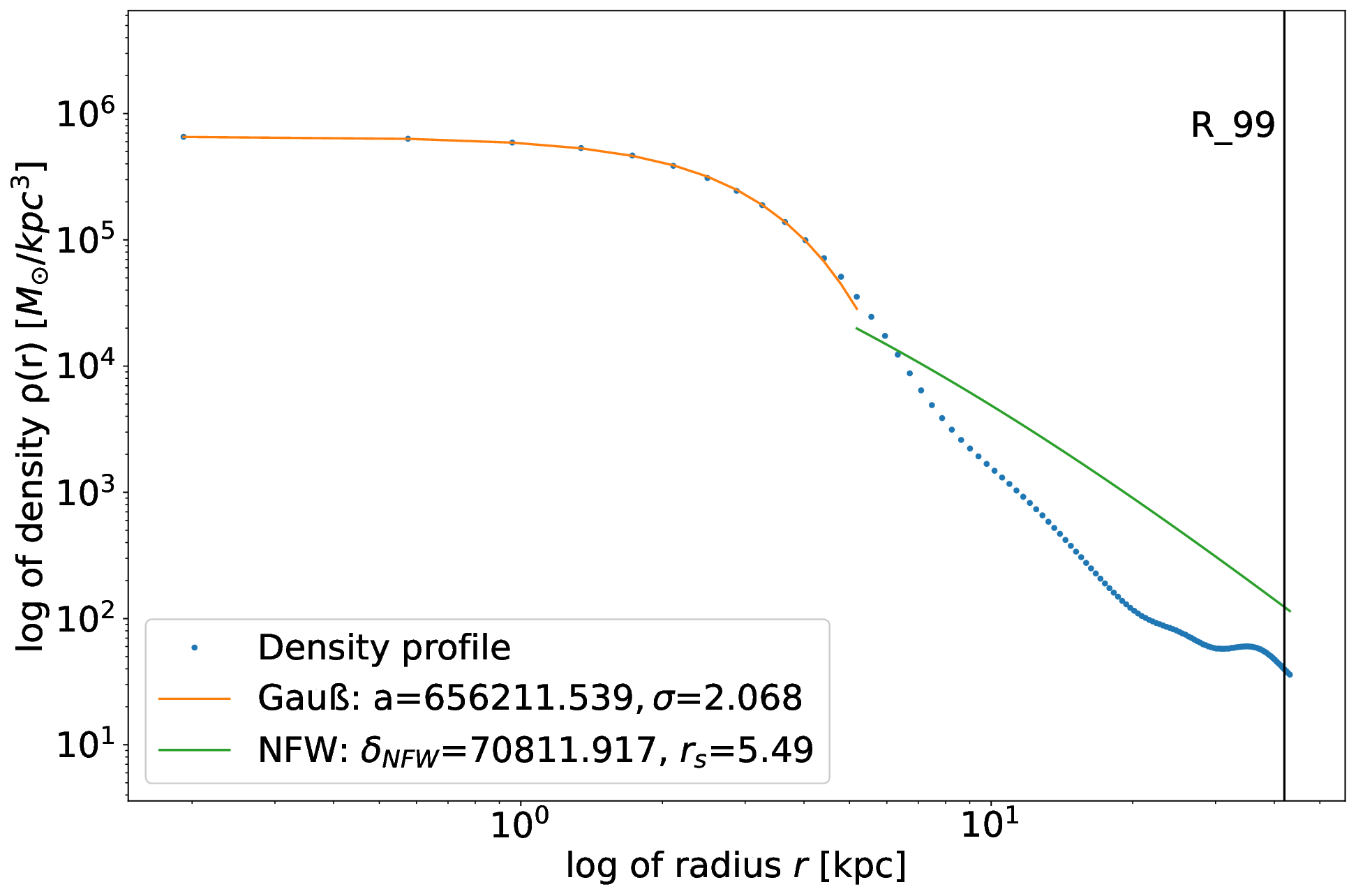}
    \caption{The density profile of the final snapshot ($\sim 13.59$\,Gyr) of the merger with mass ratio of $2$:$1$, where the solitons collided head-on along the main diagonal of the x-y-plane. $\hat{\Lambda} = 0$ and the remaining parameters can be found in tables 
    \ref{tab:SimPar_Binary_Diagonal_Unequal} and \ref{tab:SolPar_Binary_Diagonal_Unequal}. The black vertical marks that radius, which contains $99$\,\% of the mass of the post-merger object. The orange, solid curve is the fit according to equ.(\ref{eq:Gauß}), with units for $a$ in $M_\odot/\text{kpc}^3$ and $\sigma$ in kpc, while the green, solid curve is the fit according to equ.(\ref{eq:NFW}), with units for $\delta_{NFW}$ in $M_\odot/\text{kpc}^3$ and $r_{s}$ in kpc.}
    \label{fig:Profile_Binary_Diagonal_Unequal}
\end{figure}

Next, we turn our attention to the diagonal collision. Here we used the simulation and soliton parameters from tables \ref{tab:SimPar_Binary_Diagonal_Unequal} and \ref{tab:SolPar_Binary_Diagonal_Unequal}, respectively. In this case, we were able to reduce the box size in comparison to the x-axis collision, since we had more room in the diagonal direction (as intended). The final density profile of the post-merger object can be seen in figure \ref{fig:Profile_Binary_Diagonal_Unequal}. There are no big differences, compared to the x-axis case, again explained due to the rotational symmetry. Still, the limitations mentioned above prevail, leading to lower accuracy in the central region. We expect the "real" central density to be higher than what can be seen in figure \ref{fig:Profile_Binary_Diagonal_Unequal}.

\begin{table}
\caption{\label{tab:SimPar_Binary_Diagonal_Unequal} The simulation parameters for the merger without self-interaction, where the solitons had a mass ratio of $2$:$1$. They were initialized such that the two solitons collide head-on along the main diagonal of the x-y-plane. The box size is $\sim 97$\,kpc on a side, the simulation time is $\sim 13.59$\,Gyr, which results together with the saving number (gives the number of snapshots) in a temporal resolution of $1.359$\,Gyr.}
\begin{ruledtabular}
\begin{tabular}{ccc}
Parameter & Value & Unit \\ \hline
box size & $2.56$ & CU \\
resolution & $256$ & / \\
simulation time & $0.18$ & CU \\
saving number & $10$ & / \\
\end{tabular}
\end{ruledtabular}
\end{table}

\begin{table}
\caption{\label{tab:SolPar_Binary_Diagonal_Unequal} The soliton parameters for the merger without self-interaction, where the solitons had a mass ratio of $2$:$1$. They were initialized such that the two solitons collide head-on along the main diagonal of the x-y-plane, with mass of $20$\,CU $\sim 44.5\cdot10^6\,M_\odot$ and $40$\,CU $\sim 89\cdot10^6\,M_\odot$, respectively, an initial separation of $\sim 30.7$\,kpc and a relative velocity of $\sim 5$\,km/s. Note that due to the separation in the x- and y-component the values look different to the x-axis collision, but are the same if we look at the absolute value (and ignore rounding errors).}
\begin{ruledtabular}
\begin{tabular}{cccccc}
soliton nr. & mass & x & y & v$_{\text{x}}$ & v$_{\text{y}}$ \\ \hline
$1$ & $40$ & $0.28284$ & $0.28284$ & $-3.53553$ & $-3.53553$ \\
$2$ & $20$ & $-0.28284$ & $-0.28284$ & $3.53553$ & $3.53553$
\end{tabular}
\end{ruledtabular}
\end{table}

\subsection{Mergers with self-interaction}\label{sec:MergerSI}

Our main interest has been the impact of SI onto soliton and merger dynamics. However, our tests have shown that simulation runs of mergers with SI are troubled with competing requirements, in terms of physics and simulation parameters. Therefore, we present a good case of a binary merger with SI in this subsection, but consider the subject far from being settled.
As a general rule, we observe that the addition of repulsive SI makes successful mergers much more difficult, for the same dynamical reasons as discussed in the previous section: a repulsive SI works against self-gravity, in addition to the quantum pressure that is always at play. Therefore, successful binary soliton mergers for $\hat{\Lambda} = 0$ may fail to merge, once $\hat{\Lambda} > 0$. This problem gets worse for multiple ($N>2$) mergers (see next section).

For $\hat{\Lambda}=1$, we failed to produce any mergers within a Hubble time. Thus, we focus on $\hat{\Lambda}=0.02$.
Moreover, for equal-mass mergers we also failed to get mergers within a Hubble time, even for small relative velocity: if the velocity is small, the solitons will not have enough time to come close enough to interact, or the merger is delayed due to the repulsive SI which pushes the solitons apart again, after their first encounter (if one happens), preventing a successful merger, after all. We found that mergers can happen within $2-3$ Hubble times, at best. Inspecting the corresponding density profiles at earlier snapshots reveals that the ''binary-soliton-objects'' are far from any (near-)equilibrium, let alone to show the characteristics of a final merger object.
In principle, there are three knobs to turn to create successful mergers: i) increase the simulation time well beyond a Hubble time; ii) choose a tiny value for $\hat{\Lambda} \not= 0$; iii) reduce the initial separation, which enables a smaller relative velocity, in turn. Option (ii) is unattractive, given our research scope. Option (iii) is plagued by the issue mentioned in Section \ref{sec:3A1} and Appendix \ref{appendix:Problems}, and would demand a very large suite of dedicated, tailored simulations, to find the best set of parameters. This approach is also questionable from a physics' perspective, for the best set of simulation parameters are unlikely to be realized in nature. In addition, option (iii) bears the danger that solitons start overlapping too early in the process \footnote{In fact, the quantum wave nature of SFDM solitons changes the way how they interact (interfere!) and how the assembly of structure proceeds in detail. In other words, the long-range nature of gravity, which is always at play in cosmology, is compounded with the long-range nature of quantum interference effects in SFDM models. }, circumventing a real merger simulation in the first place.
This leaves option (i), a possibility which is often pursued in simulations. However, we keep our demand to seek for mergers within a Hubble time for two reasons: first, we want to compare ''apples with apples'' by choosing the same final snapshots in our simulations. Second, we deem it worthwhile to succumb SFDM to the \textit{Gretchenfrage}: if mergers (eventually structure assembly) with certain soliton masses at given $\hat{\Lambda} \not= 0$ fail to happen within a Hubble time, could SFDM models, which predict such soliton mass spectra, be disfavored as reasonable DM candidates?  
   
\begin{figure}
    \centering
    \includegraphics[scale=0.25]{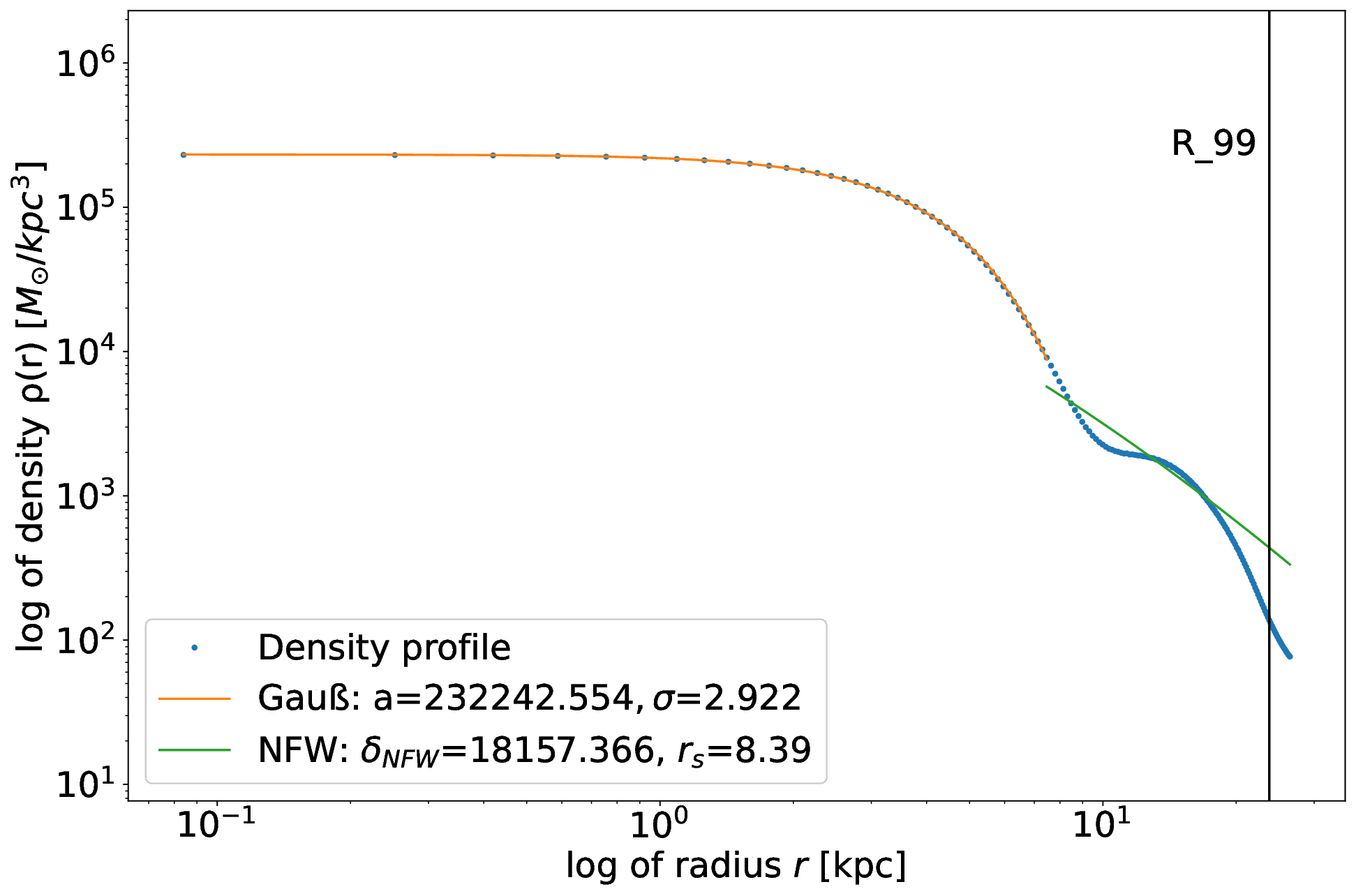}
    \caption{The density profile of the final snapshot ($\sim 13.59$\,Gyr) of the merger with mass ratio of $2$:$1$, where the solitons collided head-on along the main diagonal of the x-y-plane. 
    $\hat{\Lambda} = 0.02$ and the remaining parameters can be found in tables \ref{tab:SimPar_Binary_Diagonal_SI} and \ref{tab:SolPar_Binary_Diagonal_SI}.
      The black vertical marks that radius, which contains $99$\,\% of the mass of the post-merger object. The orange, solid curve is the fit according to equ.(\ref{eq:Gauß}), with units for $a$ in $M_\odot/\text{kpc}^3$ and $\sigma$ in kpc, while the green, solid curve is the fit according to equ.(\ref{eq:NFW}), with units for $\delta_{NFW}$ in $M_\odot/\text{kpc}^3$ and $r_{s}$ in kpc.}
    \label{fig:Profile_Binary_Diagonal_SI}
\end{figure}

In fact, this leaves the choice of soliton mass to come to the rescue. A higher total mass helps to accelerate the merging process and keeps the repulsive SI at bay. However, the issue with the initial separation does not go away with a higher mass, since the repulsive SI works in favor of exacerbating the overlap problem by expanding the solitons, compared to their brethrens without SI. 
Thus, we focus on unequal-mass mergers for their higher total mass and consider diagonal collisions.
We succeeded in producing mergers within a Hubble time and for a box size appropriate for a good (calculated) mass resolution, if we choose a low enough relative velocity.  The simulation and the soliton parameters are listed in tables \ref{tab:SimPar_Binary_Diagonal_SI} and \ref{tab:SolPar_Binary_Diagonal_SI}, respectively. The resulting density profile of the post-merger object of this successful simulation can be seen in figure \ref{fig:Profile_Binary_Diagonal_SI}.

\begin{table}
\caption{\label{tab:SimPar_Binary_Diagonal_SI} The simulation parameters for the merger with $\hat{\Lambda}=0.02$, where the solitons had a mass ratio of $2$:$1$. They were initialized such that the two solitons collide head-on along the main diagonal of the x-y-plane. The box size is $\sim 53$\,kpc on a side, the simulation time is $\sim 13.59$\,Gyr, which results together with the saving number (gives the number of snapshots) in a temporal resolution of $1.359$\,Gyr.}
\begin{ruledtabular}
\begin{tabular}{ccc}
Parameter & Value & Unit \\ \hline
box size & $1.4$ & CU \\
resolution & $320$ & / \\
simulation time & $0.18$ & CU \\
saving number & $10$ & / \\
\end{tabular}
\end{ruledtabular}
\end{table}

\begin{table}
\caption{\label{tab:SolPar_Binary_Diagonal_SI} The soliton parameters for the merger with $\hat{\Lambda}=0.02$, where the solitons had a mass ratio of $2$:$1$. They were initialized such that the two solitons collide head-on along the main diagonal of the x-y-plane, with mass of $20$\,CU $\sim 44.5\cdot10^6\,M_\odot$ and $40$\,CU $\sim 89\cdot10^6\,M_\odot$. The initial separation of the solitons is $\sim 30.7$\,kpc and the relative velocity is $\sim 0.25$\,km/s.}
\begin{ruledtabular}
\begin{tabular}{cccccc}
soliton nr. & mass & x & y & v$_{\text{x}}$ & v$_{\text{y}}$ \\ \hline
$1$ & $40$ & $0.28284$ & $0.28284$ & $-0.17678$ & $-0.17678$ \\
$2$ & $20$ & $-0.28284$ & $-0.28284$ & $0.17678$ & $0.17678$
\end{tabular}
\end{ruledtabular}
\end{table}

Similarly to the previous cases of $\hat{\Lambda} = 0$, the profile of the post-binary-merger object has a very extended, dominant core and small envelope. In fact, this overall feature is more pronounced, now that repulsive SI has been added, for SI tends to produce bigger cores. Unsurprisingly, the NFW model provides a poor fit to the numerical data in the small envelope.
Now, a wider core usefully favors the Schive model as a core fit, since this model does not fall off as rapidly as the Gauss model. Nevertheless, in this specific case, the Gauss model has been preferred by the score system of the fitting routine. We should mention, though, that in this simulation we increased the resolution, given the computational resources that go into merger simulations, in order to provide satisfactory results. The value was chosen according to test simulations, where we used lower resolutions (at smaller box size) to see what central densities we should expect. The increased spatial resolution would still not be enough to avoid the high-mass problem; for that we would have to increase the resolution above $384$. However, since this problem is common for the unequal-mass mergers we did not redo the simulation. Still, the characteristics of the final profile fit can change, if ever so slightly.

While a higher total mass helps to produce mergers (as demonstrated here), we still think there is reason to be cautious: higher soliton masses lead to higher central densities, which results in a higher impact of the repulsive SI over self-gravity. That means, SI may yet prevent a merger from happening, or at least decelerate it, such that one Hubble time is not enough to produce a (near-)equilibrium final merger state, even for high soliton masses. Therefore, we expect that mergers with repulsive SI are only possible within a certain mass range of solitons, as already discussed earlier.
How these conclusions fit with the results of some cosmological simulations of FDM (e.g. by \cite{Mocz_Solitonen2017}) as well as SFDM-TF (e.g. by \cite{Hartman2022}), which both have produced successful mergers, albeit at a much higher mass scale, remains to be seen. Certainly, more future work on SFDM structure formation in different SI regimes will be required.

\section{\label{sec:sec5}Multiple-Soliton Mergers}

In this section, we consider finally multiple mergers, where more than two solitons are involved. However, the difficulties mentioned previously for binary mergers with repulsive SI are exacerbated for multiple mergers. Therefore, we present here our results for
$\hat{\Lambda}=0$, and defer the analysis of repulsive SI to future work. The main advantage we have is that we can draw on experience gained from many simulations that we performed, and the possibility to compare our results to previous literature concerning similar studies. Our search using a variety of parameters helped to find good initial conditions for all mergers with more than two solitons. We highlight here some of our results.

\subsection{Long-time behavior}

We already addressed the generic problem in SFDM to produce successful mergers within a Hubble time, given the highly oscillatory nature of the system which makes relaxation harder to attain, compared to standard DM models.
After all, identifying a meaningful final merger end state requires some (near-) equilibrium to be achieved.
Indeed, in the course of performing many test runs for the merger parameters, we found that, even with good enough parameters where the merging process was completed after half of the simulation time, the post-merger object at the final snapshot still oscillates significantly. This oscillation is potent enough that it affects the density profiles such to yield changes over time (the central density can change, accompanied by persistent oscillations in the outer parts). In this sense, it is not quite correct to speak of a final or end state of the post-merger object - it rather signifies the state at the final chosen snapshot.
Depending on the soliton and merger parameters, this non-relaxed final state may persist during many Hubble times, a characteristic feature for any SI regime, though more pronounced for repulsive SI. So, the question arises how long it actually takes for the merger to settle down to a final (near-) equilibrium state.

As an example, we simulated a 5-soliton merger for a "short" (roughly one Hubble time) and a long simulation time and compared the results. To be specific, we used a $5$-soliton merger which is oriented in the $x-y$-plane in a pentagon (also called "star shape"), where each soliton had the same mass of $20$\,CU and the same distance to the center of $0.6$\,CU. For the velocity, we ensured that -in absolute value- all solitons move with a velocity of $5$\,CU ($\sim 2.5$\,km/s), inspired by the choice of \cite{Schwabe2016} (though their simulation times span hundreds of Hubble times, while our "long-time" simulation spans roughly ten Hubble times, see below). 
A full overview of the parameters of the solitons can be found in table \ref{tab:SolPar_Star2D}. Since we are in the $x-y$-plane, all z-components are zero.

\begin{table}
\caption{\label{tab:SolPar_Star2D} The soliton parameters for the ($N=5$)-merger for $\hat{\Lambda}=0$, where the solitons were initialized such that the five solitons collide in the center of the box, starting in pentagon/star shape, each with mass of $20$\,CU $\sim 44.5\cdot10^6\,M_\odot$, an initial distance between soliton and center of the box of $\sim 23$\,kpc, and a velocity of $\sim 2.5$\,km/s. We note that due to the separation in the x- and y-component, the values look different but are the same if we look at the absolute value.}
\begin{ruledtabular}
\begin{tabular}{cccccc}
soliton nr. & mass & x & y & v$_{\text{x}}$ & v$_{\text{y}}$ \\ \hline
$1$ & $20$ & $0$ & $0.6$ & $0$ & $-5$ \\
$2$ & $20$ & $0.57063$ & $-0.18541$ & $-4.75525$ & $-1.54508$ \\
$3$ & $20$ & $0.35267$ & $-0.48541$ & $-2.93893$ & $4.04508$ \\
$4$ & $20$ & $-0.35267$ & $-0.48541$ & $2.93893$ & $4.04508$ \\
$5$ & $20$ & $-0.57063$ & $0.18541$ & $4.75525$ & $-1.54508$ \\
\end{tabular} 
\end{ruledtabular}
\end{table}

\begin{figure}
    \centering
    \includegraphics[scale=0.25]{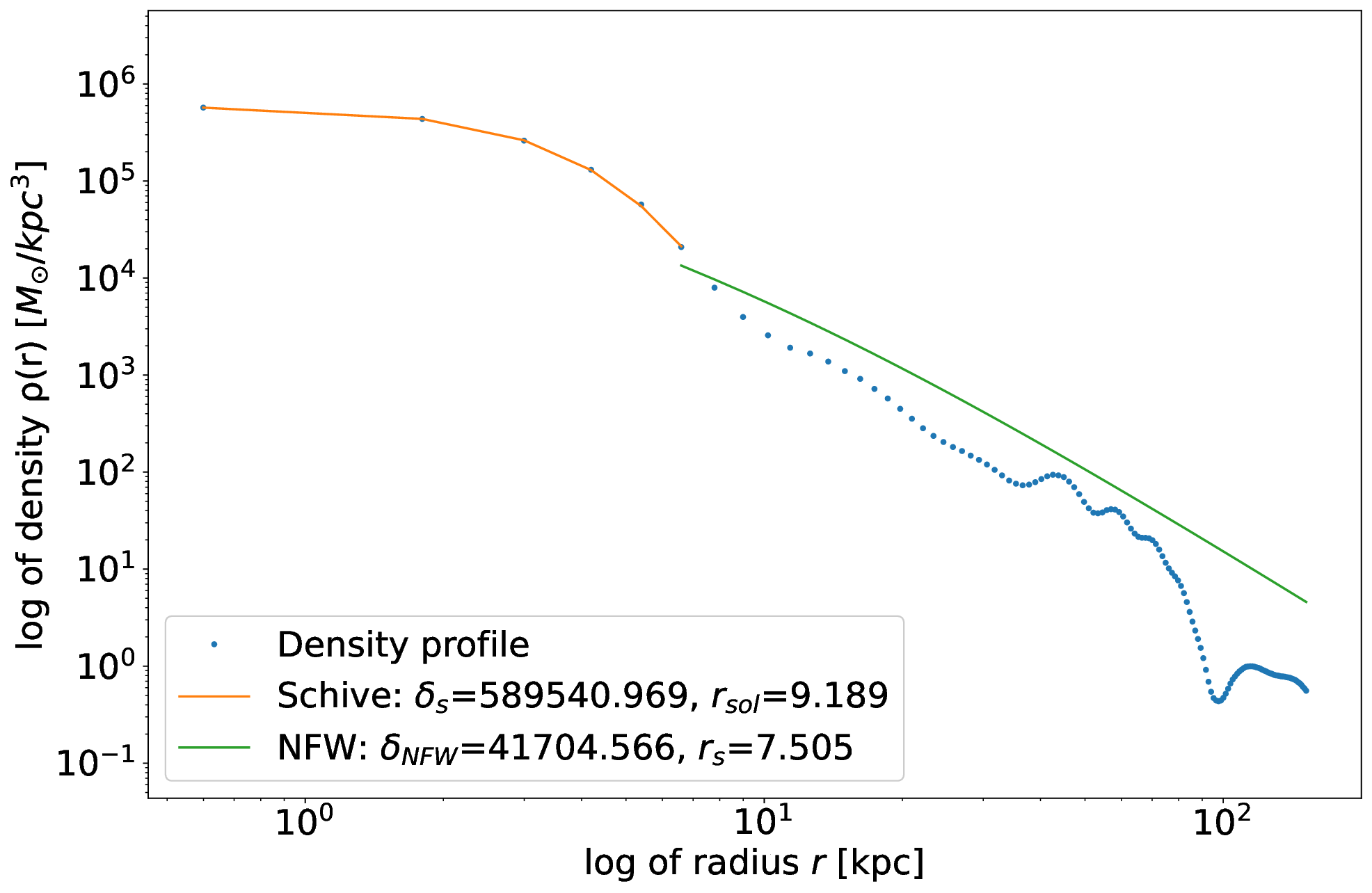}
    \caption{The density profile of the final snapshot ($\sim 13.59$\,Gyr) of the "short-time" $5$-soliton merger with $\hat{\Lambda}=0$, where the solitons collided in the center of the box, initially configured in a pentagon/star shape in the $x-y$-plane. The soliton parameters for this particular density profile can be found in table \ref{tab:SolPar_Star2D}. The box size was $4$\,CU and the resolution $384$. The orange, solid curve is the fit according to equ.(\ref{eq:Schive}), with units for $\delta_s$ in $M_\odot/\text{kpc}^3$ and $r_{sol}$ in kpc, while the green, solid curve is the fit according to equ.(\ref{eq:NFW}), with units for $\delta_{NFW}$ in $M_\odot/\text{kpc}^3$ and $r_{s}$ in kpc.}
    \label{fig:Profile_Star2D_Shortterm}
\end{figure}

For the "short-time" simulation we used a box size of $4$\,CU ($\sim 150$\,kpc) and a resolution of 384, which is enough to avoid the high-mass problem, although it is not tailored to yield an accurate total mass resolution, since this would have caused a very long calculation time and the mass accuracy is not so important for the trend of the density profile, as we discussed above \footnote{Since the mass is not accurate, we also do not include the $R_{99}$ boundary in the plot.}. The duration of the "short-time" simulation was set to the usual $0.18$\,CU ($\sim 13.59$\,Gyr), i.e. roughly a Hubble time. The resulting density profile of the post-merger object at the final snapshot can be seen in figure \ref{fig:Profile_Star2D_Shortterm}.

We clearly recognize oscillations in the tail of the profile. Apart from that, we also see that this merger developed a more pronounced envelope, due to the higher number of solitons ($N=5$, compared to $N=2$ before). As a result, and in accordance with expectations, the NFW model fits better now, in terms of the slope over most of the envelope.

\begin{figure}
    \centering
    \includegraphics[scale=0.25]{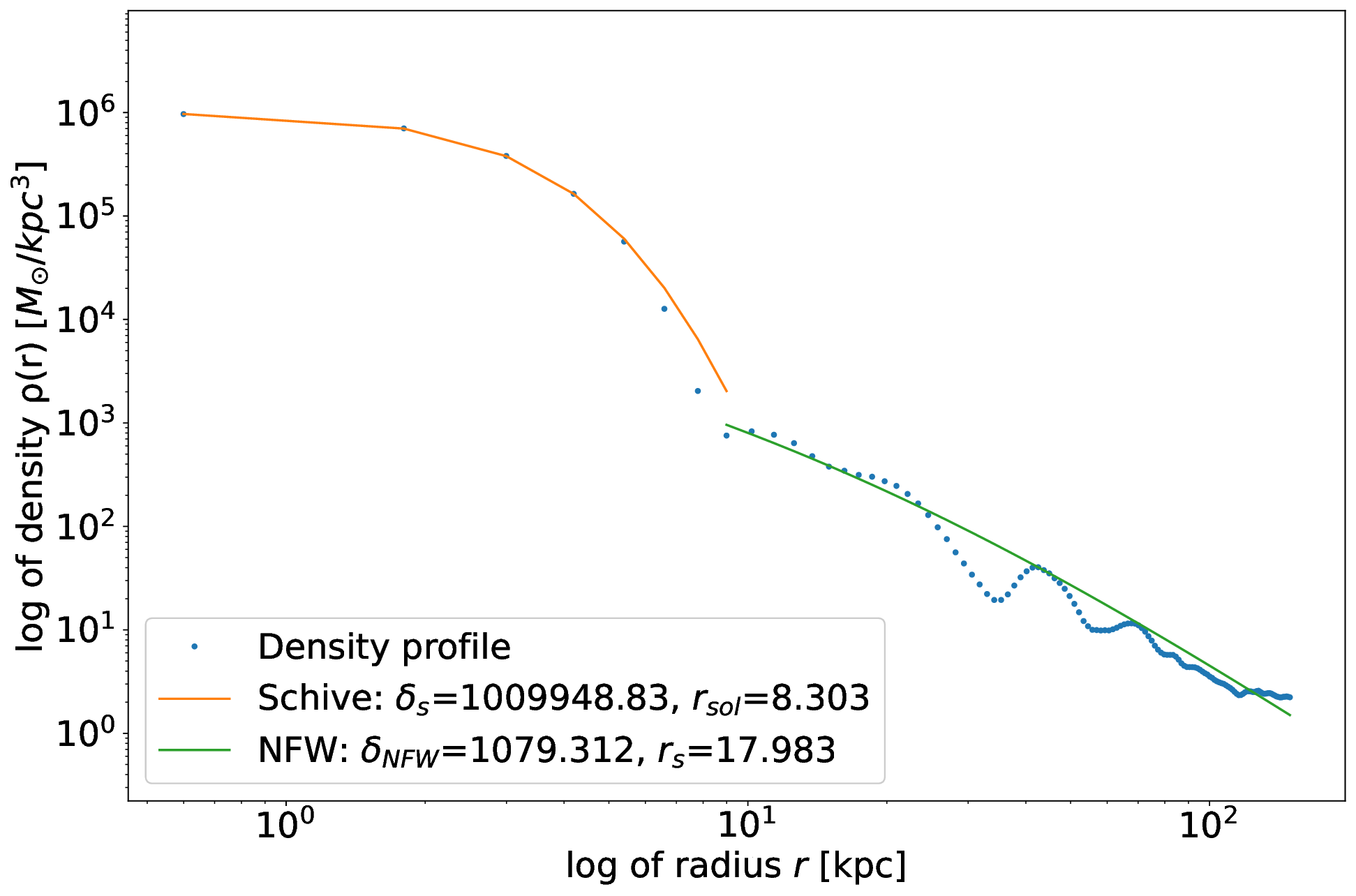}
    \caption{The density profile of the final snapshot ($\sim 150$\,Gyr) of the long-time $5$-soliton merger with $\hat{\Lambda}=0$, where the solitons collided in the center of the box, initially configured in a pentagon/star shape in the $x-y$-plane. The soliton parameters are the same as in figure \ref{fig:Profile_Star2D_Shortterm} and can be found in table \ref{tab:SolPar_Star2D}. The box size was $8$\,CU and the resolution $256$. The orange, solid curve is the fit according to equ.(\ref{eq:Schive}), with units for $\delta_s$ in $M_\odot/\text{kpc}^3$ and $r_{sol}$ in kpc, while the green, solid curve is the fit according to equ.(\ref{eq:NFW}), with units for $\delta_{NFW}$ in $M_\odot/\text{kpc}^3$ and $r_{s}$ in kpc. For this fit, the connection point in the fitting routine was moved manually three data points towards the right side.}
    \label{fig:Profile_Star2D_Longterm_3PRight}
\end{figure}

Now, for the long-time simulation we increase the duration to $2$\,CU ($\sim 150$\,Gyr), i.e. roughly ten times longer than a Hubble time. Owing to the long simulation time, we needed to reduce the spatial resolution for computational feasibility, so we used a box size of $8$\,CU and a resolution of $256$. These latter two parameters have been commonly used by the simulations of \cite{PyUltraLight}, but these authors were not concerned with soliton mergers.
In doing so, we encounter the regime of the high-mass problem, yielding a reduced central density, but since we are interested in the outer envelope/tail, this is no major restriction. The other parameters stayed the same as in the previous run. The resulting profile of the post-merger object for the long-time simulation is shown in figure \ref{fig:Profile_Star2D_Longterm_3PRight}. To fit this profile, we manually moved the connection point three data points towards the right side, such that the NFW model fits the data better. Of course, this comes with the price of loosing accuracy in the core model fit. The reason for moving the connection point is that we can see a clear distinction between core and envelope, while the automatic search for the connection point failed in recognizing this feature in this multiple-merger case. (Remember that the score system of fit criteria described in Section II D 3 is tailored for single cores.) 

Moreover, we recognize the following.
Although we expected a lower central density, due to the loss of spatial resolution, the central density is about $1.75$ times higher than in the short-time simulation. The explanation is the non-relaxed, oscillating merger: the merging object keeps oscillating well after one Hubble time, i.e. the density profile also keeps undergoing some changes. Thanks to the oscillations, the object expands and contracts, changing the central density in turn. It just so happens that figure \ref{fig:Profile_Star2D_Shortterm} displays a snapshot, where more mass happened to be in the middle part of the merger, while figure \ref{fig:Profile_Star2D_Longterm_3PRight} shows a snapshot where more mass happens to be in the central parts. 
This issue can be also seen in the middle part of the profiles: in the short-time simulation we have a less-pronounced transition between central and outer parts, while in the long-time simulation we have a more clear-cut distinction between core and envelope parts of the merger object.

Finally, with respect to the envelope, we recognize that in both runs the NFW model fits the outer data, at least the slope is well represented, as a result of the multiple merger. Nevertheless, the oscillations in the envelope have not ceased after $150$\,Gyr, instead they are persistent in terms of amplitude. Thus, we could show that not only the core region but also the envelope undergoes some significant changes, well after one Hubble time. This characteristic feature exemplified for FDM (i.e. SFDM without SI) is well in accordance with the findings of \cite{Veltmaat2018}, confirming that SFDM - in this case FDM as a special case- has a very long relaxation time. It remains to be seen whether this feature presents a potential problem for SFDM \footnote{While galactic or galaxy cluster halos are never exactly in equilibrium, due to infall of smaller-mass objects, the requirement of establishing some kind of (near-)equilibrium seems to be a necessity for successful DM models. Of course, the degree of equilibrium depends on the considered mass scale, e.g. galaxy clusters experience permanent infall, yet there are some \textit{calm} clusters like the Coma cluster whose observation may be in conflict with a DM model having high relaxation times.}.
In terms of the fitting results, we also stress that a direct comparison of the resulting fit parameters for the Schive and the NFW model, respectively, between figures \ref{fig:Profile_Star2D_Shortterm} and \ref{fig:Profile_Star2D_Longterm_3PRight} seems not very useful, given the complications of the non-relaxed state of the post-merger object (apart from the very different snapshot in time between the figures). On the contrary, the difference in the results of the fitting parameters, especially for NFW, is yet another signature of the non-relaxed state. In fact, our results suggest that neither analytical density profile proofs as "ultimate model" for post-merger SFDM objects, and should thus be used with this caution in mind, when comparing them with simulation data of SFDM structure formation. Apart from the issue of long relaxation times, it is also true that the current lack of theoretically motivated global profiles for SFDM halos is an obstacle to a proper fitting of density models to numerical data. Adding an appropriate analytical model for the "middle part", which would consistently connect the core with the envelope, would certainly improve upon density fits, such as ours presented in this paper, but even such models are currently missing, or lacking justification for the comparatively low-mass single and merger soliton objects, which we study in this paper. Turning the issue on its head, we could stress again that the fits, as performed here, provide us with a tool to recognize and demonstrate the oscillatory behavior of single and merger soliton objects, given the change of the corresponding fitting parameters.

\subsection{Impact of the initial configuration onto the density profile}\label{sec:KonfiAnalysis}

We have seen above that mergers keep oscillating well after the initial collision and mass assembly, with significant effects onto the density profile. Now, we might ask how the oscillations themselves depend upon the initial soliton configuration prior to their collision. In this subsection, we want to take a closer look on how the initial configuration, which determines the form of the collision depending on number and position of the solitons, might impact the density profiles in the final state.

In order to do so, we focus our investigation on the central densities of the profiles and analyse them, according to their initial configuration. In our merger simulations, we only used scenarios where the solitons start in a given initial position and then move toward the center of the box. In order for the merger to proceed evenly, all solitons have the same distance to the center, the same mass and move with a velocity of $5$\,CU ($\sim 2.5$\,km/s). Again, these parameters were chosen to guarantee that a merger happens within a Hubble time, i.e. the simulation time was also the same with $0.18$\,CU ($\sim 13.59$\,Gyr). To keep track of the configurations for each profile, we introduced a convention that looks like this:
\begin{equation}\label{eq:Convention}
    A (B), \text{Note}.
\end{equation}
Here is
\begin{itemize}
    \item $A$ the number of merging solitons.
    \item  $B$ denotes the mass of the solitons. If unspecified, it can be easily inferred from the total mass and the number of solitons, since all solitons have the same mass.
    \item Note: an abbreviation to indicate the configuration, where
    \begin{itemize}
        \item K denotes a frontal head-on collision along the given axis (e.g. Kx stands for a head-on collision along the x-axis). With a head-on collision we mean that two solitons approach each other on an axis by moving directly towards each other.
        \item x, y, and z: the three axes of the Cartesian coordinate system in which the solitons are initialized.
        \item a hyphen: a separation between axes (e.g. x-y means, one soliton is initialized on the x-axis and one on the y-axis).
    \end{itemize}
\end{itemize}

We also use words to describe the configuration (e.g. \textit{star shape} or \textit{triangle}). According to this convention a "$4$, Kx-y-z" stands for $4$ solitons (with a given total mass), where two collided head-on along the x-axis, while the other two were initialized on the y and z axis, respectively.
In our convention, we also use the term \textit{old} and \textit{new}. Old simulations are those where the high-mass problem had not yet been identified. Therefore, the central densities of those simulations are lower than expected. For those simulations, a box size of $8$\,CU and $256$ for the resolution was used. New simulations took care of the high-mass problem, but do not necessarily have an accurate mass resolution. As already argued, this restriction does not influence the central density much, which is why we did not pursue fine-tuned simulation runs. As for the simulation parameters, we use the ones from table \ref{tab:IdealResolBox} for the total mass.

\begin{figure*}
    \centering
    \includegraphics[scale=0.55]{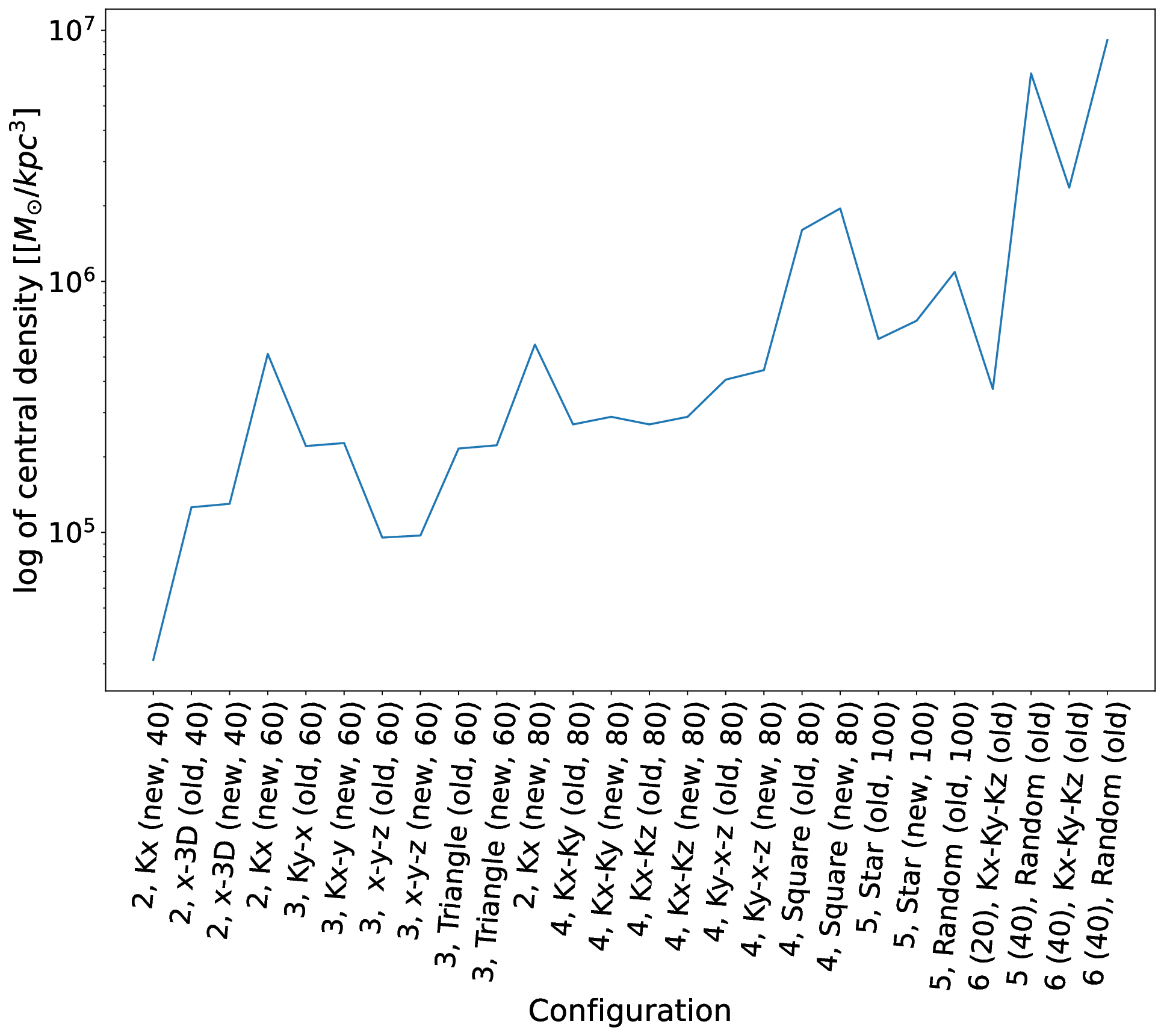}
    \caption{The central density from the final snapshots ($\sim 13.59$ Gyr) plotted versus the used geometric initial configuration. The note indicates, according to the convention in (\ref{eq:Convention}), if the simulation data stem from old or new simulations and which total mass was used. In Appendix \ref{appendix:ConfiTables}, we add more discussion.}
    \label{fig:KonfiAnalysis}
\end{figure*}

In figure \ref{fig:KonfiAnalysis}, we collect the central densities from all our multiple soliton merger simulations and plot them over the respective configuration. The configurations are sorted by total mass and, if equal, by number of solitons. In addition, an attempt was made to place the old and new simulations quite close to each other in order to illustrate the rise in density as a result of the better spatial resolution of the latter (Appendix \ref{appendix:ConfiTables} provides some more discussion, concerning the configurations.). We recognize two main trends:
    i) Mergers with fewer solitons tend to produce a higher central density, until it is outweighed by the mass increase (inside the same volume) due to higher number of solitons; see e.g. the $2-$, $3-$ and $4-$soliton mergers. Although the $3$-soliton merger has a higher total mass, the central densities are lower than for the $2-$soliton mergers. This effect can be also seen if we compare the $2-$soliton merger and the $4-$soliton merger, each with a total mass of $80$\,CU. This explains why we usually get a drop in central density, whenever another soliton is added \footnote{Of course, this has also something to do with the chosen box size and resolution, and the fact that a poor spatial resolution decreases the central density. This becomes apparent when comparing the central densities of new and old simulations (old simulations have a poorer spatial resolution).}.
    ii) Head-on collisions tend to produce lower central densities, although there are some exceptions. They result in mergers with more pronounced oscillations, and therefore an arbitrary choice of snapshot has a low probability to show a maximum central density. Higher central densities occur more frequently when the merger is "soft", since then the oscillations are weaker, leading also to a "faster" approach to (near-)equilibrium. 

To showcase some results in more detail, we present simulations of higher accuracy for dedicated cases below. Although the high-mass problem can not be avoided, since the central densities are too high, we were able to find box sizes such that the mass accuracy is satisfactory. We ran two different merger scenarios (with and without head-on collision) for the $N=3$ and $N=6$ soliton merger. These cases illustrate the findings mentioned above, in terms of the trend in the central density. To increase the "smoothness" of the final merger product, we chose individual soliton masses of $40$\,CU, which results in a sufficiently high mass for the merger to settle earlier into a (near-)equilibrium, meaning that the envelope is not so chaotic and produces not as many oscillations as for smaller-mass solitons.

\subsubsection{In detail: The $(N=3)$-case}

\begin{figure}
    \centering
    \includegraphics[scale=0.25]{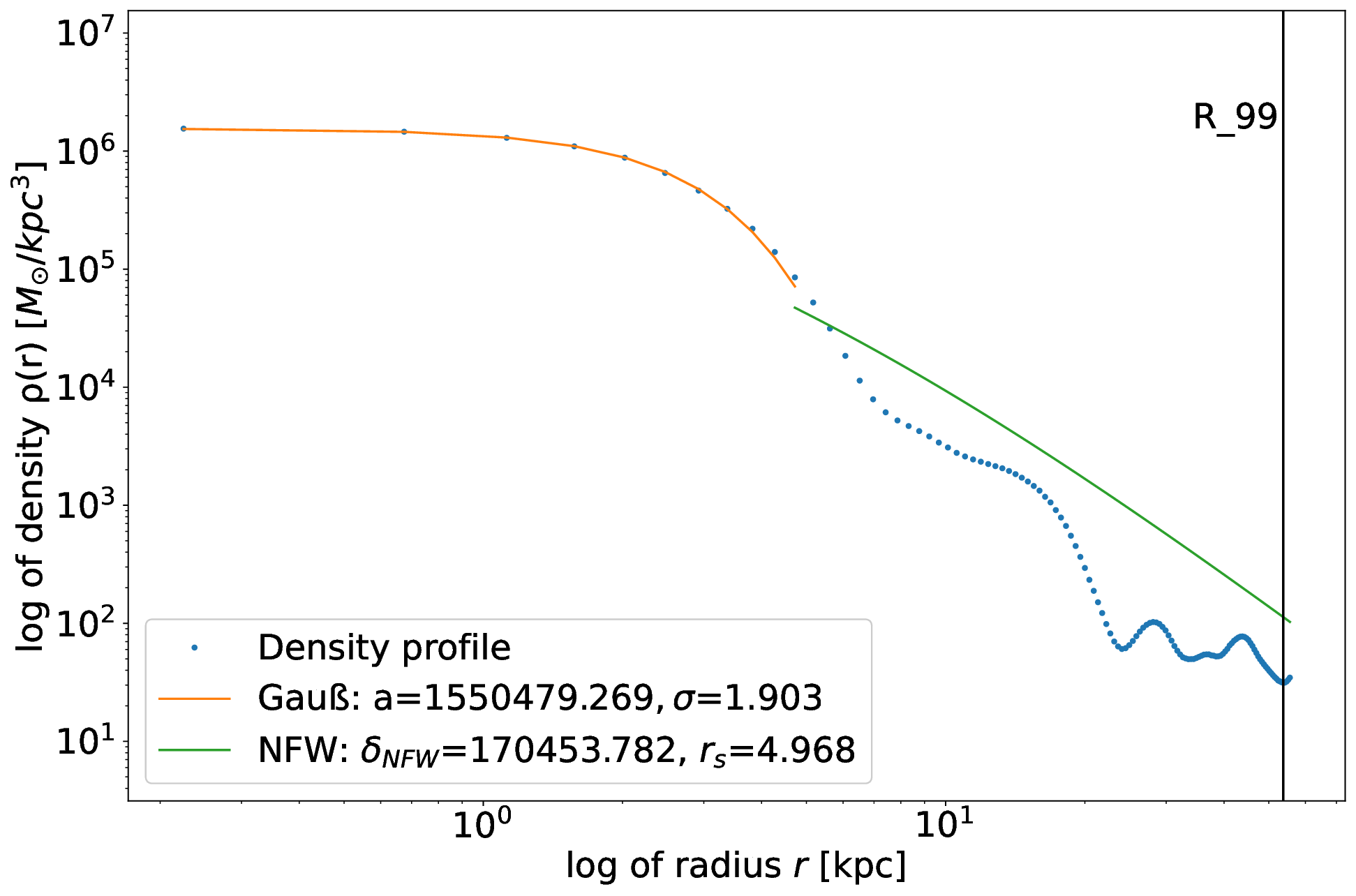}
    \caption{The density profile of the final snapshot ($\sim 13.59$\,Gyr) of the $(N=3)$ Kx-y merger with simulation parameters in tables \ref{tab:SimPar_N3_Kxy} and \ref{tab:SolPar_N3_Kxy}. The black vertical marks that radius which contains $99$\,\% of the mass of the post-merger object. The orange, solid curve is the fit according to equ.(\ref{eq:Gauß}), with units for $a$ in $M_\odot/\text{kpc}^3$ and $\sigma$ in kpc, while the green, solid curve is the fit according to equ.(\ref{eq:NFW}), with units for $\delta_{NFW}$ in $M_\odot/\text{kpc}^3$ and $r_{s}$ in kpc.}
    \label{fig:Profile_N3_Kxy}
\end{figure}

For this case, we choose the Kx-y and the triangle scenario. First, we discuss the Kx-y-scenario. Here we use the simulation parameters from table \ref{tab:SimPar_N3_Kxy} to get an accurate mass. The soliton parameters for this scenario are listed in table \ref{tab:SolPar_N3_Kxy}, while the resulting density profile of the post-merger object can be found in figure \ref{fig:Profile_N3_Kxy}.
As expected, the profile is still mostly dominated by the core. An envelope can be recognized, but the NFW model fits the numerical data poorly.

\begin{table}
\caption{\label{tab:SimPar_N3_Kxy} The simulation parameters for the $(N=3)$ Kx-y merger. The box size is $\sim 115$\,kpc on a side, the simulation time is $\sim 13.59$\,Gyr, which results together with the saving number (gives the number of snapshots) in a temporal resolution of $1.359$\,Gyr.}
\begin{ruledtabular}
\begin{tabular}{ccc}
Parameter & Value & Unit \\ \hline
box size & $3$ & CU \\
resolution & $256$ & / \\
simulation time & $0.18$ & CU \\
saving number & $10$ & / \\
\end{tabular}
\end{ruledtabular}
\end{table}

\begin{table}
\caption{\label{tab:SolPar_N3_Kxy} The soliton parameters for the $(N=3)$ Kx-y merger with individual soliton mass of $40$\,CU $\sim 89\cdot10^6\,M_\odot$, an initial distance of solitons to the center of $\sim 15$\,kpc, and the absolute value of the velocity for each soliton is $\sim 2.5$\,km/s.}
\begin{ruledtabular}
\begin{tabular}{cccccccc}
soliton nr. & mass & x & y & z & v$_{\text{x}}$ & v$_{\text{y}}$ & v$_{\text{z}}$ \\ \hline
$1$ & $40$ & $0.4$ & $0$ & $0$ & $-5$ & $0$ & $0$ \\
$2$ & $40$ & $0$ & $0.4$ & $0$ & $0$ & $-5$ & $0$ \\
$3$ & $40$ & $-0.4$ & $0$ & $0$ & $5$ & $0$ & $0$
\end{tabular}
\end{ruledtabular}
\end{table}

\begin{figure}
    \centering
    \includegraphics[scale=0.25]{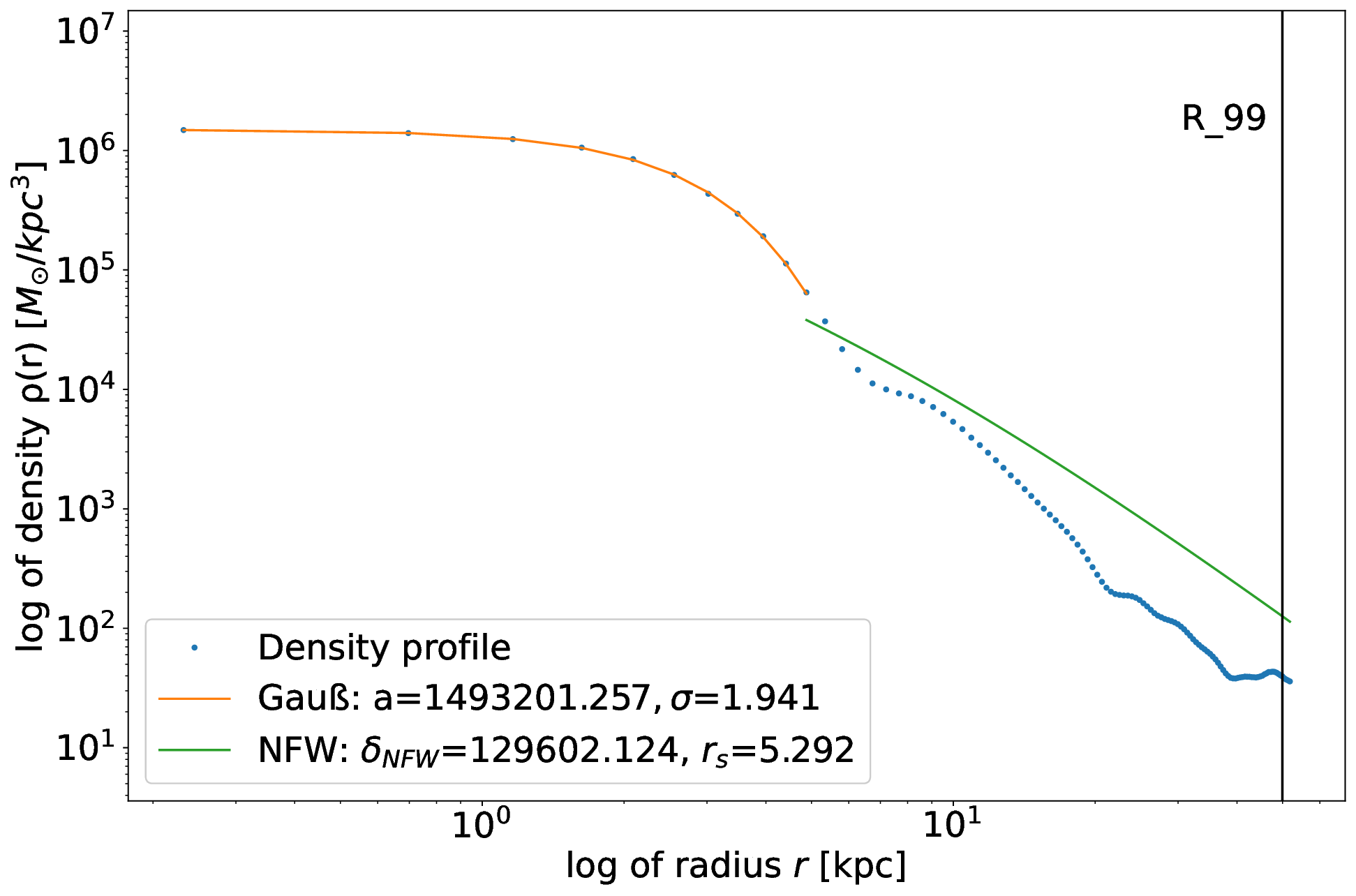}
    \caption{The density profile of the final snapshot ($\sim 13.59$\,Gyr) of the $(N=3)$ merger, where the solitons are positioned in a triangle shape with parameters in tables \ref{tab:SimPar_N3_Triangle} and \ref{tab:SolPar_N3_Triangle}. The black vertical marks that radius which contains $99$\,\% of the mass of the post-merger object. The orange, solid curve is the fit according to equ.(\ref{eq:Gauß}), with units for $a$ in $M_\odot/\text{kpc}^3$ and $\sigma$ in kpc, while the green, solid curve is the fit according to equ.(\ref{eq:NFW}), with units for $\delta_{NFW}$ in $M_\odot/\text{kpc}^3$ and $r_{s}$ in kpc.}
    \label{fig:Profile_N3_Triangle}
\end{figure}

Now, we draw our attention to the other $(N=3)$-scenario, which is the triangle shape. The simulation and soliton parameters can be found in tables \ref{tab:SimPar_N3_Triangle} and \ref{tab:SolPar_N3_Triangle}. The resulting density profile of the post-merger object for this scenario is shown in figure \ref{fig:Profile_N3_Triangle}.

\begin{table}
\caption{\label{tab:SimPar_N3_Triangle} The simulation parameters for the $(N=3)$ merger, where the solitons are positioned in a triangle shape. The box size is $\sim 119$\,kpc on a side, the simulation time is $\sim 13.59$\,Gyr, which results together with the saving number (gives the number of snapshots) in a temporal resolution of $1.359$\,Gyr.}
\begin{ruledtabular}
\begin{tabular}{ccc}
Parameter & Value & Unit \\ \hline
box size & $3.1$ & CU \\
resolution & $256$ & / \\
simulation time & $0.18$ & CU \\
saving number & $10$ & / \\
\end{tabular}  
\end{ruledtabular}
\end{table}

\begin{table}
\caption{\label{tab:SolPar_N3_Triangle} The soliton parameters for the $(N=3)$ merger, where the solitons are positioned in a triangle shape with individual mass of $40$\,CU $\sim 89\cdot10^6\,M_\odot$, an initial distance of solitons to the center of $\sim 15$\,kpc, and the absolute value of the velocity for each soliton is $\sim 2.5$\,km/s.}
\begin{ruledtabular}
\begin{tabular}{cccccccc}
soliton nr. & mass & x & y & z & v$_{\text{x}}$ & v$_{\text{y}}$ & v$_{\text{z}}$ \\ \hline
$1$ & $40$ & $0.4$ & $0$ & $0$ & $-5$ & $0$ & $0$ \\
$2$ & $40$ & $-0.27772$ & $-0.27772$ & $0$ & $3.4715$ & $3.4715$ & $0$ \\
$3$ & $40$ & $-0.27772$ & $0.27772$ & $0$ & $3.4715$ & $3.4715$ & $0$
\end{tabular}
\end{ruledtabular}
\end{table}

In this case, the profile is still dominated by the core. However, the oscillations in the envelope are much less pronounced than in the previous case. This shows that the triangle-shaped merger is softer in comparison to the Kx-y merger. The slope of the NFW model fits somewhat better to the outer data. Additionally, we see that the central density is lower than in the Kx-y-case. This suggests that we got a "lucky snapshot" in figure \ref{fig:Profile_N3_Kxy}, where the central density happens to be higher. On the other hand, in figure \ref{fig:Profile_N3_Triangle} we see a state, where the merger looks close to equilibrium, exemplified by its smooth profile.

\subsubsection{In detail: The $(N=6)$-case}

\begin{figure}
    \centering
    \includegraphics[scale=0.25]{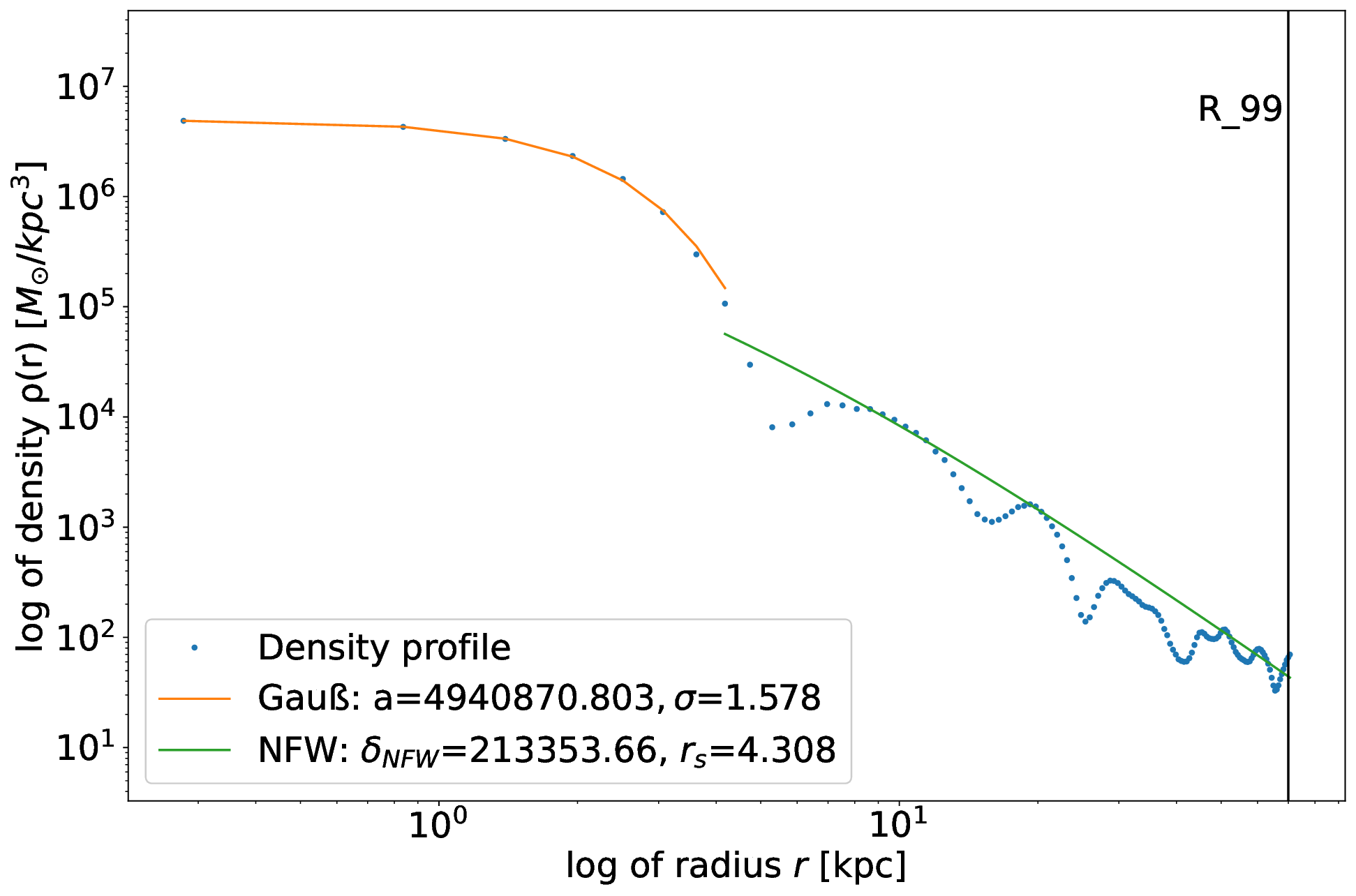}
    \caption{The density profile of the final snapshot ($\sim 13.59$\,Gyr) of the $(N=6)$ Kx-Ky-Kz merger with  parameters in tables \ref{tab:SimPar_N6_KxKyKz} and \ref{tab:SolPar_N6_KxKyKz}. The black vertical marks that radius which contains $99$\,\% of the mass of the post-merger object. The orange, solid curve is the fit according to equ.(\ref{eq:Gauß}), with units for $a$ in $M_\odot/\text{kpc}^3$ and $\sigma$ in kpc, while the green, solid curve is the fit according to equ.(\ref{eq:NFW}), with units for $\delta_{NFW}$ in $M_\odot/\text{kpc}^3$ and $r_{s}$ in kpc.}
    \label{fig:Profile_N6_KxKyKz}
\end{figure}

Here we choose also two scenarios, a pronounced head-on collision and its opposite case, so to speak. For the first we use the Kx-Ky-Kz-scenario, using the simulation parameters from table \ref{tab:SimPar_N6_KxKyKz} and the soliton parameters from table \ref{tab:SolPar_N6_KxKyKz}. With these parameters, we end up with the density profile of the post-merger object shown in figure \ref{fig:Profile_N6_KxKyKz}.

\begin{table}
\caption{\label{tab:SimPar_N6_KxKyKz} The simulation parameters for the $(N=6)$ Kx-Ky-Kz merger. The box size is $\sim 142$\,kpc on a side, the simulation time is $\sim 13.59$\,Gyr, resulting together with the saving number (gives the number of snapshots) in a temporal resolution of $1.359$\,Gyr.}
\begin{ruledtabular}
\begin{tabular}{ccc}  
Parameter & Value & Unit \\ \hline
box size & $3.72$ & CU \\
resolution & $256$ & / \\
simulation time & $0.18$ & CU \\
saving number & $10$ & / \\
\end{tabular}
\end{ruledtabular}
\end{table}

\begin{table}
\caption{\label{tab:SolPar_N6_KxKyKz} The soliton parameters for the $(N=6)$ Kx-Ky-Kz merger with individual mass of $40$\,CU $\sim 89\cdot10^6\,M_\odot$, an initial distance of solitons to the center of $\sim 23$\,kpc, and the absolute value of the velocity for each soliton is $\sim 2.5$\,km/s.}
\begin{ruledtabular}
\begin{tabular}{cccccccc}
soliton nr. & mass & x & y & z & v$_{\text{x}}$ & v$_{\text{y}}$ & v$_{\text{z}}$ \\ \hline
$1$ & $40$ & $0.6$ & $0$ & $0$ & $-5$ & $0$ & $0$ \\
$2$ & $40$ & $0$ & $0.6$ & $0$ & $0$ & $-5$ & $0$ \\
$3$ & $40$ & $0$ & $0$ & $0.6$ & $0$ & $0$ & $-5$ \\
$4$ & $40$ & $-0.6$ & $0$ & $0$ & $5$ & $0$ & $0$ \\
$5$ & $40$ & $0$ & $-0.6$ & $0$ & $0$ & $5$ & $0$ \\
$6$ & $40$ & $0$ & $0$ & $-0.6$ & $0$ & $0$ & $5$
\end{tabular}
\end{ruledtabular}
\end{table}

This profile exemplifies what we were expecting: due to the higher number of merging solitons we get a higher central density and a bigger envelope, compared to the $(N=3)$-cases. The envelope shows a lot of oscillations, thanks to the many violent head-on collisions, while its slope matches the slope of the NFW model. 

\begin{figure}
    \centering
    \includegraphics[scale=0.25]{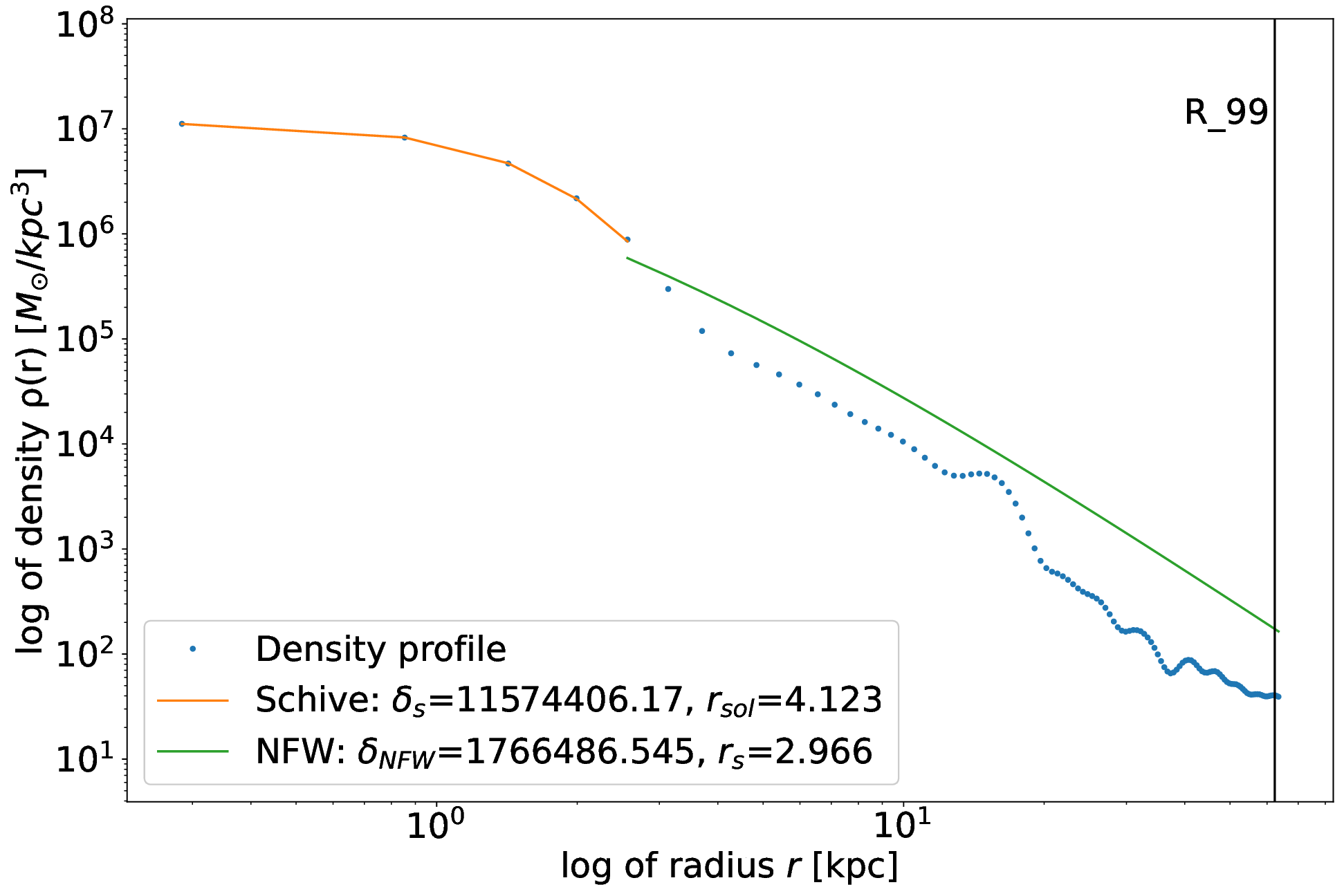}
    \caption{The density profile of the final snapshot ($\sim 13.59$\,Gyr) of the $(N=6)$ merger, where the solitons are positioned "randomly" in the simulation box with parameters in tables \ref{tab:SimPar_N6_Random} and \ref{tab:SolPar_N6_Random}. The black vertical marks that radius which contains $99$\,\% of the mass of the post-merger object. The orange, solid curve is the fit according to equ.(\ref{eq:Schive}), with units for $\delta_s$ in $M_\odot/\text{kpc}^3$ and $r_{sol}$ in kpc, while the green, solid curve is the fit according to equ.(\ref{eq:NFW}), with units for $\delta_{NFW}$ in $M_\odot/\text{kpc}^3$ and $r_{s}$ in kpc.}
    \label{fig:Profile_N6_Random}
\end{figure}

For the other $(N=6)$-case, we choose a "random configuration", by which we mean that we picked a value for one positional coordinate from a random generator and adjusted the other parameters, such that they meet our requirement of same distance and velocity of $5$\,CU. The simulation parameters for this scenario can be found in table \ref{tab:SimPar_N6_Random}, while the soliton parameters are listed in table \ref{tab:SolPar_N6_Random}. The resulting density profile of the post-merger object is shown in figure \ref{fig:Profile_N6_Random}.

\begin{table}
\caption{\label{tab:SimPar_N6_Random} The simulation parameters for the $(N=6)$ merger, where the solitons are positioned "randomly" in the simulation box. The box size is $\sim 145.5$\,kpc on a side, the simulation time is $\sim 13.59$\,Gyr, resulting together with the saving number (gives the number of snapshots) in a temporal resolution of $1.359$\,Gyr.}
\begin{ruledtabular}
\begin{tabular}{ccc}
Parameter & Value & Unit \\ \hline
box size & $3.8$ & CU \\
resolution & $256$ & / \\
simulation time & $0.18$ & CU \\
saving number & $10$ & / \\
\end{tabular}
\end{ruledtabular}
\end{table}

\begin{table*}
\caption{\label{tab:SolPar_N6_Random} The soliton parameters for the $(N=6)$ merger, where the solitons are positioned "randomly" in the simulation box with individual mass of $40$\,CU $\sim 89\cdot10^6\,M_\odot$, an initial distance of solitons to the center of $\sim 23$\,kpc, and the absolute value of the velocity for each soliton is $\sim 2.5$\,km/s.}
\begin{ruledtabular}
\begin{tabular}{cccccccc}
soliton nr. & mass & x & y & z & v$_{\text{x}}$ & v$_{\text{y}}$ & v$_{\text{z}}$ \\ \hline
$1$ & $40$ & $0.3$ & $0.1$ & $-0.51$ & $-2.5$ & $-0.83333$ & $4.25$ \\
$2$ & $40$ & $0.5$ & $-0.3$ & $0.14142$ & $-4.16666$ & $2.5$ & $-1.1785$ \\
$3$ & $40$ & $-0.2$ & $0.26458$ & $-0.5$ & $1.66666$ & $-2.20483$ & $4.16666$ \\
$4$ & $40$ & $0$ & $0.6$ & $0$ & $0$ & $-5$ & $0$ \\
$5$ & $40$ & $0.3$ & $0.4$ & $0.33166$ & $-2.5$ & $-3.33333$ & $-2.76238$ \\
$6$ & $40$ & $-0.1$ & $-0.4$ & $0.43589$ & $0.83333$ & $3.33333$ & $-3.63242$
\end{tabular}
\end{ruledtabular}
\end{table*}

As expected, the profile has a higher central density than the Kx-Ky-Kz case. Also, we see a more pronounced core-envelope structure, where the envelope itself shows less oscillatory behavior than in the previous case.

In summary, the specific cases demonstrate once more what we have seen in the trend plotted in figure \ref{fig:KonfiAnalysis}. Head-on collisions delay the merging process and its equilibration, exemplified by non-relaxed density profiles. Statistically, we will end up with a lower central density in the final snapshot. This was shown here explicitly, since we even used a higher soliton mass in order to dampen the oscillations. Thus, we conclude that a soft merging process helps to accomplish a successful merger, depending on the relative velocity and the initial geometric configuration; a result which is in good accordance with expectation and with "classic mergers". 
Also, we could show that the $(N=6)$-case produces a higher central density, even with the head-on-collision scenario. This matches our prediction based upon the trend of figure \ref{fig:KonfiAnalysis}. If we had used a smaller soliton number, we would have probably (depending on if we get a lucky or unlucky snapshot) seen a smaller central density, especially if the scenario includes head-on collisions.  Finally, we add that NFW profiles can match the overall slope of the envelopes, especially the higher the number of merging solitons, but usually they fail to provide good fits to the numerical data.

\section{\label{sec:sec6}Summary and Discussion}

Scalar field dark matter ("SFDM") has become a popular candidate for the cosmological DM, with its potential to cure several "small-scale" structure formation problems in the standard CDM paradigm. SFDM is comprised of (ultra-)light spin-zero bosons in a Bose-Einstein condensate, without or with a self-interaction (SI) included. The overwhelming astrophysics literature considers models of ultralight bosons without SI, i.e. "fuzzy dark matter" (FDM), or the opposite regime of a strongly repulsive SI, the so-called Thomas-Fermi ("TF") regime, or SFDM-TF. Whereas FDM has been modelled, using the nonlinear Schr\"odinger or Gross-Pitaevskii wave equation, along with the Poisson equation, SFDM-TF has been often modelled, using fluid approximations to the fundamental wave equation of motion. As such, the numerical algorithms, strategies and computational demands vary, but each regime poses serious challenges to realistic cosmological simulations.
Therefore, previous literature has often focused on more simplistic scenarios, where single or multiple (near-)equilibrium objects in SFDM have been investigated.
These so-called "solitons", which are also the focus of this paper, exhibit the basic features of SFDM gravitational dynamics, and have been found to approximate the central parts (the "cores") of larger, galactic-scale halos in actual cosmological simulations, i.e. "solitonic cores" in FDM, see \cite{Schive_Ursprung, Schive_CE, Mocz_Solitonen2017, Mocz2020}, or "polytropic/TF cores" in SFDM-TF, see \cite{Dawoodbhoy2021, Shapiro2022, Hartman2022, Foidl2023}.   
Given these findings and the fact, that the details of soliton dynamics and SFDM structure formation are far from being fully understood, the need for dedicated simulations of solitons and their merger dynamics remains.
Moreover, realistic cosmological simulations and their high-performance computing requests exclude the possibility to probe the boson parameter space, in terms of boson mass or SI coupling strength, in any efficient or exhaustive way. On the other hand, "smaller-scale" soliton simulations are computationally cheaper, especially for static backgrounds, and thus allow an easier sampling of SFDM parameter spaces. In fact, the various SI regimes are still in parts poorly explored, if at all, and low and intermediate SI strengths have been rarely considered. 

In this paper, we studied the dynamics of single and merging solitons for models without SI (typically FDM), and for models with a weak-to-intermediate, repulsive SI included. For this purpose, we used and modified the open-source code {\ffmfamily PyUltraLight}, calling our modified variant {\ffmfamily PyUltraLight\_SI} which is available at \footnote{\url{https://github.com/MStallovits/PyUltraLight_SI}}. The code solves the Gross-Pitaevskii-Poisson equations of motion, using a symmetric, pseudospectral Fourier-split-step algorithm. We used periodic boundary conditions for all our simulations. 

The user can specify the value of a non-dimensionalized SI coupling strength $\hat{\Lambda}$, according to the nonlinear SI term in equ.(\ref{eq:DimlessGrossPitaevskii}).
Furthermore, a prime focus of ours was the study of density profiles of single solitons, as well as of post-merger objects made of multiple solitons, because as the central cores of SFDM halos, they exhibit differences to CDM predictions, and are eventually subject to observational constraints. Therefore, we included another code feature to the original {\ffmfamily PyUltraLight} software, which allows to extract radially-averaged density profiles from the numerical density data, in order to use them as input to a fitting routine. The latter is able to fit analytical (closed-form) density profiles to the numerical averaged density data. For comparison's sake, we used in this paper a Gauss and a Schive profile, equ.(\ref{eq:Gauß}) or equ.(\ref{eq:Schive}), respectively, also used in previous literature, to fit the numerical profiles of single solitons and the cores of post-merger objects. A NFW profile equ.(\ref{eq:NFW}) is used for the envelope of post-merger objects, since the large-scale dynamics of SFDM halo envelopes has been previously shown to be similar to those of CDM halos, at least in an averaged or statistical sense (see references above). 
Applying our fitting routines to our simulation results, we find that the Schive profile is somewhat favored over the Gaussian for single solitons and the cores of post-merger objects, if $\hat{\Lambda} = 0$. In this regard, we confirm the early work by \cite{Schive_Ursprung}, who introduced what we call "Schive profile" as a good fitting model for halo central parts.
However, when it comes to the central densities, there is not much difference between the Gaussian and Schive profile. Also, we find that the size of cores does generally not only depend upon the value of $\hat{\Lambda}$, but also upon soliton mass. Overall, for $\hat{\Lambda} > 0$, we find a mild preference for the Gaussian core profile for the cases we studied, but more simulations would be required to claim a real trend. Surely, users may change or extend our choice of analytical fitting models in the future, should more theoretically motivated profiles for SFDM become available, possibly covering the full radial range.
However, even if better density approximations than the Gauss or Schive profile will be found in the future, they may have difficulties to match the oscillatory behavior of solitons, which persist over many Hubble times. In fact, our results confirm previous findings in that even single solitons at rest show persistent oscillations, which are more pronounced, the more massive the soliton. However, these oscillations are less pronounced, the higher $\hat{\Lambda} > 0$, for the same soliton mass.

Another feature of interest to us in this paper was the derivation of the scaling relation between central density $\rho_c$ and mass $M$ of single solitons. Since solitons serve as approximate halo cores, this relationship is of importance when it comes to the comparisons to galactic data or CDM predictions. For $\hat{\Lambda} = 0$, our single-soliton simulations confirmed the well-known relationship, $\rho_c \propto M^4$. Among other tests, this agreement also established that our code modifications to incorporate finite SI worked correctly, upon setting $\hat{\Lambda} = 0$ as in the original code. Performing our single-soliton simulations, we were also able to derive numerical $\rho_c - M$ relations for weak SI, $\hat{\Lambda} = 0.02$ and  $\hat{\Lambda} = 1$, respectively. In contrast to models without SI ("FDM", $\hat{\Lambda} = 0$, $\rho_c \propto M^4$), or models in the strongly repulsive TF regime ("SFDM-TF", $\hat{\Lambda} \gg 17$, $\rho_c \propto M$), there is no analytical relationship available to use as a fitting function to our numerical results, as far as we know. In lack of an analytical derivation, we refrained from presenting fits, although it appears that a low-order polynomial multiplied by a Gaussian can fit the data. This means that the initial increase of $\rho_c$ with increasing $M$ turns over after reaching a peak at some soliton mass - the value of which depends upon $\hat{\Lambda}$- into the opposite behavior, where $\rho_c$ decreases with increasing $M$, in sharp contrast to the monotonic scaling relations known for FDM or SFDM-TF solitons. 
We presented arguments to explain this phenomenon, on the basis of the mutual competition between quantum and SI pressure vs self-gravity.
However, more work will be required and, ideally, analytical $\rho_c - M$ relationships will be derived in the future, for the weak and intermediate SI regimes.

Furthermore, we performed binary and multiple (up to $6$) soliton merger simulations. While we included a repulsive SI in some binary merger simulations, we only considered multiple mergers without SI, but studied multiple geometric configurations, and their possible impact onto the final post-merger objects.
We confirmed in our work what has been seen in previous literature, namely that a NFW fit to the outer envelope only has a chance to fair well, at least in terms of the overall slope, if that envelope is well developed, as a result of the merger of many solitons. Binary soliton mergers have poorly developed envelopes and exhibit poor fits.
However, even for our multiple soliton merger simulations, the outcome of a NFW fit as being good or bad is somewhat erratic, depending on whether the chosen snapshot is still undergoing equilibration, as they mostly do. Of course, our numbers of solitons is still small, compared to the soliton merger simulations by e.g., \cite{Schwabe2016, Mocz_Solitonen2017, Kendall_etal_SolitonFormation2023}. Yet, we think caution should be observed for several reasons. First, in non-expanding backgrounds (such as used here and in most other soliton merger simulations), we actually expect outer halo density profiles to fall off as $\propto r^{-12/7}$, regardless of the DM model, see \cite{Dawoodbhoy2021} and references therein, rather than the NFW profile. Second, the fact that cosmological CDM-only simulations predict universal NFW profiles at all mass scales, does not imply that NFW should fit well for post-merger objects of arbitrary mass in SFDM. In fact, if NFW (or another CDM-inspired outer model profile) does not fit well, it will speak against using it for this regime of SFDM structure formation, and may rather be a sign that the inherent quantum and SI dynamics dominate over genuine gravitational dynamics in such post-merger objects. By extension, this same caution should be observed when actual galactic velocity data of dwarf galaxies is fitted to closed-form, "ironclad" core-envelope density profiles (see e.g., \cite{MarshPop2015, Calabrese-Spergel2016, Bernal_FDM-Rot2018}).
Generally, our density fits provided a good determination of the central density of single and multiple merger solitons, but the preference of one model over the other and their values of the fitting parameters can depend upon the considered snapshots in time. The oscillatory behavior of solitons (single and post-merger) and their long relaxation times reflect themselves in the fitting parameters, if ever so slightly, for the solitonic mass regime which we considered in this paper. The highest masses of post-merger objects amounted to $\lesssim 5 \cdot 10^{8}~M_{\odot}$, while single solitons up to $\lesssim 7 \cdot 10^{8}~M_{\odot}$ have been considered. Such masses are lower than the resolution limit of current cosmological SFDM simulations. 

For the multiple merger simulations, we found that rotationally symmetric configurations produce the same outcomes (up to rounding errors), as expected. Mergers with fewer solitons tend to produce a higher central density, until this is outperformed by the mass increase, due to a higher number of solitons, because generally the density increases for increasing total mass of the merger. The oscillations of post-merger objects can be less pronounced, if their self-gravity (i.e. their total mass) is sufficiently high to relax more rapidly. However, the initial configurations matter; e.g. head-on collisions are more violent and lead to more persistent oscillations in the post-merger objects, compared to configurations which lead to soft mergers. However, in any case, oscillations are always at play and define the approach to (near-)equilibrium.

In addition to these considerations, there is the issue of boundary conditions and box sizes, and some readers may object that our choice of periodic boundary conditions is not optimum for merger simulations. While it is true that many previous works have made use of an absorbing sponge as boundary condition in merger simulations, we stressed that we have observed much care in our simulations to pick appropriate parameters of, say, box size, in order to avoid the loss and re-entry of any SFDM mass in the box upon soliton collisions and merging. Our simulations had never more than $6$ solitons involved, but if the latter exceed $10$, an absorbing sponge is probably recommended. In fact, the recent work by \cite{RiosGuzmanShapiro2023_FDM_Randbgn} has considered this question in detail for $\hat{\Lambda}=0$, and as far as density profiles are concerned, the results are in agreement. However, more merger simulations for $\hat{\Lambda} \not= 0$ are desirable in the future. As discussed in our paper, we experienced severe issues which prevented us from creating good mergers for $\hat{\Lambda} \geq 1$. While periodic boundary conditions and limited computational times are probably part of the problem, we also suspect that other numerical algorithms might be required to tackle the intermediate SI regime in an efficient way.  

With respect to other recent work which is related to ours, we mention some of them as follows.
The authors of  \cite{Kendall_etal_SolitonFormation2023} have also simulated soliton mergers, in order to derive density profiles of post-merger objects, for $\hat{\Lambda}=0$, and to study the implications for core-halo mass relations. They used an absorbing sponge, given that their single solitons (before merging) are at least a factor of five more massive than ours, and given their higher numbers of solitons (up to $14$). As a result, they can use smaller simulation boxes than we do, which translates into much smaller radial ranges in their plotted density profiles. In order to fit their density data, they use exclusively a Schive profile for the core, along with a relation to determine the size parameter of the core. Similar to us, at some outer radius a NFW profile takes over. 

While our paper was in the process of being finished, we noted the very recent work by \cite{SI-Range_2024}, who focus on virialized single solitons by investigating the impact of (mostly repulsive) SI onto soliton shape and oscillations. Since the authors are interested in deriving analytical and scaling relations, in order to pinpoint the transitions between weak and strong SI, they use Gauss profiles, which take into account the SI parameter, for this purpose (see also \cite{Chavanis2019}), although Schive profiles (no SI) and (modified) $(n=1)$-polytropes (strongly repulsive TF regime) are also considered. As discussed in previous literature (e.g. in \cite{TRDFrontier2022}), the authors of \cite{SI-Range_2024} also find degeneracies between boson mass and SI coupling strength, such that constraints (on mass only) from observations should be reconsidered, for SFDM models with SI. Interestingly, they also find similar results from their scaling analysis than we do from our dynamical simulations of dedicated single-soliton cases: For increasing, repulsive SI strength, the solitons expand, i.e. the core radius grows, while their central density decreases, compared to solitons without SI. In this weak and intermediate SI regime, the Gaussian core fares generally better than both the Schive or the TF profile, which is in accordance with our results. A more in-depth theoretical comparison and analysis of their results with respect to ours is worthwhile, which we defer to future work. 

We might also speculate whether the "core expansion" of single solitons with finite, repulsive SI (compared to $\hat{\Lambda}=0$), which we can understand well as a result of the more prominent SI pressure, lies already at the heart of the phenomenon of post-evolved, core-plus-envelope expansion seen in the 3D halo simulations of \cite{Foidl2023}. After all, the effect is more prevalent, the higher the repulsive SI. In the TF regime considered in \cite{Foidl2023}, primordial halo cores of radius $\sim 0.1$ kpc were found to expand to $2$ kpc within dwarf-galactic SFDM-TF halos of $10^{9}~M_\odot$ by redshift zero. While we should not confuse an in-built, "static" result with an unfolding, dynamical process, it remains to be seen how genuine SFDM physics blends in detail with gravitational dynamics upon more future work.  

The authors of \cite{TFR-Solitons_2024} recently considered the regime of "TF solitons" within halos made of SFDM-TF. Their results may be compared to the appropriate papers studying the TF regime (\cite{Shapiro2022,Foidl2023}), rather than our weakly repulsive cases. In our context, we merely note that \cite{TFR-Solitons_2024} find similar phenomena with respect to a smooth central profile, with density oscillations in the envelopes, whose amplitude diminish, the larger the mass. Also, long-time fluctuations persist, such that the system never really relaxes. The authors find that solitons form earlier, if the simulation boxes are bigger, which suggests that there is more mass available to help the formation process. We also found higher masses in bigger boxes, compared to the input values, and this effect may be related (for one thing, periodic boundary conditions were also used in \cite{TFR-Solitons_2024}).
Finally, we highlight again that we also see global, persistent oscillations in our post-merger objects, which have been seen in early and recent multiple-soliton merger simulations, see e.g.\cite{Veltmaat2018, Hui_2021}, as well as in cosmological FDM simulations, such as \cite{Mocz_Solitonen2017}, along with the papers cited above. This indicates a very long relaxation time in SFDM mergers - at least for models without SI (previous and this work) and for weakly repulsive SI (this work), questioning the usefulness of CDM-inspired profiles. In general, we find less pronounced oscillations in the envelopes for higher-mass post-merger objects (different to single solitons).
In addition, a positive $\hat{\Lambda}$ damps the oscillations, as follows. In models with repulsive SI, the SI pressure is more potent in the central parts of (near-)equilibrium objects. However, during the process of equilibration our simulations suggest that the SI pressure does affect the outskirts as well, by smoothing over local wave disturbances and interferences, caused by the quantum pressure. Thus, we notice the same physical effect at the level of binary soliton mergers what has been observed in the simulations of strongly repulsive SFDM-TF models by \cite{Shapiro2022, Hartman2022, Foidl2023}.     

Nevertheless, a "gross similarity" to CDM in the envelopes of fully-fledged FDM or SFDM-TF galactic halos has been established in several previous works, and explained in \cite{Dawoodbhoy2021} as a result of the dynamical similarity (or "correspondence") between large-scale, coarse-grained quantum dynamics and large-scale velocity dispersion in collisionless CDM. In fact, even toy models respect this correspondence at the level of the pressure terms, as shown in \cite{Dawoodbhoy2021}, strengthening the coarse-graining argument. Still, the impact of the gravity term onto this correspondence seems to matter, because our results here suggest that this dynamical similarity fades below a certain mass scale of solitonic objects in gravitational (near-)equilibrium. In this light, we might question whether it is useful and justified to enforce a NFW tail upon SFDM profiles, in attempts to construct SFDM wave functions as eigenstates with this property, as e.g. pursued by \cite{Yavetz_Wavefunction_Construction2022}, although they find deviations from NFW in the outer parts of their massive model halos, upon evolving their initial states with a Schwarzschild method. Another approach to study equilibrium objects in SFDM is due to \cite{RiosGuzman_CoreTailStructure2023}, in which the cores are "fixed" as solitonic solutions, but the envelopes or tails are dynamically evolved upon the choice of outer boundary conditions (possibly subject to different physical considerations), and constrained by the total mass. 
In our paper, the radially averaged density data are the result of pure dynamics, without any in-built assumptions on how the profiles (or profile parts) should look like in the beginning or the end. The choice of analytical fitting model is subject to scrutiny, of course, as elaborated above. 

In any case, more simulations of solitons and halos in weak and intermediate SI regimes, $0 < \hat{\Lambda} \lesssim 10$, will be required in the future to make a more in-depth and fair comparison to the results in this paper.

Before we close, let us mention a different perspective, altogether, as follows.
The effect of gravity onto the dispersion of Gaussian wave packets has been also a field of study in the laboratory, using atomic or molecular systems, in quests to understand the interplay between quantum mechanics and gravity. Of course, the analogy to SFDM solitons is limited by the fact that the self-gravity of laboratory systems is highly negligible, compared to the ambient gravitational field of the Earth. Still, theoretical investigations have been pursued, as e.g. by \cite{Giulini_Grav_ind_inh_SN_2011}, where the Schr\"odinger-Newton equation is used to calculate numerically the onset of gravitationally induced inhibitions of the dispersion of wave packets and its mass dependence. For sufficiently high mass - in their case close to $10^{10}$ atomic units, the wave packet will not only resist dispersing, but actually starts contracting. On the other hand, smaller-mass wave packets will expand i.e. disperse, unhampered by self-gravity of the constituent particles (of, say, complex molecules). We mention this particular example of paper, because it demonstrates theoretically for laboratory systems of a complete different physical scale what we see in our SFDM soliton simulations, in line with our physical understanding: more massive solitons will benefit from their self-gravity to form stable, non-dispersing (near-)equilibrium objects, i.e. either remain stable if single objects, or become stable upon mergers. This is the case even if SI is added, as we showed. Of course, SFDM solitons of mass which surpass a certain maximum mass can undergo gravitational collapse (we have not studied this regime in our paper). On the other hand, SFDM solitons of small mass with comparatively little self-gravity are more prone to dispersion upon gravitational interaction with their more massive brethren. They would share a similar fate than the tidally disrupted small-mass subhalos in bigger hosts in standard CDM scenarios.

\begin{acknowledgments}
We want to thank Shaun Hotchkiss and Emily Kendall for their helpful comments regarding the main section of {\ffmfamily PyUltraLight}. A special thanks goes out to Emily Kendall, who supported us in the development of \textit{density\_profile}. 
Also, we thank Yourong Frank Wang for his support with respect to releasing
our amended code and pointing us to the new development branch of {\ffmfamily PyUltraLight}. The authors acknowledge the support by the Austrian Science Fund FWF through an Elise Richter fellowship to T.R-D., project no. V 656-N28. 
T.R-D. furthermore acknowledges the support by the FWF through the single-investigator grant (FWF-Einzelprojekt) no. P36331-N, and the support by the Wolfgang Pauli Institute in hosting this grant.
\end{acknowledgments}

\appendix

\section{Details on spatial resolution and mass accuracy, exemplified for $\hat{\Lambda}=0$}\label{appendix:Problems}

In section \ref{sec:3A1}, we mentioned the issues from the competing requests of resolution and computation time. There are basically two main problems for all simulations: i) the high-mass problem and ii) inaccurate soliton masses. For mergers there is an additional problem, namely the low-mass problem.

To reiterate, the high-mass and low-mass problem are related to the specific mass-size relationship of (single) solitons: for $\hat{\Lambda}=0$, the soliton radius (properly defined, e.g. $R_{99}$) goes inversely proportional to the soliton mass. Thus, high-mass solitons are comparatively smaller in size than low-mass solitons. This gives us an upper limit for soliton masses, since higher-mass solitons require a higher resolution, such that the central regions can be resolved, avoiding a "cut-off" of the central density. The increase in resolution leads ultimately to a significantly higher computation time. For our purposes, the highest soliton mass we simulated was $300$\,CU. 

On the other hand, the broader low-mass solitons are a problem for merger simulations, and for further reasons. Not only do mergers take significantly longer if the individual soliton masses are low, but one also runs in a vicious circle of competing demands, as follows. In order to ensure that the bigger, low-mass solitons do not overlap at the beginning of the simulation, they should be placed sufficiently far apart. Also, the velocities of the solitons have to be set sufficiently low. But this combination is a problem. The solitons now have a greater distance to travel while being slower and the merging process may not complete within a Hubble time. On the other hand, a higher velocity implies a higher kinetic energy of the solitons upon collision, which may also delay or inhibit a merger. Likewise, a smaller separation of solitons causes a (possibly significant) overlap, preventing the unfolding of a "classic merger" to begin with. All these difficulties prevented us from simulating successful mergers for all solitons with mass less than $5$\,CU, so we focused on higher-mass solitons. 

We presented a good "work-around" solution for the high-mass problem already in section \ref{sec:3A1}. However, while work on this project was well advanced, we found a "nicer" solution to determine at least a lower limit for the spatial resolution for each soliton mass, which does not depend on the numerical resolution or box size and still solves the high-mass problem by making use of the grid point length, i.e. the length between two grid points, along with the known central density for a $20$\,CU soliton. We use a $20$\,CU soliton as a benchmark, because this mass regime is not heavily affected by the issues at high or low mass (explained above). The grid point length can be calculated by dividing the box length on a side by the resolution. For example, let us assume we have a soliton with an arbitrary mass, that should have a central density of around $54\cdot10^3\,M_\odot/\text{kpc}^3$. For our $20$\,CU soliton we know we have a central density of $27\cdot10^3\,M_\odot/\text{kpc}^3$, if we use $8$\,CU for the box size, $256$ for the numerical resolution and $\hat{\Lambda}=0$. These parameters are good enough, such that we do not cut off the central region significantly, hence avoid the high-mass problem. The grid point length in this case is $8$ divided by $256$ or $1/32$\,CU. This would be the minimal spatial resolution we need. Anything above this would work, too, but is not necessary (such as not to increase the computation time). To ensure now that our soliton with arbitrary mass does not suffer from the high-mass problem, we need a spatial resolution that is at least twice the spatial resolution of the $20$\,CU soliton. This can be achieved by halving the grid point length of the $20$\,CU soliton, yielding $1/64$\,CU. The fraction is very useful, since it tells us, which relation the box size in CU has to the resolution. Thus, in this case we could simulate our soliton with arbitrary mass with a box size of $1$\,CU and a resolution of $64$. Options which require a bigger box only have to hold the relation of box size to resolution equal $1$ to $64$. This way, we can get an estimate of the minimal spatial resolution for the single-soliton case.

This is a useful approach, especially when test simulations are performed, in order to inform in advance the target simulations of higher accuracy or realism. The only disadvantage is the prior approximate knowledge of the central density of our soliton. In fact, we can use the central-density-mass relationship to infer the density. And by extension our approach would also work for cases with finite SI, $\hat{\Lambda} \not= 0$, if analytical relationships will once be derived. However, the case of mergers is more complicated, although this approach gives some indication, if the spatial resolution is high enough to be not affected by the high-mass problem. Indeed, this is why we knew in the cases of unequal mergers in sections \ref{sec:UnequalMerger} and \ref{sec:MergerSI} that the spatial resolution is too low, impacting the accuracy of the central densities that we got.

The approach of using grid point length has been also a good starting point for getting accurate masses. To reiterate, the problem of "inaccurate soliton masses" refers to the fact that the mass calculated from spherically-averaged density profiles, using e.g. the cutoff radius $R_{99}$, of post-evolved solitons can deviate significantly from the initial input soliton mass which is set in the input file \textit{soliton\_solution.py}. This deviation is quite independent of the numerical integration procedure to calculate the cumulative mass up to $R_{99}$. In general, the calculated masses turned out to be higher than the input value, although the opposite also happens, especially for $\hat{\Lambda} \not= 0$, although mass does not leave the simulation box.  

One reason for the inaccurate masses from the cumulative mass integration is the artificial radial cutoff (e.g. for $R_{99}$), given that solitons for $\hat{\Lambda} = 0$ have no compact support and the wavefunction needs to be truncated. In the code, we have to set or rather change the predefined "maximum radius", which determines the "edge" of the wavefunction profile. 
If the box size now happens to be chosen too large, the mass gets overestimated. Thus, in general we had to reduce the box size to cut off the density profile, such that the mass matches the total input mass of the soliton. In cases where the mass had been underestimated, the box size was simply too small.
In order to fix this issue, it is required to find the correct box size for each simulation case at hand. We tried many other different routes, e.g., an increase in the numerical resolution can help for certain soliton masses. However, more simulations will be required in the future to pinpoint general trends. Nevertheless, along with the use of the grid point length, we were able to find "ideal" parameters in a decent amount of time, for cases of interest. For each scenario, it is generally advisable to run some test simulations with low (numerical) resolution, to get an idea what the central density and the mass looks like. 

Now, to summarize our procedure: For the central densities which we got, we used the grid point length method from above to see which combinations of numerical resolutions and box sizes are feasible. Next, we reduced the box size to obtain a calculated soliton mass near the total input mass (in fact, a little above it, such that the increase in resolution gets us to the correct mass) of the solitons and afterwards increased the numerical resolution \footnote{We chose a tolerance of $\pm 0.5$\,CU for the calculated mass to be allowed to deviate from the input mass.}. To spare computation time, we chose the minimum required numerical resolution. In most cases, a reduction in the box size alone got us so much overall spatial resolution, that we could refrain from using a numerical resolution higher than $256$. Indeed, this resolution has been sufficient for most of our simulations, in light of the requests from the high-mass problem, mass accuracy and computation time.
Still, we caution the reader that some cases defy a good solution for all issues; the ideal parameters may simply be not achievable, if too high a spatial resolution (like in the unequal merger cases), or too big a box size (violent merger of solitons with repulsive SI) is required, for a given limited computation time.

\section{Details on boundary conditions}\label{appendix:BC} 

In this appendix, we add some discussion, concerning our choice of periodic boundary conditions (BCs). Periodic BCs come as the default setting within the original {\ffmfamily PyUltraLight} code. Since our investigations in this paper constitute a continuation of simulation works in the spirit of this code, we have refrained from changing the BCs. This is no restriction, given the scope of our work, namely the study of single solitons and mergers up to 6 solitons at a time. Mergers with a higher number of solitons should be preferably using different BCs, such as an absorbing sponge, in order to avoid the issue of re-entry of expelled scalar field mass (due to "gravitational cooling") into the simulation box; see also the comments and references in Section \ref{sec:sec6} and \footnote{During the revision phase, our attention was drawn to a further code, also based on {\ffmfamily PyUltraLight}, which does include an absorbing sponge as an option, see https://github.com/chaochen0303/WaveDM.}.
Now, with respect to our single-soliton simulations, the solitons were initialized in the center of the box, where they remained "at rest" (apart from their oscillating behavior, described in the main text). Never mind the issues described in Appendix \ref{appendix:Problems}, in these cases the choice of box size was set as to be not in conflict with the chosen BCs, with or without repulsive SI. With respect to mergers, the choice of box size, along with other demands on it, related to e.g. the mass accuracy, requires several points of caution. In any case, for single-soliton and multiple-soliton simulations, we took the outermost contour line of constant density as a reference: if it remained spherically symmetric, there is no mass near the boundary which might corrupt our simulation. If that contour was deformed, it is a sign that boundary issues could be interferring, during the evolution. In this case, the box size had to be enlarged. Moreover, the distance to the center and velocity of solitons, prior of their collision and merging, was chosen such that the interaction sphere and the post-merger object would well remain within the box, given the criterion just mentioned.
In principle, periodic BCs should always be fine, if the box size is suffiently large. However, as elaborated in Appendix \ref{appendix:Problems}, big boxes negatively affect the resolution for given computation time, especially in the central parts of solitons, from which the central density $\rho_c$ as a key quantity is derived, and with respect to the accuracy of calculated mass profiles. Now, if a repulsive SI is included, the solitons are bigger to begin with, and, in addition, the unfolding merger is even more "in danger" to expel mass outside a given box, for the repulsive SI pressure works against the self-gravity of solitons, which would otherwise keep them nearer to the common center of gravity. While this paper reports on the first explorative steps, future work will certainly be required to address the dynamical regime of $\sim 10-100$ merging solitons with repulsive SI included. In light of all the multiple computational demands, we confined ourselves to merger simulations with not more than 6 solitons, given that (central) densities and masses were of higher priority to our analysis in this paper.

\section{Configuration tables}\label{appendix:ConfiTables}

In section \ref{sec:KonfiAnalysis}, figure \ref{fig:KonfiAnalysis}, we used some abbreviations for configurations which were not explained. For illustration, we will provide the tables for the initial soliton parameters for these cases in this appendix. The mass of the individual solitons can be calculated, using their number and the total mass of the merger given in figure \ref{fig:KonfiAnalysis}.

The first scenario that was not discussed is the x-$3$D-case. This configuration describes a $2-$soliton merger, where one soliton moves along the x-axis, while the other heads from diagonally above. For both solitons the initial distance to the center of the box, where they collide, was $0.35$\,CU. The soliton parameters for this scenario are shown in table \ref{tab:x3D}.

\begin{table*}
\caption{\label{tab:x3D}The initial soliton parameters for a $(N=2)$ soliton merger, where one soliton falls diagonally onto another, which moves along the x-axis. For both solitons, the distance to the center is $\sim 13.5$\,kpc and the absolute value for the velocity is $\sim 2.5$\,km/s. The abbreviation for this scenario is "x-3D".}
\begin{ruledtabular}
\begin{tabular}{ccccccc}
soliton nr. & x & y & z & v$_{\text{x}}$ & v$_{\text{y}}$ & v$_{\text{z}}$ \\ \hline
$1$ & $0.20207$ & $0.20207$ & $0.20207$ & $-2.88675$ & $-2.88675$ & $-2.88675$ \\
$2$ & $-0.35$ & $0$ & $0$ & $5$ & $0$ & $0$
\end{tabular}
\end{ruledtabular}
\end{table*}

Another configuration for which we performed simulation runs is the \textit{square} scenario. As the name suggests, the solitons are arranged in a square shape in the x-y-plane with a distance to the center of the box of $0.4$\,CU. The associated soliton parameters can be found in table \ref{tab:Square}.

\begin{table*}
\caption{\label{tab:Square}The initial soliton parameters for a $(N=4)$ soliton merger, where the solitons are positioned in a two-dimensional square shape. The distance to the center for all solitons is $\sim 15$\,kpc and all of them move with a velocity of $\sim 2.5$\,km/s to the center. The abbreviation for this scenario is "Square".}
\begin{ruledtabular}
\begin{tabular}{ccccccc}
soliton nr. & x & y & z & v$_{\text{x}}$ & v$_{\text{y}}$ & v$_{\text{z}}$ \\ \hline
$1$ & $0.27772$ & $0.27772$ & $0$ & $-3.4715$ & $-3.4715$ & $0$ \\
$2$ & $0.27772$ & $-0.27772$ & $0$ & $-3.4715$ & $3.4715$ & $0$ \\
$3$ & $-0.27772$ & $-0.27772$ & $0$ & $3.4715$ & $3.4715$ & $0$ \\
$4$ & $-0.27772$ & $0.27772$ & $0$ & $3.4715$ & $-3.4715$ & $0$
\end{tabular}
\end{ruledtabular}
\end{table*}

Finally, we mention the "random merger" with $5$ solitons, which is similar to the random $6-$soliton merger, described in the main text. For illustration we include the soliton parameters in table \ref{tab:Random5}. The initial distance to the center for all solitons was $0.6$\,CU.

\begin{table*}[ht!]
\caption{\label{tab:Random5} The soliton parameters for a $(N=5)$ soliton merger, where the solitons are positioned "randomly" in the simulation box. The initial distance to the center for all solitons is $\sim 23$\,kpc and the absolute value of the velocity for each soliton is $\sim 2.5$\,km/s. The abbreviation for this scenario is "Random". We note that this is the same abbreviation as for the $6-$soliton merger, since the first $5$ solitons have the same initial parameters.}
\begin{ruledtabular}
\begin{tabular}{ccccccc}
soliton nr. & x & y & z & v$_{\text{x}}$ & v$_{\text{y}}$ & v$_{\text{z}}$ \\ \hline
$1$ & $0.3$ & $0.1$ & $-0.51$ & $-2.5$ & $-0.83333$ & $4.25$ \\
$2$ & $0.5$ & $-0.3$ & $0.14142$ & $-4.16666$ & $2.5$ & $-1.1785$ \\
$3$ & $-0.2$ & $0.26458$ & $-0.5$ & $1.66666$ & $-2.20483$ & $4.16666$ \\
$4$ & $0$ & $0.6$ & $0$ & $0$ & $-5$ & $0$ \\
$5$ & $0.3$ & $0.4$ & $0.33166$ & $-2.5$ & $-3.33333$ & $-2.76238$
\end{tabular}
\end{ruledtabular}
\end{table*}

All of these configurations exemplify the same behavior than the cases which we discussed in section \ref{sec:KonfiAnalysis}, as follows. The configurations here have a higher central density than their respective counterparts with one or two head-on collisions, except for the $2-$soliton merger, which can be explained (again) by having picked a "lucky" snapshot, where the oscillation yields a maximum in the center. Also, the profiles look much smoother than those of the head-on collisions, with fewer and lower-amplitude oscillations in their envelopes. As elaborated in the main text, mergers without head-on collisions develop in a more gentle manner. Additionally, mergers involving a higher number of solitons develop fewer oscillations, as well, since the higher total mass helps the merger to calm down more rapidly. Furthermore, such post-merger objects also have a more pronounced envelope, compared to those involving fewer solitons. Judging from our results, it appears that mergers involving at least four solitons mark that transition, but this statement also depends upon soliton mass and relative velocity.  
\newpage

\nocite{*}

\bibliography{References}

\end{document}